\documentclass [10pt]{article}
\usepackage{amsmath,amsthm,amsfonts,amscd,eucal,latexsym,amssymb,bm,amsbsy,mathrsfs,yhmath} 
\usepackage{epsfig}   
\input xy
\xyoption{all}
\def\sp{\hskip -5pt} 
\def\spa{\hskip -3pt} 

\newsymbol\bt 1202           

\def\cF{{\ca F}}
\def\cG{{\ca G}}

\def\cH{{\ca H}}
\def\cD{{\ca D}}
\def\cE{{\ca E}}

\def\cS{{\ca S}}

\def\cF{{\ca F}}

\def\cW{{\ca W}}

\def\sS{{\mathsf S}}

\def\bE{{\mathbb E}} 
\def\bC{{\mathbb C}}           

\def\bN{{\mathbb N}}

\def\bR{{\mathbb R}}

\def\bS{{\mathbb S}}

\def\bZ{{\mathbb Z}} 
 
\newsymbol \rest 1316         
 
\def\gH{{\mathfrak H}}

\def\mD{\mathscr{D}} 
\def\mS{\mathscr{S}}  
\def\mF{\mathscr{F}}  

\def\beq{\begin{eqnarray}}
\def\eeq{\end{eqnarray}}

\newcommand{\ca}[1]{{\cal #1}}         

\def\z{\zeta}
\def\bz{\overline{\zeta}}

\def\supp{\mbox{supp}\:}

\def\WF{\mbox{WF}}
\def\scri{\Im^+}         
\def\tg{\tilde{g}}
\def\tM{\tilde{M}}
\def\tE{\tilde{E}}

\def\tchi{\tilde{\chi}}

\def\lie{\pounds}

\def\bk{{\bf k}}

\newcounter{proposition}[section]
\newcounter{theorem}[section]
\newcounter{lemma}[section]
\newcounter{definition}[section]
\newcounter{remark}[section]

\def\theproposition{\thesection.\arabic{proposition}}
\def\thetheorem{\thesection.\arabic{theorem}}
\def\thelemma{\thesection.\arabic{lemma}}
\def\thedefinition{\thesection.\arabic{definition}}
\def\theremark{\thesection.\arabic{remark}}

\def\s #1 {\section{#1}}

\def\ssa #1 {\ifhmode{\par}\fi\refstepcounter{subsection}
  \noindent {\bf\thesubsection}. {\em #1}.\quad
  \addcontentsline{toc}{subsection}{\protect\numberline{\thesubsection} #1}%
  }

\def\ssb #1 {\ifhmode{\par}\fi\refstepcounter{subsection}
  \noindent {\bf\thesubsection.} {\em #1.}\quad
  \addcontentsline{toc}{subsection}{\protect\numberline{\thesubsection} #1}%
  }

\def\proposizione {\ifhmode{\par}\fi\refstepcounter{proposition}
  \noindent {\bf Proposition \theproposition}. \quad}
\def\teorema {\ifhmode{\par}\fi\refstepcounter{theorem}
  \noindent {\bf Theorem \thetheorem}. \quad}
\def\lemma {\ifhmode{\par}\fi\refstepcounter{lemma}
  \noindent {\bf Lemma \thelemma}. \quad}
\def\definizione {\ifhmode{\par}\fi\refstepcounter{definition}
  \noindent {\bf Definition \thedefinition}. \quad}
\def\remark {\ifhmode{\par}\fi\refstepcounter{remark}
  \noindent {\bf Remark \theremark}. \quad}

\begin{document} 
 
\hfill{\sl November 2006  Preprint  UTM-704, 3nd Revised version June 2007} \\

 
\par 
\LARGE 
\noindent 
{\bf  Quantum out-states holographically induced by asymptotic flatness: 
 Invariance under spacetime symmetries, energy positivity and Hadamard property.} \\
\par 
\normalsize 
  
 

\noindent {\bf Valter Moretti\footnote{E-mail: moretti@science.unitn.it}} 

\par
\small

\noindent  
Dipartimento di Matematica, Facolt\`a di Scienze M.F.N., Universit\`a di Trento, \\
 \& Istituto Nazionale di Alta Matematica ``F.Severi'' - Unit\`a Locale di Trento,\\
 \&  Istituto Nazionale di Fisica Nucleare - Gruppo Collegato di Trento,\\
  via Sommarive 14  
I-38050 Povo (TN), Italy. \smallskip \smallskip

 \normalsize

\small 
\noindent {\bf Abstract}. {This paper continues the analysis of the quantum states introduced 
in previous works and determined by the universal asymptotic structure of four-dimensional
asymptotically flat vacuum spacetimes at null infinity $M$. 
It is now focused on the quantum state $\lambda_M$, of a massless conformally coupled 
scalar field $\phi$ propagating in $M$. $\lambda_M$ is ``holographically'' induced in the bulk  by the universal
BMS-invariant state $\lambda$ 
defined on the future null 
infinity $\scri$ of $M$.  It is done by means of the correspondence
between observables in the bulk and those on the boundary at 
future 
null infinity discussed in previous papers.
This induction is possible when some requirements are fulfilled, in particular whenever 
the spacetime $M$ and the associated unphysical one, $\tM$, are globally hyperbolic and $M$ admits future time infinity $i^+$. 
$\lambda_M$ coincides with Minkowski vacuum if $M$
is Minkowski spacetime. It is now proved that, in the general case of a curved spacetime $M$, the state  $\lambda_M$ enjoys the following further remarkable properties.\\
{\bf (i)} $\lambda_M$ is invariant under the (unit component of the Lie) group of isometries of the bulk spacetime $M$.\\
{\bf (ii)} $\lambda_M$ fulfills a natural energy-positivity condition with respect to every notion of Killing time (if any) in the 
bulk spacetime $M$:
If $M$ admits a  time-like Killing vector, the associated one-parameter group of isometries is represented by 
a strongly-continuous unitary group in the GNS representation of $\lambda_M$. The unitary group has
positive self-adjoint generator without zero modes
in the one-particle space. In this case $\lambda_M$ is a so-called regular ground state.\\
{\bf (iii)}  $\lambda_M$ is (globally) Hadamard in $M$ and thus it can be used as starting point for perturbative renormalisation
procedure of QFT of $\phi$ in $M$.}

\normalsize

\section{Introduction} 
In this paper we continue the analysis of the states determined by the asymptotic structure of four-dimensional
asymptotically flat spacetimes at null infinity started in \cite{DMP} and fully developed in \cite{CMP5}.
Part of those results will be summarised in Sec.2.
In \cite{DMP} and \cite{CMP5} it has been established that the null boundary at future infinity $\scri$
of an asymptotically flat spacetime admits a natural formulation of bosonic linear QFT living therein. 
A preferred quasifree pure state $\lambda$ has been picked out in the plethora of algebraic states 
defined on the algebra  of Weyl observables $\cW(\scri)$ of the QFT  on $\scri$. 
That state enjoys remarkable properties, in particular 
it is invariant under the action of the natural (infinite-dimensional) group of symmetries of $\scri$ -- the so-called  
BMS group -- describing the asymptotic symmetries of the physical spacetime $M$.
$\lambda$ is the vacuum state for BMS-massless particles if one analyses the unitary representations of the 
BMS group within the Wigner-Mackey approach \cite{DMP}. $\lambda$ is uniquely determined by a positive BMS-energy
requirement in addition to the above-mentioned BMS invariance \cite{CMP5} (actually the latter requirement can be 
weakened considerably). Finally, in the folium of $\lambda$ there are no further pure BMS-invariant (not necessarily quasifree
or positive energy)
states. 
$\lambda$ is universal: it does not depend on the particular asymptotically flat spacetime under consideration
but it is defined in terms of the asymptotic extent which is the same for all asymptotically flat spacetimes.\\
However, in every fixed asymptotically flat spacetime $M$ and under suitable hypotheses on $M$ and $\tM$, $\lambda$
it induces a preferred quasifree state $\lambda_M$ on the algebra $\cW(M)$ of the Weyl observables of a bosonic massless 
conformally-coupled
linear field propagating in the bulk $M$.
This is because, due to a sort of holographic correspondence discussed in \cite{DMP,CMP5}, the algebra of bulk 
observables $\cW(M)$ is one-to-one mapped to a subalgebra of boundary
observables $\cW(\scri)$ by means of a (isometric) $*$-algebra homomorphism $\imath : \cW(M) \to \cW(\scri)$.
In this way $\lambda_M$ is defined as  $\lambda_M(a):= \lambda(\imath(a))$ for all $a\in \cW(M)$.
 The existence of $\imath$ is assured if some further conditions
defined in \cite{DMP} are fulfilled for the spacetime $M$ and the associated unphysical spacetime $\tM$ 
(see (b) of  Proposition \ref{holographicproposition}).
In particular both $M$ and $\tM$ are required to be globally hyperbolic.
 Those requirements are valid when $M$
is Minkowski spacetime and, in that case, $\lambda_M$ turns out to coincide with Minkowski vacuum \cite{DMP}.
However, it has been established in \cite{CMP5}, that the conditions are verified in a wide class of spacetimes 
(supposed to be globally hyperbolic with the unphysical spacetime $\tM$) individuated 
by Friedrich \cite{Friedrich}: the  asymptotically flat vacuum spacetimes admitting future time infinity $i^+$. \\
This paper is devoted to study the general features of the state $\lambda_M$.
In particular, is Sec.3 we focus on isometry-invariance properties of $\lambda_M$ 
 and on 
the energy positivity condition with respect to timelike Killing vectors in any bulk spacetime $M$ (Theorem \ref{teoinvariance}):
we show that 
$\lambda_M$ is invariant under the unit component of the Lie group of isometries of the bulk spacetime $M$. 
 This fact  holds true also replacing $\lambda_M$ with any other state $\lambda'_M$
uniquely defined by assuming that  $\lambda'_M(a):= \lambda'(\imath(a))$ for all $a\in \cW(M)$,
where $\lambda'$ is any (not necessarily quasi free or pure) BMS invariant state defined on $\cW(\scri)$.
Furthermore
$\lambda_M$ fulfils a natural energy-positivity condition with respect to every notion of Killing time in the 
bulk spacetime $M$:
If $M$ posses a time-like Killing vector $\xi$, the associated one-parameter group of isometries is 
represented by a strongly continuous unitary group in the GNS representation of $\lambda_M$, that group
admits a positive self-adjoint generator $H$  
and $H$ has no zero modes in the one-particle space.
In this sense the quasifree state $\lambda_M$ is a {\em regular ground state} for $H$ \cite{KW}.
Actually these properties are proved to hold also if $\xi$ is causal and future directed, but not
necessarily timelike.\\
Sec.4 is devoted to discuss the validity of the Hadamard condition for the state $\lambda_M$.
First  we show  that the two-point function of $\lambda_M$ is a proper distribution of $\mD'(M\times M)$
(Theorem \ref{propD'})
when the asymptotically flat spacetime $M$ admits future time infinity $i^+$. 
It is interesting noticing that the explicit expression of the two-point function of $\lambda_M$ we present
strongly resembles that of Hadamard states in manifolds with bifurcate Killing horizons studied by Kay and Wald
\cite{KW} when restricted to the algebra of observables localised on a Killing horizon.\\
The last result we establish in this work is that 
 $\lambda_M$ is Hadamard (Theorem \ref{lambdaH}).
 In this case  the kernel of the two-point function of $\lambda_M$ satisfies
  the {\em global Hadamard condition} \cite{KW}. The proof -- performed within
  the microlocal framework taking advantage of some well known result established by Radzikowski \cite{Rada,Radb} --
   is based on a
``from local to global'' argument and the analysis of the wavefront sets of the involved distributions.\\

\remark After the submission of this paper one of the referees pointed out to the author that 
a result of a very similar nature as Proposition \ref{propAA} was presented in Hollands' Ph.D thesis \cite{Hollands}. 
In that work, the definition of local
vacuum states near a point $x$ in spacetime is considered. The idea of the construction is
to define the state by specifying the "initial data" of its $2$-point
function on the (say, past-) lightcone throught $x$. The formula for this
initial datum is the formula (\ref{formulaN}) when the lightcone in
\cite{Hollands} is replaced by $\Im^-$ in as in our work. In \cite{Hollands} it is also established
that the individuated state is of Hadamard form (in the sense of the microlocal
condition (\ref{QLHN})) employing a proof which is very analogous to our proof.\\

The rest of this section is devoted to remind the reader the basic geometric structures in asymptotically flat spacetimes. \\

\ssa{Notations, mathematical conventions} \label{secgauge} 
Throughout $\bR^+:= [0,+\infty)$, $\bN:= \{0,1,2,\ldots\}$. For  smooth manifolds $M,N$, 
$C^\infty(M;N)$
(omitting $N$ whenever $N=\bR$) is the space of smooth functions $f: M\to N$.
$C^\infty_0(M;N)\subset C^\infty(M;N)$ is the subspace of compactly-supported functions.  
If $\chi : M\to N$ is a diffeomorphism, $\chi^*$ is the natural extension to tensor bundles 
(counter-, co-variant and mixed) from $M$ to $N$ (Appendix C in \cite{Wald}).
A spacetime is a smooth four-dimensional semi-Riemannian 
connected manifold $(M,g)$, whose metric has signature $-+++$,
and it is assumed to be oriented and time oriented.
We adopt definitions of causal structures of Chap. 8 in \cite{Wald}.
  If $S\subset M\cap \tM$, $(M,g)$ and $(\tM,\tg)$ being spacetimes, $J^\pm(S;M)$ ($I^\pm(S;M)$) and $J^\pm(S;\tM)$
 ($I^\pm(S;\tM)$) indicate the causal (chronological) sets associated to $S$ and respectively referred to the spacetime 
 $M$ or $\tM$.  Concerning distribution and wavefront-set theory we essentially adopt standard definitions and notation
 used in \cite{Hor,Hor1} and also in \cite{Rada,Radb}.\\

\ssa{Asymptotic flatness at future null infinity and $\scri$} \label{intro} Following \cite{AH,AO,Wald},
a smooth spacetime $(M,g)$ is called {\bf asymptotically flat vacuum spacetime at future null infinity} if there is 
a second smooth spacetime  $(\tilde{M},\tilde{g})$ such that $M$ can be viewed as an open
embedded submanifold of $\tilde{M}$ with boundary $\scri \subset \tM$.  $\scri$ is an embedded 
submanifold of $\tM$ satisfying $\scri \cap J^-(M; \tilde{M}) = \emptyset$.
$(\tM,\tg)$ is required to be strongly causal in a neighbourhood of $\scri$ and 
it has to hold $\tilde{g}\spa\rest_M= \Omega^2 \spa\rest_M g\spa\rest_M$ where $\Omega \in C^\infty(\tM)$
is strictly positive on $M$. On $\scri$ one has $\Omega =0$ and $d\Omega \neq 0$. 
Moreover, defining $n^a := \tg^{ab} \partial_b \Omega$, 
there must be a smooth function, $\omega$, defined in $\tM$ with $\omega >0$ on $M\cup \scri$, such that 
$\tilde{\nabla}_a (\omega^4 n^a)=0$ on $\Im$ and the integral lines of $\omega^{-1} n$ are complete on $\scri$.
The topology of  $\scri$ has to be that of $\bS^2\times \bR$. 
 Finally vacuum Einstein 
equations are assumed to be fulfilled for $(M,g)$ in a neighbourhood of $\scri$ or, more weakly, 
``approaching'' $\scri$ as discussed on p.278 of \cite{Wald}.\\
Summarising  $\scri$ is a 
$3$-dimensional submanifold of $\tilde{M}$ which is the union of  integral lines of the nonvanishing  null field 
$n^\mu:= \tilde{g}^{\mu\nu}\nabla_\nu \Omega$, these lines are complete
for a certain regular rescaling of $n$,
 and $\scri$ is equipped with a degenerate metric $\tilde{h}$ induced by $\tilde{g}$. $\scri$ is called {\em future infinity} of $M$.  
\remark  For  brevity, from now on 
{\bf asymptotically flat spacetime} means {\em  asymptotically flat vacuum spacetime at future null infinity}. \\

\noindent $\scri$ is the {\em conic surface} indicated by $\:\mbox{I}^+\:$ in the figure. 
The tip is {\em not} a point of $\tM$.
\begin{figure}[th]
\begin{center}
\includegraphics[bb=0 0 500 270, scale=.7]{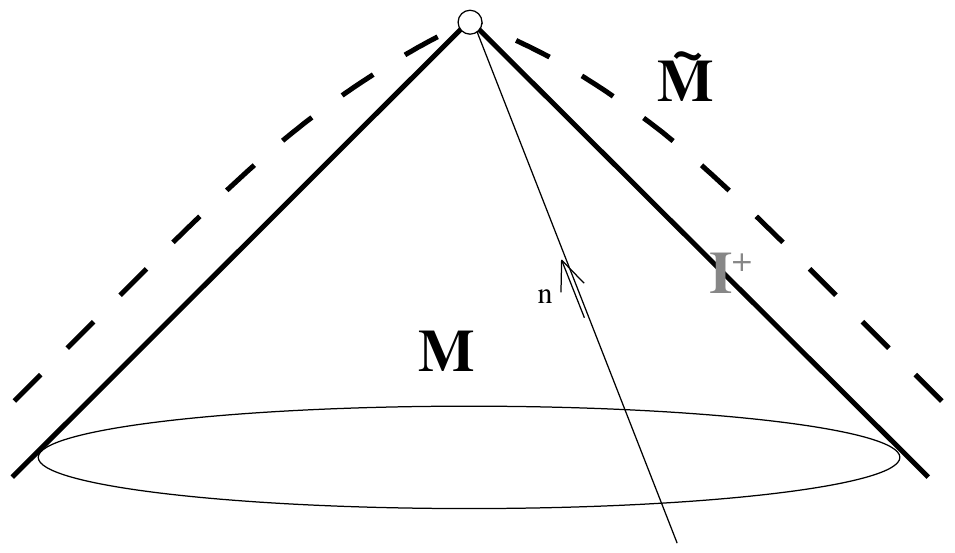}
\end{center}
\caption{Asymptotically flat spacetime. ($\scri$ is indicated by $I^+$ in the figure.)}
\end{figure}
 Minkowski spacetime and Schwarzschild spacetime are well-known examples of asymptotically flat spacetimes.
It is simply proved -- for instance reducing to the Minkowski space case --
 that, with our conventions, the null vector  $n$ is always {\em future directed} with respect to the time-orientation of $(\tM,\tg)$ induced 
 from that of $(M,g)$.\\
As far as the geometric structure on $\scri$ is concerned,  
changes of the unphysical spacetime $(\tilde{M}, \tilde{g})$, associated with a {\em fixed} asymptotically flat spacetime $(M,g)$, are
completely encompassed
by {\bf gauge transformations} $\Omega \to \omega \Omega$ valid in a neighbourhood of $\scri$,  
with $\omega$ smooth and 
strictly positive. Under these gauge transformations the triple $(\scri,\tilde{h}, n)$ transforms as
\beq
\scri \to \scri \:,\:\:\:\:\: \tilde{h} \to \omega^2 \tilde{h} \:,\:\:\:\:\: n \to \omega^{-1} n \label{gauge}\:.
\eeq
If $C$ is the class of  the triples $(\scri,\tilde{h}, n)$ transforming as in (\ref{gauge})
for a fixed asymptotically flat spacetime,
there is no general physical principle to single out a preferred element in $C$.
On the other hand, $C$ is {\em universal} for all asymptotically flat spacetimes \cite{Wald}:
If $C_1$ and $C_2$ are the classes of 
triples associated respectively to $(M_1,g_2)$
and $(M_2,g_2)$, there is a diffeomorphism $\gamma: \scri_1 \to \scri_2$ such that for suitable $(\scri_1,\tilde{h}_1, n_1)\in C_1$
and $(\scri_2,\tilde{h}_2, n_2)\in C_2$: $\quad \gamma(\scri_1) = \scri_2 \:,\:\:\:\:\: \gamma^* \tilde{h}_1=\tilde{h}_2 \:,\:\:\:
\:\:\gamma^* n_1=n_2$.\\
Choosing $\omega$ such that $\tilde{\nabla}_a (\omega^4 n^a)=0$
-- this choice is permitted in view of  the very definition of 
asymptotically flat spacetime -- and using the fact that vacuum Einstein's equations are fulfilled in a neighbourhood of $\scri$,
the tangent vector $n$ turns out to be that 
of {\em complete null geodesics} with respect to $\tg$ (see Sec. 11.1 in \cite{Wald}).  
$\omega$ is completely fixed by requiring that, in addition,
the non-degenerate metric on the transverse section of $\scri$
is the standard metric of $\bS^2$ in $\bR^3$ constantly along geodesics. We indicate by $\omega_B$ and
$(\scri,\tilde{h}_B,n_B)$ that value of $\omega$ and the associated triple respectively. 
For $\omega = \omega_B$,
a {\bf Bondi frame} on $\scri$ is a coordinate system   $(u,\z,\bz)$ on $\scri$, 
where $u\in \bR$ is an affine parameter of the complete 
null $\tilde{g}$-geodesics whose union is $\scri$ ($n= \partial/\partial u$ in these coordinates) 
and $\z,\bz$ are standard complex coordinates on the cross section $\bS^2$ of 
$\scri$ when $\bS^2$ is identified with the Riemann complex sphere : 
$\z= e^{i\varphi}\cot(\theta/2)$ with $\theta, \varphi$ usual spherical coordinates of $\bS^2$
(see \cite{KND,Mc} for the use of these complex coordinates in the specific case of $\scri$).  Obviously $\z$ and $\bz$ bring the same information as two 
{\em real} independent
coordinates as in {\em Wirtinger notation} in the theory of complex-variable functions.
With these choices, the metric on the transverse section of $\scri$ reads $2(1+\z\bz)^{-2}(d\z\otimes d\bz+ d \bz \otimes d\z)
= d\theta \otimes d\theta + \sin^2 \theta \: d\varphi\otimes d\varphi$.\\
 By definition $\chi: \scri \to \scri$ belongs to the {\bf BMS group}, $G_{BMS}$ \cite{Penrose, Penrose2, Geroch, AS},  if $\chi$ is a diffeomorphism and 
 $\chi^*\tilde{h}$ and $\chi^*n$ differ from  $\tilde{h}$ and $n$ by a rescaling (\ref{gauge})  at most. 
Henceforth, whenever it is not explicitly stated otherwise,
{\em we consider as admissible realisation
of the unphysical metric on $\scri$ only those metrics $\tilde{h}$ which are accessible 
from a metric with associate
 triple $(\scri,\tilde{h}_B,n_B)$, by means of 
a transformations in $G_{BMS}$ }.

In coordinates of a fixed Bondi frame $(u,\z,\bz)$, the group  $G_{BMS}$ is realised as semi-direct group product 
$SO(3,1)\sp\uparrow \sp \ltimes C^\infty(\bS^2)$, where
 $(\Lambda, f) \in SO(3,1)\sp\uparrow \times C^\infty(\bS^2)$ acts as
\begin{eqnarray}
u &\to & u':= K_\Lambda(\z,\bz)(u + f(\z,\bz))\:,\label{u}\\
\z &\to & \z' :=\Lambda\z:= \frac{a_\Lambda\z + b_\Lambda}{c_\Lambda\z +d_\Lambda}\:, \:\:\:\:\:\:
\bz \: \to \: \bz' :=\Lambda\bz := \frac{\overline{a_\Lambda}\bz + \overline{b_\Lambda}}{\overline{c_\Lambda}\bz +\overline{d_\Lambda}}\:.
\label{z}
\end{eqnarray}
$K_\Lambda$ is the smooth positive function on $\bS^2$
 \begin{eqnarray}
 K_\Lambda(\z,\bz) :=  \frac{(1+\z\bz)}{(a_\Lambda\z + b_\Lambda)(\overline{a_\Lambda}\bz + \overline{b_\Lambda}) +(c_\Lambda\z +d_\Lambda)(
 \overline{c_\Lambda}\bz +\overline{d_\Lambda})}
\label{K}\:\: \: \mbox{and}\:\:\:\:
 \left[
\begin{array}{cc}
  a_\Lambda & b_\Lambda\\
  c_\Lambda & d_\Lambda 
\end{array}
\right] = \Pi^{-1}(\Lambda)\:.
\end{eqnarray}
Above $\Pi$ is the well-known surjective covering homomorphism $SL(2,\bC) \to SO(3,1)\sp\uparrow$ (see \cite{DMP} for further
details). 
Two Bondi frames are connected each other through the transformations 
(\ref{u}),(\ref{z}) with $\Lambda \in SU(2)$. 
Conversely, any coordinate frame $(u',\z',\bz')$ on $\scri$ connected to a Bondi frame $(u,\z,\bz)$
 by means of an arbitrary
BMS transformation (\ref{u}),(\ref{z}) is {\em physically equivalent} to the Bondi frame 
from the point of view of General Relativity, but it is not necessarily a Bondi frame in turn.
A global reference frame $(u',\z',\bz')$ on $\scri$
related with a Bondi frame $(u,\z,\bz)$ by means of a  BMS transformation (\ref{u})-(\ref{z}) will be called
{\bf admissible frame}. 
By construction, the action of $G_{BMS}$ takes the form (\ref{u})-(\ref{z}) in admissible fames too.
The notion of Bondi frame is useful but conventional.  
{\em Any physical object must be invariant under 
the whole BMS group and not only under the subgroup of $G_{BMS}$ connecting Bondi frames}. 

 The local one-parameter group of diffeomorphisms generated by a (smooth) vector field $\xi$ defined in 
an asymptotically flat spacetime $(M,g)$
is called {\bf asymptotic Killing symmetry} if
(i) $\xi$ extends smoothly to a field $\tilde{\xi}$ tangent to $\scri$
and (ii) $\Omega^2 \lie_\xi g$ has a smooth extension to $\scri$ which vanishes there. 
This is the best approximation of a Killing symmetry for a generic asymptotically
flat spacetime which does {\em not} admits proper Killing symmetries (see e.g. \cite{Wald}). 
The following well-known result illustrates how $G_{BMS}$ describes asymptotic Killing symmetries
valid for every asymptotically flat spacetime.
 \cite{Geroch,Wald}.\\

\proposizione\label{prop1} {\em Let $(M,g)$ be asymptotically flat. 
The one-parameter group of diffeomorphisms generated by a vector field $\tilde{\xi}$ tangent to $\scri$  is a subgroup of $G_{BMS}$ if and only if  
$\tilde{\xi}$ is the smooth extension to $\scri$ of some vector field of $(M,g)$ defining an asymptotic Killing symmetry of $(M,g)$.}

\section{Summary of some previously achieved results.}

\ssa{Quantum fields on $\scri$}\label{ssa1} Let us summarise how a natural linear QFT can be defined on $\scri$
employing the algebraic approach and the GNS reconstruction theorem.
Motivations for the following theoretical construction and more details can be found in \cite{DMP,CMP5}.\\
Referring to a fixed Bondi frame on $\scri$, consider the real symplectic space
 $(\sS(\scri),\sigma)$, where
\beq \sS(\scri) := \left\{ \left.\psi \in C^\infty(\scri)\: \:\right|\:\: \psi\:, \partial_u 
\psi \in  L^2(\bR\times \bS^2, du \wedge \epsilon_{\bS^2}(\z,\bz)) \right\} \:,
\quad  \epsilon_{\bS^2} (\z,\bz):= 
\frac{2d\z \wedge d\bz}{i(1+\z\bz)^2} \:,
\eeq
$\epsilon_{\bS^2} (\z,\bz)$ being the standard volume form of the unit $2$-sphere,
and the nondegenerate symplectic form $\sigma$ is given by, if $\psi_1,\psi_2 \in \sS(\scri)$
\beq \sigma(\psi_1,\psi_2) := \int_{\bR\times \bS^2} 
\left(\psi_2 \frac{\partial\psi_1}{\partial u}  - 
\psi_1 \frac{\partial\psi_2}{\partial u}\right) du \wedge \epsilon_{\bS^2}(\z,\bz)\:.\eeq

There is a natural representation of $G_{BMS}$ acting on $(\sS(\scri),\sigma)$ discussed in \cite{DMP,CMP5}. 
Start from the representation $A$ of $G_{BMS}$ made of transformations on functions $\psi \in C^\infty(\scri)$
\beq (A_g\psi)(u,\z,\bz):=K_\Lambda\left(g^{-1}(u,\z,\bz)\right)^{-1}\:\psi\left(g^{-1}(u,\z,\bz)\right)\:,\quad 
\mbox{where $g= (\Lambda,f)$.} \label{ABMS}\eeq
It turns out that $A_g(\sS(\scri)) \subset \sS(\scri)$. Moreover, 
due to the weight $K_\Lambda^{-1}$, the $G_{BMS}$ representation $A$ {\em preserves} the symplectic form $\sigma$.
As a consequence the space $(\sS(\scri),\sigma)$ does not depend on the used Bondi frame. In this context it is convenient to assume 
that the elements of $\sS(\scri)$ are densities which transform under the action of $A$ when one changes admissible frame.
In the following the restriction $A_g\spa\rest_{\sS(\scri)}$ will be indicated by $A_g$ for the sake of simplicity.
Naturalness and relevance of the representation $A$ follow from the content of Proposition \ref{asympprop} below as discussed in \cite{DMP,CMP5}. 

As is well known \cite{BR,BR2}, it possible to associate canonically any symplectic space, for instance  $(\sS(\scri),\sigma)$, with a
{\bf Weyl $C^*$-algebra}, $\cW(\sS(\scri),\sigma)$. This is the, unique up to (isometric) $*$-isomorphisms,
 $C^*$-algebra with generators $W(\psi)\neq 0$, $\psi \in \sS(\scri)$, satisfying {\bf Weyl commutation relations} (we use here
 conventions adopted in \cite{Wald2})
\beq W(-\psi)= W(\psi)^*\:,\quad\quad W(\psi)W(\psi') = e^{i\sigma(\psi,\psi')/2} W(\psi+\psi')\:.\eeq 
Here $\cW(\sS(\scri)) :=  \cW(\sS(\scri),\sigma)$ has the natural interpretation of the algebra of observables for a linear bosonic QFT
defined on $\scri$ as discussed in \cite{DMP,CMP5} (see also the appendix A of \cite{CMP5}).\\
The representation $A$ induces \cite{BR2} a $*$-automorphism $G_{BMS}$-representation
 $\alpha : \cW(\sS(\scri)) \to \cW(\sS(\scri))$, uniquely individuated (by linearity and continuity) 
  by  the requirement
$\alpha_g(W(\psi)):= W(A_{g^{-1}}\psi)$ for all $\psi \in \sS(\scri)$ and $g\in G_{BMS}$.

Since we expect that physics is BMS-invariant we face the issue about the existence 
of  $\alpha$-invariant algebraic states on $\cW(\sS(\scri))$. To this end it has been established in \cite{DMP} that there is
 at least one algebraic 
  quasifree\footnote{We adopt the definition of {\bf quasifree state} given in \cite{KW}, and also adopted in \cite{DMP,CMP5}, summarised
  in the appendix A of \cite{CMP5}.} pure state $\lambda$ defined on $\cW(\sS(\scri))$ which is invariant under $G_{BMS}$. It is that
  uniquely induced by linearity and continuity from:
\beq 
\lambda(W(\psi)) = e^{-\mu_\lambda(\psi,\psi)/2}\:,\quad \mu_\lambda(\psi_1,\psi_2):=  -i \sigma(\overline{\psi_{1+}},\psi_{2+})\:,
\quad\psi \in \sS(\scri) \label{scriiii}\eeq
the bar over $\psi_+$ denotes 
 the complex conjugation,
$\psi_+$ being the {\bf positive $u$-frequency part} of $\psi$ computed with respect to the Fourier-Plancherel
transform discussed in the Appendix \ref{APPfourier}:
$$\psi_+(u,\z,\bz) := \int_{\bR} \frac{e^{-i ku}}{\sqrt{2\pi}} \Theta(k) \widehat{\psi}(k,\z,\bz) 
dk \:, 
\quad (k,\z,\bz)\in \bR \times \bS^2\:.$$ 
Everything is referred to an arbitrarily fixed Bondi frame $(u,\z,\bz)$ and
$\Theta(k)=0$ for $k< 0$ and $\Theta(k) =1$ for $k\geq0$.
Consider the GNS representation of $\lambda$, $(\gH, \Pi, \Upsilon)$.
Since $\lambda$ is quasifree,  $\gH$ is a bosonic Fock space $\cF_+(\cH)$ 
with cyclic vector $\Upsilon$ given by the Fock vacuum
and $1$-particle Hilbert $\cH$ space generated by  
the positive-frequency parts of $u$-Fourier-Plancherel
 transforms $\widehat{\psi}_+ := \Theta \widehat{\psi}$. In other words one has that 
$\cH \equiv L^2(\bR^+\times \bS^2; 2kdk \wedge \epsilon_{\bS^2})$. Indeed it arises from (\ref{scriiii}):
 \beq
 \langle \psi_+, \psi'_+\rangle = \int_{\bR\times \bS^2} 2k\Theta(k)\overline{\widehat{\psi}(k,\z,\bz)} \widehat{\psi'}(k,\z,\bz) dk\wedge
 \epsilon_{\bS^2}(\z,\bz)  \label{added}\:.
 \eeq
 In \cite{DMP,CMP5} we used a different, {\em but unitarily equivalent}, definition of positive frequency part
 in Fourier variables (we used, in fact, positive frequency parts defined, in Fourier variables, as 
 $\widetilde{\psi}_+(E,\z,\bz) := \sqrt{2E}\widehat{\psi}_+(E,\z,\bz)$
 so that $\cH \equiv L^2(\bR^+\times \bS^2; dE \wedge \epsilon_{\bS^2})$.)

  $\lambda$ is a {\em regular} state, that is self-adjoint  {\bf symplectically-smeared field operators}  
$\sigma(\Psi,\psi)$  are defined via Stone's theorem:
 $\Pi(W(t\psi)) = e^{-it\sigma(\Psi,\psi)}$ 
with $t\in \bR$ and $\psi\in \sS(\scri)$.

As a remarkable result, it has been established in \cite{DMP} that, equipping $G_{BMS}$  with a suitable Fr\'echet topology,
the unique unitary representation $U$ of $G_{BMS}$
leaving $\Upsilon$ invariant and implementing $\alpha$ 
is strongly continuous. Its restriction to $\cH$ (which is invariant under $U$)
is an irreducible and strongly continuous
 Wigner-Mackey  representation associated with a $0$-elicity representation of the 
 little group given by the double covering of $2D$ Euclidean group.
 The little group is the same as in the case of massless Poincar\'e particles.
As a matter of facts,  in the space of characters of $G_{BMS}$, where a generalisation of Mackey machinery works
\cite{Mc,AD,Da04,Da05,Da06} (notice that $G_{BMS}$ is not locally compact),
 there is a a notion of {\em mass}, $m_{BMS}$, which is invariant under the action of $G_{BMS}$.
 It turns out that  the found $G_{BMS}$ representation is defined over an orbit in the space of characters 
 with $m_{BMS}= 0$. So we are dealing with BMS-invariant massless particles.
 
 $\lambda$ enjoys some further properties, in particular a uniqueness property, which will be re-visited later with a 
 point of view different from that adopted in \cite{CMP5}. As it should be clear, the existence of $\lambda$ has no 
 relation with the presence or the absence of Killing 
 symmetries in $M$. This $BMS$-invariant state exists on $\scri$ which admits the BMS group as symmetry group in any cases
since we are dealing with asymptotically flat spacetimes. The interplay of $\lambda$ and the  symmetries of $M$ (if any)
will be discussed in section \ref{SM}.\\

\ssa{Interplay with massless particles propagating in the bulk spacetime} We want now to summarise some achieved results in
\cite{DMP,CMP5}) on the interplay of QFT defined on $\scri$ and that defined in the bulk $M$, for a massless conformally coupled scalar field.
Consider an asymptotically flat spacetime $(M,g)$ with associated unphysical spacetime 
$(\tM,\tg= \Omega^2 g)$. In addition to asymptotic flatness assume also that both $M,\tM$ are {\em globally hyperbolic}. 
Consider standard bosonic QFT in $(M,g)$ 
based on the symplectic space $(\cS(M),\sigma_M)$, where $\cS(M)$ is the space of real
smooth, compactly supported on Cauchy surfaces, solutions $\phi$ of massless, conformally-coupled,  Klein-Gordon equation in $M$:
\beq  P\phi =0\:, \quad \mbox{where $P$ is the Klein-Gordon operator  $P=\Box_g  - \frac{1}{6} R$}\:.\label{PKG}\eeq
The Cauchy-surface independent symplectic form $\sigma_M$ is:
\beq \displaystyle \sigma_M(\phi_1,\phi_2) := \int_S \left(\phi_2 \nabla_N \phi_1 - \phi_1 \nabla_N \phi_2\right)\: 
d\mu^{(S)}_g \:,\eeq 
$S$ being any
Cauchy surface of $M$ with normal unit future-directed vector $N$ and $3$-volume measure $d\mu^{(S)}_g$ induced by $g$.  
Henceforth  the {\bf Weyl algebra associated with the symplectic space $(\cS(M),\sigma_M)$}, whose 
{\bf Weyl generators} are indicated by $W_M(\phi)$, $\phi \in \cS(M)$, will be denoted by $\cW(M)$. That $C^*$-algebra represents
the basic set of quantum observables associated with the bosonic field $\phi$ propagating in the bulk 
spacetime $(M,g)$.  The generators $W_M(\phi)$ are formally interpreted as the exponentials $e^{-i\sigma_M(\Phi,\phi)}$
where $\sigma_M(\Phi,\phi)= -\sigma_M(\phi,\Phi)$ is the {\bf field operator  symplectically smeared} 
with a solution $\phi\in \cS(M)$ of field equations
(concerning the sign of $\sigma$ we employ conventions used in \cite{Wald2} which differ from those adopted in \cite{KW}).
The interpretation has a rigorous meaning referring to a GNS representation of $\cW(M)$. If the considered state $\omega$
is {\bf regular}, the unitary group $\bR \ni t\mapsto \Pi_\omega(W(t\psi))$ is strongly continuous and
$-i\sigma_M(\Phi,\phi)$ can be defined as the self-adjoint generator of it.
The more usual field operator $\Phi(f)$ {\bf smeared with functions} $f\in C^\infty_0(M)$
is related with $\sigma_M(\Phi,\phi)$ by means of $\Phi(f) := \sigma_M(\Phi,E(f))$, where the {\bf causal propagator}
$E: C_0^\infty(M) \to C^\infty(M)$
is the difference of the advanced and retarded fundamental solutions of  Klein-Gordon equation
which exist in every globally hyperbolic spacetime \cite{Leray,Dimock,BGP}.
$\Phi$ solves Klein-Gordon equation in distributional 
sense: $\Phi(Pf)=0$ because $E\circ P=0$ by definition.\\
The relation between QFT in $M$ and that defined on $\scri$ can be now illustrated as follows (simplified form of 
Proposition 1.1 in \cite{CMP5}) joined to Proposition 2.5 in \cite{DMP}.\\

\proposizione \label{holographicproposition}
{\em  Assume that both the asymptotically flat spacetime $(M,g)$  and the unphysical spacetime 
 $(\tM,\tg)$ are globally hyperbolic. The following holds.\\
 {\bf (a)} Every $\phi \in \cS(M)$ vanishes approaching $\scri$, but 
 $(\omega\Omega)^{-1}\phi$ extends to a smooth field,   $\omega$ being any (arbitrarily fixed) positive 
function defined in a neighbourhood of $\scri$ associated with a gauge transformation of the geometry 
on $\scri$ (see section \ref{secgauge}). \\ For the special case $\omega=\omega_B$ we define
the $\bR$-linear map 
\begin{center}$\Gamma_M : \cS(M) \ni \phi \mapsto \left((\omega_B \Omega)^{-1}\phi\right)\rest_{\scri}$.\end{center}
 {\bf (b)} If $\Gamma_M$ fulfils both the following requirements:\\
{\bf (i)} $\Gamma_M(\cS(M))\subset \sS(\scri)$ and {\bf (ii)} symplectic forms are preserved by $\Gamma_M$, that is, for all $\phi_1,\phi_2 \in \cS(M)$, it holds 
$\sigma_{M}(\phi_1,\phi_2) = \sigma(\Gamma_M\phi_1,\Gamma_M\phi_2)$,\\
then $\cW(M)$ can be identified with a sub $C^*$-algebra of $\cW(\scri)$
by means of a $C^*$-algebra isomorphism $\imath$ uniquely determined by the requirement
\beq
\imath(W_{M}(\phi)) = W(\Gamma_M \phi)\:, \:\:\:\:\mbox{for all $\phi\in \cS(M)$} \label{lc}\:.
\eeq}
\noindent In other words, if (i) and (ii) are valid, the field observables of the bulk $M$ can be identified 
with certain observables of the boundary $\scri$. This is a sort of holographic correspondence.\\
If $(M,g)$ is Minkowski spacetime (so that $(\tM,\tg)$ is Einstein closed universe), hypotheses
(i) and (ii) are fulfilled so that $\imath$ exists \cite{DMP}. However there is a large 
class of asymptotically flat spacetimes which fulfil hypotheses (i) and (ii) as proved in Theorem 4.1 in \cite{CMP5}. 
They are the asymptotically flat spacetimes, which are globally hyperbolic together
with the associated unphysical spacetime and such that {\em admit future time infinity} $i^+$ in the sense of 
Friedirich \cite{Friedrich}.
Roughly speaking we may define an {\bf asymptotically flat vacuum spacetime with future time infinity $i^+$}
as an asymptotically flat vacuum spacetime at future null infinity $(M,g)$ such that
there is a point $i^+$ in the chronological future of $M$ in $\tilde{M}$,  $i^+ \not \in \scri$,
such that the geometric extent of $\scri \cup \{i^+\}$ about $i^+$ ``is the same as 
that in a region about the tip $i^+$ of a light cone in a (curved) spacetime''. The precise definition is stated 
in the appendix B (see also the discussion in \cite{CMP5}).
\begin{figure}[th]
\begin{center}
\includegraphics[bb=0 0 500 270, scale=.7]{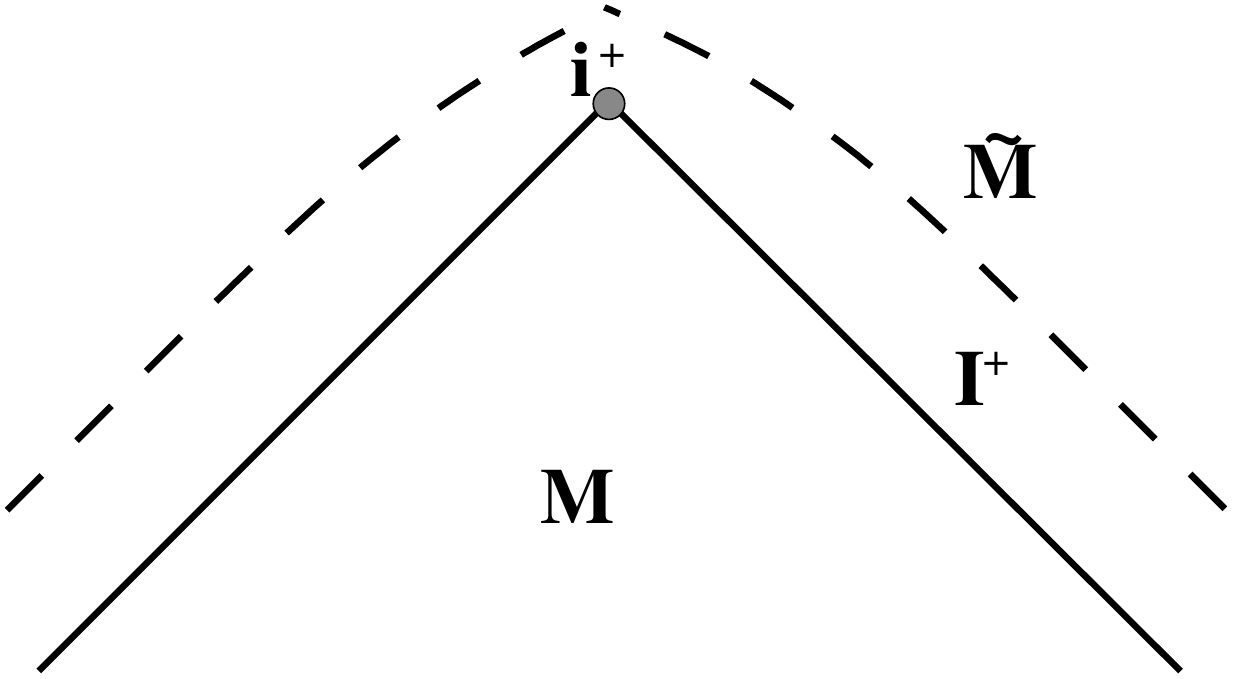}
\end{center}
\caption{Asymptotically flat spacetime with future time infinity $i^+$. ($\scri$ is indicated by $I^+$ in the figure.)}
\end{figure}

\section{The state $\lambda_M$: invariance under isometries and energy positivity.} \label{SM}
A straightforward but very important consequence of Proposition \ref{holographicproposition} is that, whenever
(i) and (ii) are fulfilled,  {\em the $G_{BMS}$-invariant quasifree pure state $\lambda$ defined on $\scri$ 
can be pulled back 
to a  state 
$\lambda_M$ (quasifree by construction) acting on bulk observables} defined by:
\beq\lambda_M(a):= \lambda(\imath(a))\quad \mbox{for all $a\in \cW(M)$\:.} \label{lambdaM}\eeq
If $(M,g)$ is Minkowski spacetime,  it turns out that {\em $\lambda_M$ coincides with Minkowski vacuum}.
The main goal of this paper is to study the general properties of  $\lambda_M$ whenever it can be defined.\\
\remark The assignment of the state $\lambda_M$ has actually quite a little overlap  with the difficult problem 
of finding 
the solution of a hyperbolic field equation in $\tM$ with general Cauchy data assigned on the null surface 
$\scri$ (the so-called {\em Goursat problem}). Here we  are instead dealing with fields on $\scri$ obtained by extending to 
$\scri$ solutions of field equations in $M$, determined by (compactly supported) Cauchy data assigned on
 Cauchy surfaces in $M$.  The central point is that the symplectic form of the solutions in $M$ is preserved passing 
 from $M$ to $\scri$ (under suitable hypotheses in particular the presence of $i^+$). 
 This only fact individuates the $*$-homomorphism $\imath$ of field algebras which associates the observable
 algebra of the bulk $M$ with a subalgebra of the observables on the boundary $\scri$ as stated in Proposition 
\ref{holographicproposition}. 
 Since algebraic states are functionals on the corresponding algebras, a state 
 on the algebra on $\scri$ individuates automatically a state of the algebra in the bulk as in (\ref{lambdaM}).\\
 
  In the rest of this section we prove, in particular, that the state $\lambda_M$ is invariant under the 
 Killing symmetries of $M$ if any and that the self-adjoint generator associated with any Killing time
 of the bulk must have positive energy (Theorem \ref{teoinvariance} below). The proof is partially based on the known nice interplay between bulk 
 isometries and BMS symmetries, 
 established by several authors in the past and discussed into details in the next section (Proposition \ref{prop2} below). Summarising, 
 the procedure is the following.
 If $M$  admits a Killing symmetry $g_M$ (associated with  a Killing field everywhere defined on $M$), 
 this field extends to $\scri$ 
and individuates an associated transformation, say $g$, of the BMS group. This correspondence is injective. 
Now one notices that 
the isometries of the bulk 
have a natural action $\beta^{(M)}_{g_M}$ on the algebra of observables of the bulk and, 
on the other hand, the {\em associated}
 BMS symmetries
on $\scri$ have a natural action $\beta_{g}$ on the algebra of observables defined on $\scri$. The remarkable fact, 
 partially established in
\cite{DMP}, is that (Proposition \ref{asympprop} below) the actions of these corresponding transformations, respectively
in the bulk $M$ and on $\scri$, commute with the map $\imath$ which identifies the elements $a$ of the bulk algebra with 
the elements $\imath(a)$ of  the boundary algebra: $\imath(\beta^{(M)}_{g_M} (a)) = \beta_g(\imath(a))$.
Since $\lambda(\imath(a)) = \lambda_M(a)$ by definition and since 
$\lambda$ is invariant under the whole $BMS$ group, $\lambda_M$ must in turn be invariant under the bulk isometries:
$\lambda_M(\beta^{(M)}_{g_M}(a))= \lambda(\imath (\beta^{(M)}_{g_M}(a)) ) = \lambda(\beta_g \imath(a)) = 
\lambda (\imath(a)) = \lambda_M(a)$.\\
In this context, the energy positivity property of $\lambda_M$ with respect to timelike Killing vectors in the bulk
descends from the fact that isometries generated by timelike Killing vectors of $M$ corresponds to BMS symmetries of particular type 
(timelike $4$-translations studied in Section \ref{4T}).
It has been proved in \cite{CMP5} that $\lambda$ admits positive energy with respect to that type of transformations.
Essentially by means of the previously mentioned machinery, it arises that the state $\lambda_M$ admits positive energy with respect
to any notion of Killing time in the bulk.\\

\ssa{The spaces of supertranslations, $4$-translations and interplay with bulk symmetries}\label{4T}
In this section we introduce some notions and results, missed in \cite{DMP,CMP5}, which will play a central role 
in studying the properties of $\lambda_M$ mentioned in the previous section.
We focus on the {\bf internal action} of $G_{BMS}$
$G_{BMS} \ni \alpha \mapsto g\circ \alpha \circ g^{-1}$, for any fixed $g\in
 G_{BMS}$.
The decomposition of $h \in G_{BMS}$ as a pair  $(\Lambda,f)\in SO(3,1)\sp\uparrow  \times C^\infty(\bS^2)$ depends on the used 
admissible frame. However
the factor $\Sigma:= C^\infty(\bS^2)$ is $BMS$-{\bf invariant}, i.e. invariant under the above-mentioned
internal action for every fixed $g\in G_{BMS}$\footnote{In other words
 $\Sigma$ and the subsequent subgroup $T^4$ are {\em normal} subgroups
of $G_{BMS}$.} and thus it is well-defined 
independently from the used  admissible frame,  since admissible frames are connected to each other by  BMS transformations:
If $h \in G_{BMS}$ belongs to $C^\infty(\bS^2)$ (i.e. has the form $(I,\alpha)$ with $\alpha \in \Sigma$) when referring to an admissible frame, 
the same result holds referring to any other admissible frame.
$\Sigma$ is called the {\bf group of supertranslations}.
However there is another, more important normal subgroup of both $G_{BMS}$ and $\Sigma$. 
 That is the {\bf group of $4$-translations}:
 \beq T^4 := \left\{\left.\alpha = \sum_{j=0,1}\sum_{|m|\leq j} c_{jm} Y_{jm}\:\:\right|\:\: c_{jm} \in \bC\:, \alpha(p) \in \bR\:,
 \forall p\in \bS^2 \right\}\:,\eeq
 $Y_{jm}$ being the standard spherical harmonics normalized with respect to the measure of the unit sphere $\bS^2$.
 $T^4$ turns out to be --once-again -- $BMS$-{\bf invariant} and thus, like $\Sigma$, it is well-defined 
independently from the used admissible frame.
Notice that $T^4$ enjoys the structure of real vector space in addition to that of additive group.
By direct inspection one sees that the internal  action of $G_{BMS}$ on $T^4$ defines in fact a representation 
of $G_{BMS}$ made of {\em linear} transformations with respect to the real-vector-space structure of $T^4$.
It is possible to pass from the complex basis of $T^4$, $\{Y_{jm}\}_{j=0,1, |m| \leq J}$ to a real basis
$\{Y_{\mu}\}_{m=0,1,2,3}$ (see \cite{DMP} and references cited therein for more details\footnote{In Eq. (3.19) in \cite{DMP} the statement
 ``if $1<k\leq l$'' has to be corrected to 
``if $1 \leq k\leq l$'', whereas in the right-hand side of subsequent 
Eq. (3.20), $Y_{l\:-k}$ and  $Y_{l\:k}$ have to be corrected to $Y_{l\:-(k-l)}$ and  $Y_{l\:(k-l)}$ respectively.}),
with $Y_0:= \sqrt{2/\pi}$ , $Y_1= -\sqrt{2/\pi}\sin\theta \cos \varphi$,
 $Y_2= -\sqrt{2/\pi}\sin\theta \sin \varphi$, $Y_3= -\sqrt{2/\pi}\cos \theta$. Referring to that basis,
if $\alpha := \sum_\mu \alpha^\mu Y_\mu$ and $\tilde{\alpha} := \sum_\mu \tilde{\alpha}^\mu Y_\mu$
 the Lorentzian scalar product $<\alpha,\tilde{\alpha}>_{BMS} := -\alpha^0\tilde{\alpha}^0 + \alpha^1\tilde{\alpha}^1+
 \alpha^2\tilde{\alpha}^2+\alpha^3\tilde{\alpha}^3$, turns out to be 
$BMS$-{\bf invariant} with respect to the above-mentioned internal (and linear) action of $G_{BMS}$.
As a consequence $T^4$ results to be equipped with a {\em light cone structure}: 
there is a $BMS$-{\bf invariant} decomposition of $T^4\setminus\{0\}$ into {\bf spacelike}, {\bf timelike} and
{\bf null} $4$-translations. Every fixed admissible frame $(u,\z,\bz)$ individuates a time-orientation of $T^4$.
Indeed, consider the $BMS$ diffeomorphism  associated with a
positive rigid translations of $\scri$,  $\alpha_\tau:  \bR\times \bS^2 \ni (u,\z,\bz) \mapsto (u+\tau,\z,\bz)$, ($\tau > 0 $ fixed).
Looking at (\ref{u})-(\ref{z}) one finds that, trivially, 
$\alpha_\tau$ identifies with $\tau \sqrt{\pi/2} Y_{0}\in T^4$. Since $\tau>0$,
$\alpha_\tau$ picks out the same half of the light-cone not depending on $\tau$. 
This choice for time-orientation is not affected by changes in the used admissible frame.  
This is because $n= \partial/\partial u$ is always future-directed
with respect to the time-orientation of $(\tM,\tg)$ induced by that of $(M,g)$ when working in a Bondi frame. The action 
of $BMS$ group, to pass to a generic admissible frame, does not changes the extent as a consequence of (\ref{u})-(\ref{z})
as one can check by direct inspection.
Therefore the light cone in $T^4$ has a natural preferred time-orientation. With our definition
of time-orientation of $T^4$, if $\alpha \in T^4$ is causal and future-directed, its action on $\scri$ displaces the points toward the very 
future defined in $(\tM,\tg)$ by
the time orientation of $(M,g)$.

 The $G_{BMS}$-subgroup $SO(3,1)\sp\uparrow \sp \ltimes T^4$ is 
 isomorphic to the proper orthochronous Poincar\'e group. However, differently from $T^4$, that group is not normal  and  different 
 admissible frames select different copies of $G_{BMS}$-subgroup isomorphic to the proper orthochronous Poincar\'e group.
 
  We are now ready to state a key result concerning the interplay of BMS group and symmetries. 
The following proposition is obtained by collecting together several known results but spread in the literature.
In the Appendix \ref{AWF} there is a proof of the statement (c). The results in (a)-(b) can be made much more 
strong as established in \cite{AX}. However
we do not need here stronger statements than (a)-(b).\\

\proposizione\label{prop2} {\em Let $(M,g)$ be an asymptotically flat spacetime. The following facts hold.\\
{\bf (a)} \cite{Geroch}  If $\xi$ is a Killing vector field of $(M,g)$, then $\xi$ 
smoothly  extends to a vector field on $\tM$. The restriction to $\scri$, $\tilde{\xi}$, of such an extension 
is  tangent to $\scri$,  is uniquely determined by $\xi$, and  generates a one-parameter subgroup of $G_{BMS}$. \\
{\bf (b)} \cite{AX}  The  linear map $\xi \mapsto \tilde{\xi}$ defined in (a)
fulfils the following properties:

(i) it is injective ($\tilde{\xi}$ is the zero vector field on $\scri$ only if $\xi$ is the zero vector field in $M$);

(iii) if, for a fixed $\xi$, the one-parameter $G_{BMS}$-subgroup generated by $\tilde{\xi}$ lies in $\Sigma$ then, more strictly,
 it must be a subgroup of $T^4$.\\
{\bf (c)} Consider an one-parameter subgroup of $G_{BMS}$, $\{g_t\}_{t\in \bR} \subset \Sigma$. Suppose that
$\{g_t\}_{t\in \bR}$ arises from the integral curves of a smooth vector $\tilde{\xi}$ tangent to $\scri$.
 Then, in any fixed Bondy frame one has:
 $$g_t : \bR \times \bS^2 \ni (u,\z,\bz) \mapsto \left(u+ tf(\z,\bz), \z, \bz\right)\:,$$
where the function $f \in C^\infty(\bS^2)\equiv \Sigma$ individuates completely the subgroup\:.}\\ 
 
\noindent Notice that the fields $\tilde\xi$ associated with one-parameter subgroup of $G_{BMS}$ are always complete 
since the parameter of the generated one-parameter subgroup
ranges in the whole real line by definition. This would be false in case of incompleteness of the field $n$.\\
  The following proposition can be established  by direct inspection from (c) in Proposition \ref{prop2}
 and (\ref{u})-(\ref{z}).\\

\proposizione \label{prop3} {\em  Consider
 a nontrivial one-parameter subgroup of $G_{BMS}$, $\{g_t\}_{t\in \bR} \subset T^4$
generated by a smooth complete vector $\tilde{\xi}$ tangent to $\scri$. The following facts
 hold true referring to the  time-oriented light-cone structure of $T^4$  defined above.\\
  {\bf (a)}  $\{g_t\}_{t\in \bR}$ is made of future-directed timelike $4$-translations
  if and only if there is an admissible frame $(u,\z,\bz)$ such that the action of $\{g_t\}_{t\in \bR}$ reduces there to:
 $g_t: (u,\z,\bz) \mapsto (u+t,\z,\bz)\:, \quad \mbox{$\forall t\in \bR$\:.}$\\
 {\bf (b)} $\{g_t\}_{t\in \bR}$ is made of future-directed causal $4$-translations
  if and only if there is a Bondi frame $(u,\z,\bz)$ and constants $c>0$, $a\in \bR$ with $|a|\leq 1$,
  such that the action of $\{g_t\}_{t\in \bR}$ reduces there to
 \beq g_t: (u,\z,\bz) \mapsto \left(u+t c \left(1 -a\frac{\z\bz-1}{\z\bz+1}\right) ,\z,\bz\right)\:, 
 \quad \mbox{$\forall t\in \bR$\:.}\label{gentr}\eeq
These translations  are null if and only if  $|a| = 1$. 
  They are timelike for $|a| < 1$.\\
{\bf (c)} A $4$-translation of $T^4 \setminus\{0\}$  viewed as 
 a function $f \in C^\infty(\bS^2)$  in any arbitrarily fixed admissible frame: 
 
 (i)  is spacelike if and only if $f$ attains both signs,
 
 (ii) is timelike and future-directed  if and only if $f$ is strictly positive, 
 
 (iii) is null and future-directed  if and only if $f$ is positive and vanishes on a single point of $\bS^2$.}\\

\noindent  Propositions \ref{prop3} and \ref{prop2} have the following technical consequence relevant for our goal.\\

\proposizione\label{prop2bis} {\em Let $\xi$ be a Killing vector of an asymptotically flat spacetime $(M,g)$. Then:\\
{\bf (a)} $\xi$ individuates an asymptotic Killing symmetry as expected;\\
{\bf (b)} If $\xi$ is everywhere causal future-oriented, the associated one-parameter subgroup
of $G_{BMS}$ is made of causal future-directed elements of $T^4$.}\\
  
\noindent{\em Proof}. The proof is in the Appendix \ref{AWF}. $\Box$\\

\ssa{Isometry invariance and energy positivity of $\lambda_M$} We go to prove that the state $\lambda_M$
is invariant under any isometry generated by the Killing vector $\xi$ of the bulk spacetime $M$. Moreover we prove that  
the spectrum of the self-adjoint generator associated with $\xi$ is positive whenever $\xi$ is timelike and thus 
the generator may be interpreted as an Hamiltonian with positive energy as it is expected from physics.
Positivity of the spectrum of the Hamiltonian is a {\em stability requirement}: it guarantees that, under small (external) perturbations, the system 
 does not collapse to lower and lower energy states.\\ 
The proof of invariance of $\lambda_M$ is based 
on the following remarkable result.\\

\proposizione \label{asympprop} {\em Assume that both the asymptotically flat spacetime $(M,g)$  and the unphysical spacetime 
 $(\tM,\tg)$ are globally hyperbolic and consider the linear map  $\Gamma_M : \cS(M) \to C^\infty(\scri)$ 
 in Proposition 
 \ref{holographicproposition} and the $BMS$ representation $A$ defined in (\ref{ABMS}).\\
If a complete vector field $\xi$ on $(M,g)$ smoothly extends to $\tM$ and defines the asymptotic 
Killing symmetry
 $\{g^{(\xi)}_t\}_{t\in \bR}$, then
the action of that asymptotic symmetry on the field $\phi$ in $M$ is equivalent to 
the action of a BMS-symmetry on the  
associated field $\psi := \Gamma_M \phi$ on $\scri$ via the representation $A$: 
\beq   \Gamma_M(\phi \circ g^{(\xi)}_{-t})  = A_{g^{(\tilde{\xi})}_t}(\psi)
\quad \mbox{for all $t\in \bR$ if $\psi = \Gamma_M\phi$ with $\phi \in \cS(M)$}\:, \label{centro}
\eeq
where  $\{g^{(\tilde{\xi})}_t\}_{t\in \bR}$ 
is the one-parameter subgroup of $G_{BMS}$ generated by the smooth extension $\tilde{\xi}$ to $\scri$ of $\xi$.}\\

\noindent{\em Proof}. The proof is in the Appendix \ref{AWF}. $\Box$\\

\noindent Notice that, in general,  $\phi \circ g^{(\xi)}_{-t}$ does not belong to $\cS(M)$ if $\phi$ does. However it happens
when $g^{(\xi)}_{t}$ is an isometry, since Klein-Gordon equation and thus $\cS(M)$ are invariant under isometries of $(M,g)$. \\
We now prove one of the main results of this work. As is known the {\em identity component} $\cG_1$ of a Lie 
group $\cG$ is the subgroup made of the connected component of $\cG$ containing the unit element of $\cG$.\\

\teorema \label{teoinvariance} {\em Assume that both the asymptotically flat spacetime $(M,g)$  and the unphysical spacetime 
 $(\tM,\tg)$ are globally hyperbolic and conditions (i) and (ii) in (b) of Proposition \ref{holographicproposition} are fulfilled.
 Consider the quasifree state $\lambda_M$ canonically induced on $\cW(M)$ from the BMS-invariant quasifree pure 
 state $\lambda$ defined on $\scri$. The following
 facts are valid.\\
 {\bf (a)} $\lambda_M$ 	coincides with (free) Minkowski vacuum if $(M,g)$ is Minkowski spacetime.\\
 {\bf (b)} $\lambda_M$ is invariant under the identity component $\cG_1$ of the Lie group $\cG$ of isometries of $M$:
 \beq
 \lambda_M(\beta_g a) = \lambda_M(a)\:, \quad \mbox{for all $a\in \cW(M)$ and every $g\in \cG_1$\:,}
 \eeq
 where $\beta$ is the (isometric) $*$-isomorphism representation of $\cG$ uniquely induced 
(imposing linearity and continuity) by the requirement on Weyl generators
$$\beta_g(W(\phi)) := W(\phi\circ g^{-1})\:, \quad \mbox{for every $\phi \in \cS(M)$ and $g\in \cG$}\:.$$
 Thus, in particular the Lie-subgroup $\cG_1$ admits unitary implementation in the
  GNS representation of $\lambda_M$.\\
{\bf (b)'} (b) holds for any state $\lambda'_M$ with
$\lambda'_M(a):= \lambda'(\imath(a))$, $\forall a\in \cW(M)$,
where $\lambda'$ is any BMS invariant state (not necessarily quasifree or pure or satisfying some positivity-energy condition) 
defined on $\cW(\scri)$. \\
 {\bf (c)} Assume that $(M,g)$ admits a complete causal future-directed  Killing vector $\xi$. 
The  unitary one-parameter group which leaves the cyclic vector fixed and 
 implements the one-parameter group of isometries generated by $\xi$ 
 in the GNS (Fock) space of $\lambda_M$ satisfies the following properties:
 
(i) it is strongly continuous, 

(ii) the associated self-adjoint generator, $H^{(\xi)}$,  has nonnegative spectrum,

(iii) the restriction of $H^{(\xi)}$ to the one-particle space 
has no zero modes.}\\

\remark 
{\bf (1)}  Concerning in particular the statement (c),  when  $\xi$ is timelike and future-directed,  $H^{(\xi)}$ provides
a natural (positive) notion of {\em energy},
 associated with $\xi$ displacements.\\
Since $\lambda_M$ is quasifree, its GNS representation is a Fock representation. 
When $\xi$ is timelike, the collection of properties (i), (ii)  are summarised \cite{KW} by saying that $\lambda_M$ is a 
{\bf ground state}. Then property (iii) states that $\lambda_M$ 
is a {\bf regular} ground state if adopting terminology in \cite{KW}. \\
{\bf (2)} The notion of ``completeness'' adopted in Proposition \ref{holographicproposition}
and in (c) of Theorem \ref{teoinvariance} has the standard meaning: the  vector field 
(which is defined everywhere on $M$,  and satisfies 
Killing's constraints in the second case) generates a {\em global} one-parameter  group of isometries, i.e. 
the parameter of the group  must range from $-\infty$ to $+\infty$ along every integral line. 
This requirement allows one to avoid problems with the associated one-parameters group of transformations 
acting on functions with domain in $M$: The functions are extended objects 
and one would prevent problems concerning the domains of the parameter of integral
lines since, in principle they could depend on the considered integral curve.\\

\noindent {\em Proof of Theorem \ref{teoinvariance}}. (a) It was proved in Theorem 4.5 \cite{DMP}. (b) It is well-known \cite{O'Neill} 
that there is a unique way to assign a Lie-group
structure to the group of isometries $\cG$ of a (semi-)Riemannian manifold $(M,g)$ such that the action of the one-parameter
subgroup is jointly smooth when acting on the manifold. Moreover the Lie algebra of $\cG$ is that of the
complete (i.e. generating global group of isometries) Killing vectors of
$(M,g)$. Finally using the exponential map one sees that every element of the identity component $\cG_1$ can be obtained as 
a finite product 
of elements which belong to one-parameter subgroups. As a consequence, to establish the validity of (b) it is sufficient 
to prove that $\lambda_M$ is invariant under the one-parameter subgroups generated by Killing vectors of $(M,g)$.
Let us prove 
it.
Let $\xi$ be complete Killing vector of $(M,g)$ and $\tilde{\xi}$ the associated generator of $G_{BMS}$ on $\scri$
in view of Proposition \ref{prop2}. 
Employing the same notation as in Proposition \ref{asympprop} and using the definition (\ref{lambdaM}), one achieves:
$$\lambda_M\left(W_M(\phi \circ g^{(\xi)}_{-t})\right) = \lambda\left( W( A_{g^{(\tilde{\xi})}_t}(\psi))\right)\:.$$
The right hand side is, by definition,
$$\lambda\left( \alpha_{g^{(\tilde{\xi})}_t} (W(\psi))\right) = \lambda\left(W(\psi)\right)\:,$$
where, in the last step we have used the invariance of $\lambda$ under the representation $\alpha$ 
of BMS-group defined in Sec. \ref{ssa1}. Since $\psi= \Gamma_M \phi$ and using (\ref{lambdaM}) again we finally obtain that
$$\lambda_M\left(W_M(\phi \circ g^{(\xi)}_{-t})\right) = \lambda_M\left(W_M(\phi)\right)\:.$$
By linearity and continuity this result extends the the whole algebra $\cW(M)$:
$$\lambda_M(\beta_{g^{(\xi)}_t}(a)) = \lambda_M(a)\:, \quad\mbox{for every $a\in \cW(M)$\:.}$$
Since $\lambda$ is invariant  there is a unique unitary implementation of the representation $\beta$ in the GNS representation of $\lambda$
which leaves fixed the cyclic vector (e.g. see \cite{Araki}).   The proof of (b)' is 
the same as that given  for (b), replacing $\lambda_M$ with $\lambda'_M$.\\
(c) As $\lambda_M$ being quasifree, its GNS representation is a Fock representation 
(e.g. see the appendix A of \cite{CMP5}
and references cited therein, especially \cite{KW}). As a consequence it is sufficient to prove the positivity 
property for the restriction of the unitary group which represents the group of isometries in the one-particle space  $\cH_M$.
The GNS triple of $\lambda_M$ is obtained as follows. Consider the GNS triple of $\lambda$, $(\gH,\Pi, \Upsilon)$ where
$\gH = \cF_+(\cH)$ is the bosonic Fock space with one-particle space $\cH$. As said above, that space, is isomorphic to
the space of (Fourier transforms of the) $u$-positive frequency parts  
$L^2(\bR_+\times \bS^2, 2kdk \wedge \epsilon_{\bS^2}(\z,\bz))$ referring to a fixed Bondi frame $(u,\z,\bz)$, 
$k$ being the Fourier variable associated with $u$. 
Consider the Hilbert subspace $\cH_M$ of $\cH$ obtained by taking the closure of the complex span of the $u$-positive-frequency parts 
of the wavefunctions $\Gamma_M\phi$, for every $\phi \in \cS(M)$. Let $\gH_M = \cF_+(\cH_M)$ be the Fock space generated by 
$\cH_M$ which, in turn, is a Hilbert subspace of $\gH$. Notice that we are assuming that the vacuum vectors
 $\Upsilon_M$ and $\Upsilon$
coincide. By construction $\gH_M$ is invariant under $\Pi$ and $\Pi_M := \Pi\spa \rest_{\gH_M}$
is a $*$-representation of $\imath(\cW(M))$. Moreover $\Pi_M(\imath(\cW(M))) \Upsilon_M = \Pi_M(\imath(\cW(M))) \Upsilon$
is dense in $\gH_M$ by construction. By the uniqueness (up to unitary maps) property of the GNS triple, we conclude that
$(\gH_M,\Pi_M, \Upsilon_M)$ is the GNS triple of $\lambda_M$\footnote{Notice that $\Pi_M$ may be reducible also if $\Pi$ is irreducible: 
in other words $\lambda_M$ may be a mixture also if $\lambda$ is pure.}.\\
 Consider the unique unitary $G_{BMS}$ representation $U$ which acts on $\gH$ implementing $\alpha$ and leaving
 $\Upsilon (= \Upsilon_M)$ fixed. It is the unitary $BMS$ representation defined by linearity and continuity
 by the requirement:
 \beq U_g \Pi (W(\psi)) \Upsilon := \Pi(\alpha_g(W(\psi)))\Upsilon\:, \quad\mbox{for all $\psi \in \sS(\scri)$ and $g\in G_{BMS}$}\:.\label{rep}\eeq
 Since the space $\cS(M)$ is invariant under the group of isometries  $g^{(\xi)}_t$ and (\ref{centro}) holds true, it arises that
 $\alpha_{g^{(\tilde{\xi})}_t}(W(\psi)) \in \imath(\cW(M))$ if $W(\psi) \in  \imath(\cW(M))$ 
 and thus  (\ref{rep}) entails
 \beq U_{g^{(\tilde{\xi})}_t} \Pi_M (W_M(\phi)) \Upsilon_M := \Pi(\beta_{g^{(\xi)}_t}(W_M(\phi)))\Upsilon_M\:, 
 \quad\mbox{for all $\phi \in \cS(M)$ and $t\in \bR$}\:.\label{rep2}\eeq
  As a consequence of (\ref{rep2}) we can conclude that the unique unitary representation $U^{(\xi)}$ of $\{g^{(\xi)}_t\}_{t\in \bR}$ on $\gH_M$
 which  leaves $\Upsilon_M$ invariant, is nothing but the restriction of $U$ to $\{g^{(\tilde{\xi})}_t\}_{t\in \bR} \subset G_{BMS}$
 and $\gH_M \subset \gH$. This result allows us to compute explicitly the self-adjoint generator of the unitary
 representation of $\{g^{(\xi)}_t\}_{t\in \bR}$. The representation $U$ is obtained by tensorialization of a unitary 
 representation ${^{(1)}U}$ of $G_{BMS}$ working in the one particle space  \cite{DMP,CMP5} (notice that in those papers, as one-particle space,
  we 
 used the unitarily isomorphic space 
 $L^2(\bR^+\times \bS^2; dk \wedge \epsilon_{\bS^2})$ instead of $L^2(\bR^+\times \bS^2; 2kdk \wedge \epsilon_{\bS^2})$
 therefore the expression below looks different, but it is equivalent to that given in \cite{DMP,CMP5}):
\beq
  \left({^{(1)}U}_{(\Lambda,f)}\varphi\right)(k,\z,\bz) = e^{ikK_{\Lambda}(\Lambda^{-1}(\z,\bz))f(\Lambda^{-1}(\z,\bz))}
 \varphi\left(kK_{\Lambda}\left(\Lambda^{-1}(\z,\bz)\right),\Lambda^{-1}(\z,\bz)\right) \:, \label{repX}
 \eeq
 for every $\varphi \in L^2(\bR^+\times \bS^2; 2kdk \wedge \epsilon_{\bS^2})$ and $G_{BMS}\ni g \equiv (\Lambda,f)$.
The restriction of ${^{(1)}U}$ to $\{g^{(\tilde{\xi})}_t\}_{t\in \bR}$ and $\cH_M$ defines a unitary representation
of $\{g^{(\xi)}_t\}_{t\in \bR}$ whose tensorialization on $\cF_+(\cH_M)$ is the very representation ${^{(1)}U}$.
Notice that ${^{(1)}U}$ restricted to $\{g^{(\tilde{\xi})}_t\}_{t\in \bR}$ leaves invariant $\cH_M$ by construction because
$\cH_M$ is the closure of the span of vectors  $d/dt|_{t=0}\Pi (W(t\psi)) \Upsilon$ for $\psi \in \Gamma_M(\cS(M))$
(the derivative being computed using the Hilbert topology).
To conclude the proof it is sufficient to prove that the self-adjoint generator of 
$\left\{{^{(1)}U}\null_{g^{(\tilde{\xi})}_t } \rest_{\cH_M}\right\}_{t\in\bR}$ exists and has positive spectrum.\\
In our hypotheses $\{g^{(\tilde{\xi})}_t\}$ is a one-parameter group of causal future-directed $4$-translations. As a consequence,
selecting the Bondi frame as in (b) in Proposition \ref{prop3}, we have that
there are a real $a$ with $|a|\leq 1$ and a real $c>0$ such that
$$g^{(\tilde{\xi})}_t : (u,\z,\bz) \mapsto \left(u+t c \left(1 -a\frac{\z\bz-1}{\z\bz+1}\right) ,\z,\bz\right)\:, \quad \mbox{for every $t\in
\bR$.}$$
Therefore, if $\varphi \in \cH \equiv L^2(\bR^+\times \bS^2; 2kdk \wedge \epsilon_{\bS^2})$,
\beq
  \left({^{(1)}U}_{g^{(\tilde{\xi})}_t }\varphi\right)(k,\z,\bz) = e^{it kc \left(1 -a\frac{\z\bz-1}{\z\bz+1}\right)}
 \varphi\left(k ,\z,\bz\right) \:. \label{repX2}
 \eeq
Strong continuity is obvious
(also after restriction to $\cH_M$). Finally, using Lebesgue's dominate convergence to evaluate 
 the strong-operator topology derivative at $t=0$
of ${^{(1)}U}_{g^{(\tilde{\xi})}_t }$, one obtains that this derivative is $ih^{(\xi)}$ where
 the self-adjoint operator  $h^{(\xi)}$ reads
\beq (h^{(\xi)} \varphi) (k,\z,\bz) :=  kc \left(1 -a\frac{\z\bz-1}{\z\bz+1}\right)\varphi(k,\z,\bz)\:,\label{H}\eeq
with dense domain $\cD(h^{(\xi)})$ made of the vectors $\varphi \in L^2(\bR^+\times \bS^2; 2kdk \wedge \epsilon_{\bS^2})$
such that the right-hand side of (\ref{H}) belongs to $L^2(\bR^+\times \bS^2; 2kdk \wedge \epsilon_{\bS^2})$
again. In view of Stone Theorem $h^{(\xi)}$ is the self-adjoint generator of ${^{(1)}U}_{g^{(\tilde{\xi})}_t }$. Passing to work in polar coordinates:
\beq kc \left(1 -a\frac{\z\bz-1}{\z\bz+1}\right) =  kc \left(1 -a\cos\theta \right)\geq 0\label{pos}\eeq
because $k\in [0+\infty)$, $c>0$ and $a\in \bR$ with $|a|\leq 1$. Therefore
interpreting the integral below as a Lebesgue integral in  $F:= [0,+\infty) \times [0,\pi] \times [-\pi,\pi]$:
\beq\left\langle \varphi,h^{(\xi)} \varphi\right\rangle =
2c \int_{F} |\varphi(k,\theta,\phi)|^2  \left(1 -a\cos \theta\right) k^2  \sin^2 \theta \:dk d\theta d\phi 
\geq 0 \:, \quad \mbox{for all $\varphi \in \cD(h^{(\xi)})$.}\label{ssss}\eeq
This fact entails that the spectrum of $h^{(\xi)}$ is included in $[0,+\infty)$ via spectral theorem. The result remains unchanged when restricting
 ${^{(1)}U}_{g^{(\tilde{\xi})}_t}$ (and thus $h^{(\xi)}$) to the invariant Hilbert-subspace $\cH_M$.\\
Suppose that there is a zero mode of $h^{(\xi)}$, that is $\varphi \in \cH_M\setminus \{0\}$ with $h^{(\xi)} \varphi =0$.
 By (\ref{ssss}),
$$ \int_{F} |\varphi(k,\theta,\phi)|^2  \left(1 -a\cos \theta\right) k^2  \sin^2 \theta \:dk d\theta d\phi =0 \:,$$
The integrand is nonnegative on $F$ since (\ref{pos}) is valid, therefore
the integrand  must vanish almost everywhere in the Lebesgue measure of $\bR^3$. 
Since the function  $(k,\theta, \phi)\mapsto \left(1 -a\cos \theta\right) k^2  \sin^2 \theta$ is almost-everywhere
strictly positive  on
$F$, it has to hold $\varphi=0$ almost-everywhere. Thus $\varphi=0$ as an element of $L^2(\bR^+\times \bS^2; 2kdk \wedge \epsilon_{\bS^2})$. In other words  $h^{(\xi)}$ has no zero modes. $\Box$\\

 \ssa{Reformulation of the uniqueness theorem for $\lambda$} It is clear that there  are asymptotically flat spacetimes which 
 do not admit any isometry. In those cases 
the invariance property stated in  (a) and the positivity energy condition (c) of Theorem \ref{teoinvariance} are meaningless. 
However those statements remain valid if referring to the asymptotic theory based on QFT on $\scri$ and 
the universal state $\lambda$. Indeed $\lambda$ is invariant under the whole $G_{BMS}$ group 
-- which represents {\em asymptotic symmetries} of every asymptotically flat spacetime --
and $\lambda$ satisfies a positivity energy condition with respect to every smooth one-parameter subgroup 
 of $G_{BMS}$ made of future-directed  timelike or null $4$-translations -- which correspond to Killing-time evolutions
 whenever the spacetime admits a timelike Killing field, as established above. \\
 As proved in Theorem 3.1 in \cite{CMP5}, the energy positivity condition 
 with respect to timelike $4$-translations
  determines  $\lambda$ uniquely. We may restate the theorem into a more invariant form as follows.
 The possibility of such a re-formulation was already noticed in a comment in \cite{CMP5}, 
 here, using to the introduced machinery, we are able to do it explicitly\footnote{The author 
 is grateful to A. Ashtekar for suggesting this improved formulation of the theorem.}.\\

 \teorema \label{mainrev} {\em Consider a nontrivial one-parameter subgroup of $G_{BMS}$, $G :=\{g_t\}_{t\in \bR}$ 
 made of future-directed timelike $4$-translations, associated with a smooth complete vector tangent to $\scri$
 and let $\alpha^{(G)}$ be the one-parameter group of $*$-isomorphisms induced by $G$ on $\cW(\scri)$. \\
{\bf (a)} The BMS-invariant state $\lambda$ 
is the unique pure quasifree state on $\cW(\scri)$ satisfying both:

(i) it is invariant 
under $\alpha^{(G)}$,

(ii) the unitary group which implements $\alpha^{(G)}$ leaving fixed the cyclic GNS vector is strongly continuous with nonnegative
generator.\\
 {\bf (b)}  Let $\omega$ be a pure (not necessarily quasifree) state on $\cW(\scri)$ which is $BMS$-invariant 
 or, more weakly,  $\alpha^{(G)}$-invariant.
$\omega$ is the unique state on $\cW(\scri)$ satisfying both:

(i) it is invariant 
under $\alpha^{(G)}$, 

(ii) it belongs to the folium of $\omega$.}\\  
  
\noindent {\em Proof}. The proof is that given for Theorem 3.1 in \cite{CMP5} working 
  in the admissible frame, individuated in (a) of Proposition \ref{prop3}, where the action of $G$ reduces to
 $$g_t: (u,\z,\bz) \mapsto (u+t,\z,\bz)\:, \quad \mbox{$\forall t\in \bR$\:.} \quad \Box$$

 \section{The Hadamard property.}
 \ssa{Hadamard states} \label{Radres} It is well known that Hadamard states \cite{KW,Wald2} have particular physical interest in relation with
 the definition of physical quantities which, as the stress-energy tensor operator (e.g. see \cite{stress,HW2}),  
 cannot be represented in terms of elements of the Weyl algebra or the associated $*$-algebra of products of smeared field operators. 
In the last decade the deep and strong relevance of Hadamard states in local generally covariant 
QFT in curved spacetime 
has been emphasised from different points of view (e.g. see \cite{BFK,BF,HW1,BFV}).
Consider a quantum scalar real bosonic field $\phi$ propagating in a globally hyperbolic spacetime $(M,g)$ satisfying
 Klein-Gordon equation with  Klein-Gordon operator $P:= \Box + V(x)$ ($V$ being any fixed smooth real function) and
consider the quantisation procedure based on Weyl algebra approach.
The rigorous definition of Hadamard state $\omega$ for $\phi$
 has been given in \cite{KW} in terms of a requirement on the behaviour of the 
 singular part of the integral kernel of {\bf two-point function} of $\omega$. 
 The two-point function of $\omega$ is the bi-linear functional:
 \beq
\omega(f,g) := -\left.\frac{\partial^2}{\partial s \partial t}\left\{
 \omega\left(W_M(sEf + tEg) \right) e^{ist \sigma_M(Ef,Eg)/2}\right\}\right|_{s=t=0}\sp\sp \:, \quad \mbox{$f,g \in  C_0^\infty(M)\times
 C_0^\infty(M)$.}\label{tp}
 \eeq
  Above $E: C_0^\infty(M) \to \cS(M) \subset C^\infty(M)$ is the previously mentioned {\em causal 
  propagator}. The two-point function $\omega(f,g)$ exists if and only if the right-hand side of (\ref{tp}) 
  makes sense for every pair $f,g$. This happens in particular whenever the GNS representation of $\omega$
  is a Fock representation, that is when $\omega$ is quasifree \cite{KW} (see also appendix A in \cite{CMP5}).
   The proof is straightforward and it provides a heuristic motivation for the definition (\ref{tp}).  For a quasifree state $\omega$ 
   one finds
   \beq \omega(f,g)  = \langle \Upsilon_\omega,  \Phi(f)  \Phi(g) \Upsilon_\omega \rangle 
   \label{tp2}\:, \eeq
  where $\Upsilon_\omega$ is the cyclic GNS vector which, in this case, coincides with the Fock vacuum vector
and $\Phi(f)$ denotes the self-adjoint field operator smeared with the smooth function $f$ defined in the 
GNS Hilbert space $\gH_\omega$. 
  Notice that, since $E\circ P = P\circ E =0$ if the two-point function exists  one gets \cite{Wald2}:
  $$\omega(Pf,g) = \omega(f,Pg)  =0 \quad \mbox{(KG)}$$
  and, directly from (\ref{tp}),
  $$\omega(f,g) - \omega(g,f)= -i \int_M f(x) \left(E(g)\right)(x)\: d\mu_g(x)
  \quad \mbox{(Com)}\:.$$
  If the a two-point function $\omega(f,g)$ exists on $C_0^\infty(M)\times C_0^\infty(M)$,
   the integral kernel $\omega(x,y)$ is defined (if it exists at all) as the  function, generally singular and affected 
  by some $\epsilon \to 0^+$ prescription,
  such that
  \beq
  \int_{M\times M} \omega(x,y) f(x)g(y)\: d\mu_g(x) d\mu_g(y) = \omega(f,g) \:, \quad \mbox{for all $f,g \in
  C^\infty_0(M)$\:.}
  \eeq
   Referring to a quantum scalar real bosonic field $\phi$ propagating in a globally hyperbolic spacetime $(M,g)$ satisfying
 Klein-Gordon equation,  
  a quasifree state $\omega$ which admits two-point function 
  is said to be {\bf Hadamard} if its integral kernel $\omega(x,y)$ exists and satisfies the 
  {\bf global Hadamard prescription}.
  That prescription
  requires that  $\omega(x,y)$ takes a certain -- quite complicated -- form in a neighbourhood 
  of a Cauchy surface of the spacetime, as discussed in details in Sec. 3.3 of \cite{KW}. The global Hadamard condition
  implies the validity of the  {\bf local Hadamard condition} which states that, every point $p \in M$ 
  admits a (open geodesically convex normal) neighbourhood $\cG_p$, such that 
\begin{align}
\omega (x,y) = \mbox{w-}\sp\lim_{\epsilon \to 0^+} \left\{\frac{U(x,y)}{\sigma(x,y) + 2i\epsilon(T(x)-T(y)) + \epsilon^2} 
 +  V(x,y) 
 \ln (\sigma(x,y) + 2i\epsilon(T(x)-T(y)) + \epsilon^2)\right\}\nonumber \\
 + \omega_{\scriptsize reg}(x,y)\:, \quad \mbox{if $(x,y) \in \cG_p\times \cG_p$,}
 \nonumber \quad\quad\quad\quad\quad \mbox{(LH)}
\end{align}
where $\sigma(x,y)$ is the ``squared geodesic distance'' of $x$ from $y$, $T$ is any, arbitrarily fixed, 
time function increasing to the future 
 and $U$ and $V$ are locally well-defined quantities depending on the local geometry only. 
 Finally $\omega_{\scriptsize reg}$ is smooth and is, in fact,
 the only part of the two-point function determining the state.
 w-$\lim_{\epsilon \to 0^+}$ indicates that the limit as $\epsilon \to 0^+$ as to be understood 
 in weak sense, i.e. {\em after} the integration of $\omega(x,y)$ with smooth compactly supported functions $f$ and $g$. \\ 
  In a pair of very remarkable papers \cite{Rada,Radb} Radzikowski established several important results 
  about Hadamard states, in particular he found out a {\em microlocal characterisation} 
  of Hadamard states (part of Theorem 5.1 in \cite{Rada}, Henceforth $\mD'(N)$ is the space of distributions on $C_0^\infty(N)$): \\
  
  \proposizione \label{GA} {\em  In a globally hyperbolic spacetime $(M,g)$, consider a quasifree state 
  with two-point function $\omega$ (so that (KG) and (Com) are valid) defining a distribution of $\mD'(M\times M)$.
   The state is Hadamard if and only if  the wavefront set $WF(\omega)$  of the distribution 
  is 
  \beq
  WF(\omega) = \left\{ ((x,k_x),(y,-k_y)) \in T^*M\setminus {\bf 0} \times  T^*M\setminus {\bf 0} \:\:|\:\:
  (x,k_x) \sim (y,k_y)\:, k_x \vartriangleright 0\right\}
  \eeq
 where $(x,k_x) \sim (y,k_y)$  means that there is a null geodesic joining $x$ and $y$ with co-tangent vectors at $x$ and $y$
 given by $k_x$ and $k_y$ respectively, whereas $k \vartriangleright 0$ means that $k$ is causal and future directed. ${\bf 0}$
 is the zero section of the cotangent bundle.\\}
 
 \noindent A second result by Radzikowski, which in fact proved a conjecture by Kay, establishes that (immediate 
 consequence of Corollary 11.1 in \cite{Radb}):  \\

 \proposizione \label{LA}
{\em In a globally hyperbolic spacetime $(M,g)$, if the two-point function  of a quasifree state 
 is a distribution  $\omega \in \mD'(M\times M)$ satisfying (LH) when restricted to $\cG_p\times \cG_p$, for some open neighbourhood $\cG_P$ 
 of every point $p\in M$, then the state is Hadamard.} \\
 
 \noindent In the following we shall prove that, in the presence of $i^+$, the following results hold true.
  (i) $\lambda_M$ is a distribution of $\mD'(M\times M)$ and, making use of Radzikowski results,  (ii)  $\lambda_M$ is  Hadamard. 
 To tackle the item (i) we have to introduce some notions concerning a straightforward
 extension of Fourier-Plancherel transform theory for functions and distributions defined on $\scri \equiv \bR\times \bS^2$. This is done in the Appendix \ref{APPfourier}.  \\

 \ssa{The integral kernel of $\lambda_M$ is a distribution when  $(M,g)$ admits $i^+$}   
 Since the considered spacetimes are equipped, by definitions, with metrics and thus preferred volume measures, 
 here we assume that distributions of $\mD'(M\times M)$ work on smooth compactly-supported scalar fields 
of $\mD(M):= C_0^\infty(M)$ as in \cite{friedlander} 
 instead of smooth compactly-supported scalar densities as in \cite{Hor1}. As is well known this choice is pure matter of convention
since the two points of view are equivalent.\\
 First of all we prove that $\lambda_M$ individuates a distribution in $\mD'(M\times M)$, i.e. it is continuous in the relevant weak 
 topology \cite{friedlander}
 whenever the spacetime $(M,g)$ is a vacuum asymptotically flat at null infinity spacetime and admits
 future temporal infinity $i^+$. We give also a useful explicit expression for the distribution.\\

\teorema \label{propD'}
{\em  Assume that  the spacetime $(M,g)$ 
is an asymptotically flat vacuum spacetime with future time infinity $i^+$ (Definition \ref{dlast}) and that 
both $(M,g)$ and the unphysical spacetime 
 $(\tM,\tg)$ are globally hyperbolic.
 Let $E: C_0^\infty(M) \to \cS(M)$ be the causal propagator associated with the real  massless conformally-coupled
  Klein-Gordon operator $P$  defined by Eq.(\ref{PKG}) on $(M,g)$. 
  Then the following facts are valid concerning the state $\lambda_M$ defined in Eq.(\ref{lambdaM}).\\
 {\bf (a)} Referring to a Bondi frame $(u,\z,\bz)$ one has
 \beq
 \lambda_M(f,g) = \lim_{\epsilon \to 0^+} -\frac{1}{\pi}\int_{\bR^2\times \bS^2} \frac{\psi_f(u,\z,\bz)
\psi_g(u',\z,\bz)}{\left(u-u' - i\epsilon \right)^2} \: du \wedge du' \wedge \epsilon_{\bS^2}(\z,\bz) \label{formula}\:,
 \eeq
 where $\psi_h := \Gamma_M (Eh)$ for all $h\in C^\infty_0(M)$ with $\Gamma_M : \cS(M) \to \sS(\scri)$
  defined in Proposition \ref{holographicproposition}.\\
  {\bf (b)} The two-point function of the state $\lambda_M$ individuates 
 a distribution of $\mD'(M\times M)$.\\}\\

 \remark It is intriguing noticing that the expression (\ref{formula}) is  the same as
  that  for two-point functions of quasifree {\em Hadamard} states obtained in \cite{KW} (Eq. (4.13))
 in globally hyperbolic spacetimes with bifurcate Killing horizon. In that case
  the null $3$-manifold $\scri$  is replaced by a bifurcate Killing horizon, the $2$-dimensional
 cross section $\bS^2$  with spacelike 
  metric  
   corresponds to the bifurcation surface $\Sigma$ with spacelike metric and, finally, 
  the null geodesics forming $\scri$, parametrised by the affine parameter $u$, 
  correspond to the null geodesics forming the Killing horizon parametrised by the affine parameter $U$.\\
 
 \noindent {\em Proof of Theorem \ref{propD'}}.
From now on we use the notations and the theory about Fourier-Plancherel transform presented in the Appendix \ref{APPfourier}.
In particular $\mF : \mS'(\scri) \to \mS'(\scri)$ denotes the extension to distributions of $\mF_+$
as stated in (d) in Theorem \ref{fourier} whose inverse, $\mF^{-1}$, is the analogous extension of $\mF_-$. 
We call $\mF$ {\bf Fourier-Plancherel transformation}, also if, properly speaking this name should be reserved to its 
restriction to $L^2(\bR\times \bS^2, du\wedge \epsilon_{\bS^2}(\z,\bz))$ defined in (e) in Theorem \ref{fourier}.
We also use the formal distributional notation for $\mF$ (and the analog for $\mF^{-1}$) 
$$\mF[\psi](k,\z,\bz) := \int_{\bR} \frac{e^{i ku}}{\sqrt{2\pi}} \psi(u,\z,\bz) du \:, $$ 
 regardless if $f$ is  a function or a distribution.  Throughout the notation
$\widehat{\psi}(k,\z,\bz)$ is also used for the Fourier(-Plancherel, extension to distributions) transform 
$\mF[\psi](k,\z,\bz)$. \\
 We start with a useful lemma.\\
 
 \lemma \label{lemmaLemme} 
 {\em In the hypotheses of theorem above, if $h\in C_0^\infty(M)$, the following holds.\\
 {\bf (a)} $\psi_h$ can be written  in terms of the causal propagator $\tilde{E}$ for the massless conformally 
 coupled Klein-Gordon operator $\tilde{P}$ in $(\tM, \tg= \Omega^2 g)$ and the smooth function $\omega_B>0$
 defined on $\scri$ (see Section \ref{intro}):
   \beq  {\psi_h}(u,\z,\bz) =  \omega_B(u,\z,\bz)^{-1}\tilde{E}(\Omega^{-3} h)\spa\rest_\scri(u,\z,\bz)\:,
   \quad \mbox{for $u\in \bR$ and $(\z,\bz) \in \bS^2$.}\label{CENTRAL}\eeq 
{\bf (b)} For any compact $K\subset M$ there is $u_0\in \bR$  such that, if $\supp h \subset K$,
 ${\psi_h}(u,\z,\bz)=0$ for $u<u_0$  and all $(\z,\bz) \in \bS^2$.}\\
 
 \noindent{\em Proof}. The proof is in the Appendix \ref{AWF}. $\Box$\\

\noindent  Let us pass to the main proof.\\
 (a) We start from the fact that, as found in the proof of Theorem \ref{teoinvariance},
 the Fock GNS triple of $\lambda_M$, $(\gH_M, \Pi_M, \Upsilon_M)$ is such that
 $\Upsilon_M = \Upsilon$ and $\Pi_M(W_M(\psi)) = \Pi_M(W(\Gamma_M(\psi)))$.
In our hypotheses, since $\lambda_M$ is quasifree, one has referring to its GNS representation
$(\gH_M,\Pi_M,\Upsilon_M)$:
 $$\lambda_M(f,g) =  \langle \Upsilon_M,  \Phi(f)  \Phi(g) \Upsilon_M \rangle = 
 \langle \Upsilon, \sigma(\Psi, \Gamma_M(Ef)) \sigma(\Psi, \Gamma_M(Eg))\Upsilon \rangle = \langle \psi_{f+}, \psi_{g+} \rangle\:.$$
 where $\psi_{h+}$ is the $u$-positive frequency part of $\Gamma_M(Eh)$.
 Using (\ref{added}), if $\widehat{\psi_{f}}$ is the Fourier-Plancherel transformation of $\psi_h$ one has finally:
  $$\lambda_M(f,g) = \int_{\bR_+\times \bS^2} 2k \overline{\widehat{\psi_{f}}(k,\z,\bz)}\widehat{\psi_{g}}(k,\z,\bz)
 dk\wedge\epsilon_{\bS^2}(\z,\bz)\:.$$
 If $\Theta(k)=0$ for $k\leq 0$ and $\Theta(k)=1$ for $k>0$,  the identity above can be rewritten as
 \beq \lambda_M(f,g) = \int_{\bR\times \bS^2} \overline{\widehat{\psi_{f}}(k,\z,\bz)}2k\Theta(k)\widehat{\psi_{g}}(k,\z,\bz)
 dk\wedge\epsilon_{\bS^2}(\z,\bz)\:. \label{help}\eeq
 We remind the reader that, by definition of $\sS(\scri)$, $\psi_f$ and $\psi_g$ are real, smooth 
 and $\psi_f,\psi_g,\partial_u \psi_f
 ,\partial_u \psi_g$ belong to $L^2(\bR\times\bS^2, dk \wedge \epsilon_{\bS^2}(\z,\bz))$.
 Using the fact that Fourier-Plancherel transformation defined on the real line is unitary one gets:
 \beq \int_{\bR}  \overline{\widehat{\psi_{f}}(k,\z,\bz)}2k \Theta(k)\widehat{\psi_{g}}(k,\z,\bz) dk\wedge\epsilon_{\bS^2}(\z,\bz) =
 \int_\bR \psi_f(u,\z,\bz) \mF^{-1}[2\Theta k \widehat{\psi_{g}}](u,\z,\bz) du\wedge\epsilon_{\bS^2}(\z,\bz)
 \:.\label{quasi} \eeq
Notice that the identity above makes sense because both $\psi_f, \partial_u \psi_g 
\in L^2(\bR\times\bS^2, dk \wedge \epsilon_{\bS^2}(\z,\bz))$, by definition of the space $\sS(\scri)$,
so that the Fourier-Plancherel transform of $\partial_u \psi_g $, which is $k\widehat{\psi_{g}}$ up to a constant factor, 
and the restriction to the latter to $k\in \bR^+$ are $L^2$ as well.
Now, since $\Theta(k) e^{-k\epsilon} \widehat{\psi_g}(k, \z,\bz)$ converges, as $\epsilon \to 0^+$, to 
$\Theta(k) \widehat{\psi_g}(k, \z,\bz)$
in the sense of $L^2(\bR\times\bS^2, dk \wedge \epsilon_{\bS^2}(\z,\bz))$, and using the fact that the (inverse) 
Fourier-Plancherel transform
is continuous, one has
\beq \mF^{-1}[\Theta e^{-k\epsilon} k\widehat{\psi_g}] \to \mF^{-1}[\Theta k \widehat{\psi_g}]\:, 
\quad \mbox{as $\epsilon\to 0^+$ in the topology of
$L^2(\bR\times\bS^2, dk \wedge \epsilon_{\bS^2}(\z,\bz))$\:.}\label{quasi2} \eeq
The left-hand side can be computed by means of convolution theorem (the convolution restricted to the variable $u$)
since both functions $k \mapsto  \widehat{\psi_g}(k,\z,\bz)$ and $k \mapsto \Theta(k)e^{-\epsilon k}$ belong to 
$L^2(\bR\times\bS^2, dk)$ by construction almost everywhere in $(\z,\bz)$ fixed (for the former function 
it follows from Fubini-Tonelli's theorem
using the fact that $\widehat{\psi_g} \in L^2(\bR\times\bS^2, dk \wedge \epsilon_{\bS^2}(\z,\bz))$ since
$\psi_g \in L^2(\bR\times\bS^2, du \wedge \epsilon_{\bS^2}(\z,\bz))$). In this way, by direct inspection one finds
$$\mF^{-1}[\Theta e^{-k\epsilon} k\widehat{\psi_g}](u,\z,\bz) = 
\frac{1}{2\pi}\int_{\bR} \frac{\partial_{u'} \psi_g(u',\z,\bz)}{u-u'-i\epsilon} du'\:.$$
Inserting it in (\ref{quasi2}) we have achieved that, as $\epsilon\to 0^+$ in the topology of
$L^2(\bR\times\bS^2, dk \wedge \epsilon_{\bS^2}(\z,\bz))$,
$$\frac{1}{2\pi}\int_{\bR} \frac{\partial_{u'} \psi_g(u',\z,\bz)}{u-u'-i\epsilon} du' \to \mF^{-1}[\Theta k \widehat{\psi_g}]\:.$$
Inserting it in the right-hand side of (\ref{quasi}) we have:
\beq
\lambda_M(f,g) =  \frac{1}{\pi}\int_{\bR \times \bS^2} du \wedge \epsilon_{\bS^2}(\z,\bz) \lim_{\epsilon \to 0^+}\int_{\bR} 
 \frac{\psi_f(u,\z,\bz)\partial_{u'}\psi_g(u',\z,\bz)}{u-u'-i\epsilon} du' \:, \label{aggiunta}
\eeq
then, using the continuity of the scalar product of the Hilbert space
$L^2(\bR\times\bS^2, du \wedge \epsilon_{\bS^2}(\z,\bz))$ one obtains:
\beq
\lambda_M(f,g) = \lim_{\epsilon \to 0^+} \frac{1}{\pi}\int_{\bR \times \bS^2} du \wedge \epsilon_{\bS^2}(\z,\bz) \int_{\bR} 
\frac{\psi_f(u,\z,\bz)\partial_{u'}\psi_g(u',\z,\bz)}{u-u'-i\epsilon} du'\:.
\eeq
Since, by hypotheses, both $\psi_g, \partial_u \psi_g$ belong to $C^\infty(\bR)\cap L^2(\bR,du)$
almost everywhere in $(\z,\bz)$, one has $\psi_g(u,\z,\bz) \to 0$ 
for $u\to \pm \infty$\footnote{Work at fixed $(\z,\bz)$. Using elementary calculus, by continuity of $\partial_u\psi_g$
and Cauchy-Schwarz inequality, one has
 $|\psi_g(u') -\psi_g(u)| \leq ||\psi_g||_{L^2(\bR,du)}|u-u'|$ so that $u\mapsto \psi_g(u)$ is uniformly continuous. If were 
 $\psi_g \not \to 0$ as $u\to +\infty$ (the other case is analogous) one would find a sequence of intervals $I_k$
 centred on $k= 1,2,\ldots$ with $\int_{I_k} du > \epsilon$ and
 $|\psi_g|\spa\rest_{I_k} > M$ for some $M>0$ and $\epsilon >0$. As a consequence it would be $\int_\bR |\psi_g(u)|^2 du = +\infty$.}. 
  Integrating by parts the last integral one obtains in that way
\beq
\lambda_M(f,g) = \lim_{\epsilon \to 0^+} -\frac{1}{\pi}\int_{\bR \times \bS^2} du \wedge \epsilon_{\bS^2}(\z,\bz)\int_{\bR} 
\frac{\psi_f(u,\z,\bz)\psi_g(u',\z,\bz)}{(u-u'-i\epsilon)^2} du'\:. \label{inter}
\eeq
To conclude the proof it is sufficient to show that, for $\epsilon>0$ the function
$$(u,u'\z,\bz) \mapsto \frac{|\psi_f(u,\z,\bz)||\psi_g(u',\z,\bz)|}{(u-u')^2+\epsilon^2} =: H(u,u',\z,\bz)$$
 is integrable in the joint measure of $\bR\times \bR \times \bS^2$. Since the function is positive, it is equivalent to prove that
 the function is integrable under iterated integrations, first in $du'$ and then with respect to $du \wedge \epsilon_{\bS^2}(\z,\bz)$.
 We decompose the iterated integration into four terms:
 \beq \int_{[0,u_1) \times \bS^2} \sp\sp \sp\sp du \wedge \epsilon_{\bS^2}(\z,\bz)\int_{[0,u_1)} \sp\sp du'  H(u,u',\z,\bz) +
 \int_{[u_1,+\infty) \times \bS^2} \sp\sp \sp\sp du \wedge \epsilon_{\bS^2}(\z,\bz) \int_{[0,u_1)}\sp\sp  du' H(u,u',\z,\bz)\nonumber \eeq
 \beq +
 \int_{[0,u_1) \times \bS^2} \sp\sp \sp\sp du \wedge \epsilon_{\bS^2}(\z,\bz)\int_{[u_1,+\infty)}\sp\sp  du' H(u,u',\z,\bz)+
 \int_{[u_1,+\infty) \times \bS^2}  \sp\sp \sp\sp du \wedge \epsilon_{\bS^2}(\z,\bz) \int_{[u_1,+\infty)} \sp\sp du' H(u,u',\z,\bz)\:.
 \label{decompose}\eeq
Above we have fixed the origin of $u$ and $u'$ in the past of the support of $\psi_f$ and $\psi_g$ on $\scri$. This is possible 
due to the last statement in Lemma \ref{lemmaLemme}. The point $u_1$ is taken as specified in the following lemma. \\

\lemma \label{bastard}
  {\em Assume that  the spacetime $(M,g)$ 
is an asymptotically flat vacuum spacetime with future time infinity $i^+$ (Definition \ref{dlast}).
Referring to a Bondi frame, for every $\beta \in [1,2)$ there are
 $u_1>0$, a compact  ball $B$  centred in $i^+$
 defined with respect a suitable coordinate patch $x^1,x^2,x^3,x^4$ in $\tM$ centred on $i^+$, and constants $a, b > 0$ 
 such that 
 if $u\geq u_1$, $(\z,\bz) \in \bS^2$:
 \beq \left|\omega_B^{-1}\Psi\rest_\scri(u,\z,\bz)\right| \leq \frac{a M_\Psi}{|u-b|}, 
 \quad \left|\partial_u(\omega_B^{-1}\Psi\rest_\scri)(u,\z,\bz)\right| 
 \leq \frac{a M_\Psi}{|u-b|^\beta}\:, 
\label{trick00}
\eeq
 for every $\Psi \in C^\infty(\tM)$ and where:
 \beq 
 M_\Psi := \max \left(\sup_{B} |\Psi |, \sup_B |\partial_{x^1} \Psi|,\cdots  ,\sup_B |\partial_{x^4}\Psi|\right)\:.
 \label{Max0}
\eeq}\\

\noindent{\em Proof}. The proof is in the Appendix \ref{AWF}. $\Box$\\

\noindent If $h\in C_0^\infty(M)$, the lemma above entails that (with $\beta=1$), for some constants $a,b>0$:
 \beq |\psi_h(u,\z,\bz)|, |\partial_u \psi_h(u,\z,\bz)| \leq \frac{a M_h}{u-b}\:, \label{trick0}\eeq
  where 
 \beq 
 M_h := \max \left(\sup_{B} |\tilde{E}(h)|, \sup_B |\partial_{x^1} \tilde{E}(h)|,\cdots  ,\sup_B |\partial_{x^4} \tilde{E}(h)|\right)\:.
 \label{Max}\eeq
 Enlarging $u_1$ if necessary, we can always assume that $u_1>u_0,b$. In the decomposition (\ref{decompose}) we use that value for $u_1$.
  Therein the first integral converges trivially. Concerning the last integral, due to Eq. (\ref{trick0}),
  we have the estimation
 in its domain of integration
 $$H(u,u',\z,\bz) \leq \frac{a^2}{(u-u')^2 + \epsilon^2} \frac{M_f M_g}{(u-b)(u'-b)}\:.$$
 Using that and the fact that the volume of $\bS^2$ is finite, by direct computation one finds
 $$\int_{[u_1,+\infty) \times \bS^2} \sp\sp \sp\sp \sp\sp du \wedge \epsilon_{\bS^2}(\z,\bz) \int_{[u_1,+\infty)} \sp\sp du' H(u,u',\z,\bz)$$
 $$\leq \int_{u_1}^{+\infty}   \sp\sp du 
\frac{4\pi a^2M_f M_g}{(u-b)(1+u-b)} \left\{
 \frac{\frac{\pi}{2} + \tan^{-1}\left(\frac{u-u_1}{\epsilon}\right)}{\epsilon}
 - \frac{1}{2(u-b)} \ln \frac{(u_1-b)^2}{(u_1-u)^2 +\epsilon^2}
 \right \} <+\infty\:.$$ 
 By Fubini-Tonelli's theorem ($H$ is positive) the second iterated integral in (\ref{decompose}) converges if the third does. Concerning
 the third we have the estimation
 in its domain of integration (notice that $\psi_f$ is smooth in $[0,u_1]\times \bS^2$ and thus bounded
 and $u'\geq u_1>b$.)
 $$H(u,u',\z,\bz) \leq \frac{C}{[(u-u')^2 + \epsilon^2](u'-b)} \leq  \frac{C'}{(u-u')^2 + \epsilon^2}$$
 for some constants $C,C'\geq 0$. Therefore, computing the integral in $u'$ and using the finite volume of $\bS^2$ we find:
 $$ \int_{[0,u_1) \times \bS^2} \sp\sp \sp\sp du \wedge \epsilon_{\bS^2}(\z,\bz)\int_{[u_1,+\infty)}\sp\sp  du' H(u,u',\z,\bz)
 \leq C' \int_0^{u_1} du \: \frac{\frac{\pi}{2} + \tan^{-1}\left(\frac{u-u_1}{\epsilon}\right)}{\epsilon} <+\infty\:. $$
 We conclude that   the function $H$
 is integrable in the joint measure of $\bR\times \bR \times \bS^2$ so that (\ref{inter}) entails (\ref{formula}).\\
 (b) Due to {\em Schwartz kernel theorem} \cite{Hor}, the statement (b) is equivalent to prove that 
 (i) for every $g\in C_0^\infty(M)$, $C_0^\infty(M) \ni f\mapsto \lambda_M(f,g)$
 is continuous in the topology of $C_0^\infty(M)$ and
 (ii) the linear  map
 $C_0^\infty(M) \ni g \mapsto \lambda_M(\cdot, g) \in \mD'(M)$
 is weakly continuous. (ii) means that, for every fixed $f\in  C_0^\infty(M)$, if $\{g_n\}_{n\in \bN} \subset C^\infty_0(M)$ converges to 
 $0$, as $n\to +\infty$, in the topology of $C_0^\infty(M)$, then $\lambda_M(f, g_n) \to 0$ as $n\to +\infty$.
 To prove that the couple of requirements is fulfilled notice that, by Cauchy-Schwarz inequality and (\ref{help}) one finds
 \begin{align}
\left|\lambda_M(f,g)\right| &\leq \left|\left|\widehat{\psi_f}\right|\right|_{L^2(\bR \times \bS^2, dk \wedge \epsilon_{\bS^2})} 
 \left|\left| k \Theta \widehat{\psi_g}\right|\right|_{L^2(\bR \times \bS^2, dk \wedge \epsilon_{\bS^2})}\:\:
 \leq C_g \left|\left|\psi_f\right|\right|_{L^2(\bR \times \bS^2, du \wedge \epsilon_{\bS^2})}\\
\left|\lambda_M(f,g)\right| &\leq \left|\left| k \Theta\widehat{\psi_f}\right|\right|_{L^2(\bR \times \bS^2, dk \wedge \epsilon_{\bS^2})} 
 \left|\left|\widehat{\psi_g}\right|\right|_{L^2(\bR \times \bS^2, dk \wedge \epsilon_{\bS^2})}\:\:
 \leq C_f \left|\left|\psi_g\right|\right|_{L^2(\bR \times \bS^2, du \wedge \epsilon_{\bS^2})}
 \end{align}
 where, in the last passages $C_f := || k \Theta\widehat{\psi_f}||_{L^2(\bR \times \bS^2)}$,
 $C_g := || k \Theta\widehat{\psi_g}||_{L^2(\bR \times \bS^2)}$
  and
 we have used the fact that Fourier-Plancherel transform is isometric.
 Thus, the statement (b) is true if $||\psi_{g_n}||_{L^2(\bR \times \bS^2)} \to 0$
 for  $g_n \to 0$ in the topology of $C_0^\infty(M)$. Let us prove this fact exploiting (\ref{CENTRAL}) and 
 (\ref{trick0}) for $h=g_n$.
 It is known that the causal propagator defined in a  globally hyperbolic spacetime
 $\tilde{E} : C_0^\infty(\tM) \to C^\infty(\tM)$ is continuous in the standard compactly-supported test-function 
 topology in the domain and the natural Fr\'echet topology in $C^\infty(\tM)$ (see 
 \cite{Leray,Dimock,BGP}).
  Fix $f\in C_0^\infty(M)$,
 a compact set $K \subset M$ and a sequence $\{g_n\}_{n\in \bN} \subset C^\infty_0(M)$ supported in $K$.
 From (b) in Lemma \ref{lemmaLemme} there is $u_0 \in \bR$ such that the support of 
 every $\psi_{g_n}$ is included in the set $u\geq u_0$.
  Moreover form Lemma \ref{bastard}, we know 
 that, if $u_1>0$ is sufficiently large, there is a compact  ball $B$  centred in $i^+$
 defined with respect a suitable coordinate patch centred on $i^+$, and constants $a, b > 0$ such that 
 if $u\geq u_1$, $(\z,\bz) \in \bS^2$  (\ref{trick0}) hold for $u= g_n$ (for every $n$), where
 $$M_{g_n} := \max \left(\sup_{B} |\tilde{E}(g_n)|, \sup_B
  |\partial_{x^1} \tilde{E}(g_n)|,\cdots  ,\sup_B |\partial_{x^4} \tilde{E}(g_n)|\right)\:.$$
 Enlarging $u_1$ if necessary, we can always assume that $u_1>u_0,b$.\\
 Continuity of $\tilde{E}$
implies that $M_{g_n} \to 0$ as $n\to +\infty$. If $B'\subset \tM$ is another compact set such that 
$B'\supset \{(u,\z,\bz)\in \scri \:|\: u_1 >u>u_0\}$, since $\omega^{-1}_B$ is bounded therein, continuity 
of $\tilde{E}$ entails by (\ref{CENTRAL}) that $\psi_{g_n}$  vanishes uniformly as $n\to +\infty$ in $B'$.
Now
 $$||\psi_{g_n}||_{L^2(\bR \times \bS^2, du \wedge \epsilon_{\bS^2})}^2
 = \int_{[u_0,+\infty)\times \bS^2}  \left|\psi_{g_n}(u,\z,\bz)\right|^2   du\wedge \epsilon_{\bS^2}(\z,\bz)
 $$
 Decompose the last integral into two terms, the former corresponding to the integration from $u_0$ to $u_1$ and the latter
 from $u_1$ to $+\infty$. Both parts vanish as $n\to +\infty$. The former vanishes because 
 $\psi_{g_n}$  vanishes  uniformly on  $\{(u,\z,\bz)\in \scri \:|\: u_1 > u> u_0\}$ as $n\to +\infty$,
 the latter vanishes as consequence of (\ref{trick0})  with $h=g_n$, since $M_{g_n} \to 0$ as $n\to +\infty$ and
 $$\int_{[u_1,+\infty)\times \bS^2}  \left|\psi_{g_n}(u,\z,\bz)\right|^2   du\wedge\epsilon_{\bS^2}(\z,\bz)
 \leq a^2 M^2_{g_n} {4\pi}\int_{u_1}^{+\infty}\frac{1}{(u-b)^2} du\:.\quad \quad \quad $$
 $\Box$\\

 \ssa{$\lambda_M$ is Hadamard when $(M,g)$ admits $i^+$} 
We are in place to state and prove the main result of this 
 section. We prove that the two-point function of $\lambda_M$ -- which we know 
 to be a bi-distribution due to Theorem \ref{propD'} -- has global Hadamard form.\\
  The idea of the proof is the following. 
  Using the very definition of $\lambda_M$ in terms of $\lambda$ 
  one sees that the two-point function of $\lambda_M$, restricted to any
suitably-defined open neighbourhoods $N \subset M$, 
is obtained by means of a suitable convolution of the two-point function of $\lambda$ and 
a couple  of causal propagators in 
$\tM$ (this is nothing but a re-arranged version of (\ref{formula})):
 \beq
 \lambda^{(N)}_M(f,g) =  -\frac{1}{\pi}\int_{\bR^2\times \bS^2} \frac{\tilde{E}(\Omega^{-3}f)(u,\z,\bz)
\tilde{E}(\Omega^{-3}g)(u',\z,\bz)}{\omega_B(u,\z,\bz)\omega_B(u',\z',\bz)\left(u-u' - i0^+ \right)^2} 
\: du \wedge du' \wedge \epsilon_{\bS^2}(\z,\bz) \:,\nonumber 
 \eeq
where $f,g \in C^\infty_0(M)$ are supported in $N$.
Next one evaluates the wavefront set of the bi-distribution defined by the right-hand side using 
H\"ormander theorems of composition of wavefront sets (taking advantage also of H\"ormander  theorem of propagation of singularities). This is the most complicated step of the whole proof and it is explicitly implemented in the last subsection before the section of final comments.
The final result is stated in Proposition 4.3:
The wave front set of $\lambda^{(N)}_M$ has Hadamard shape. 
Finally, varying the set $N$ in $M$ and ``collecting  together'' all the wavefront sets by means of ``local-to-global''
 Radzikowski's 
theorem (our Proposition \ref{LA}), one achieves the global Hadamard form for $\lambda_M$.\\
 
 \teorema \label{lambdaH} {\em  Assume that  the spacetime $(M,g)$ 
is an asymptotically flat vacuum spacetime with future time infinity $i^+$ (Definition \ref{dlast}) and that 
both $(M,g)$ and the unphysical spacetime 
 $(\tM,\tg)$ are globally hyperbolic.
 Consider the  quasifree state $\lambda_M$ -- on the Weyl algebra $\cW(M)$  of the massless conformally coupled real scalar
 field propagating in $M$ -- canonically induced by the BMS-invariant state $\lambda$  on $\scri$. Under those hypotheses
 $\lambda_M$ is Hadamard.}\\

 \noindent{\em Proof}.  We start by considering  properties of the restriction of $\lambda_M$ to sets $I^-(p;M)\cap I^+(q;M)$, with $p,q \in M$.
Since the class of all the sets $I^-(r;M)\cap I^+(s;M)$ 
define a topological base 
of the strongly causal spacetime $(M,g)$, and since the geodesically convex normal neighbourhoods define an analogous base,
 $I^-(p';M)\cap I^+(q';M)$ must be contained in a geodesically convex normal neighbourhood $U$ provided that $p'$ 
is sufficiently close
to $q'$. Taking $p,q \in I^-(p';M)\cap 
I^+(q';M)$ 
with $p\in I^+(q;M)$ we have  $J^-(p;M)\cap J^+(q;M) \subset  I^-(p';M)\cap I^+(q';M) \subset U$. \\
In the following, a set $I^-(p;M)\cap I^+(q;M) \subset M$ with $p,q \in M$  such that both $I^-(p;M)\cap I^+(q;M) $ 
and $J^-(p;M)\cap J^+(q;M)$ are contained in a geodesically convex normal neighbourhood of $(M,g)$ will be called a {\bf standard domain}.
Standard domains form a base of the topology of every strongly causal spacetime.
A strongly causal spacetime $(M,g)$ is globally hyperbolic if and only if  every set $J^-(p;M)\cap J^+(q;M)$ is compact \cite{Wald}.
In that case $\overline{I^-(p;M)\cap I^+(q;M)} = J^-(p;M)\cap J^+(q;M)$ holds as well \cite{Wald}, the closure being referred to $M$.
Since compactness is  topologically invariant and compact sets are closed, the closure  with respect to 
$M$ coincides with that refereed to $\tM$.
Summarizing, in the hypotheses of theorem \ref{lambdaH}, the standard domains $N$ of $(M,g)$ are
open, form a base of the topology of $M$ and $\overline{N}\subset M$ where $\overline{N}$ is compact and the closure can be interpreted 
indifferently as referred to $M$ or $\tM$ since they coincide. Finally notice that, by construction, both spacetimes $(N, g\spa\rest_N)$ 
and $(N, \tg\spa\rest_N)$ are globally hyperbolic (the latter because the former is globally hyperbolic and $\tg = \Omega^2 g$ with
$\Omega^2$ smooth and strictly positive on $N$).

In the hypotheses of Theorem \ref{lambdaH}, consider a standard  domain $N\subset M$ and the restriction of the two point
function of $\lambda_M$ to $C_0^\infty(N)\times C_0^\infty(N)$. Since we know that $\lambda_M$ is a distribution by Theorem \ref{propD'},
this is equivalent to restrict the distribution $\lambda_M \in \mD'(M\times M)$ to $C_0^\infty(N\times N)$ producing
a distribution of $\mD'(N\times N)$.
We have the central result whose proof, given in the Appendix \ref{AWF}, relies on  the know wavefront set of the causal propagator $\tilde{E}$,
 on several pieces of information on the wavefront set of $\lambda_M$ extracted from (\ref{formula}) and
on standard results about composition of wavefront sets \cite{Hor}.\\

\proposizione \label{propAA} {\em In the hypotheses of Theorem \ref{lambdaH}, consider a standard domain $N\subset M$ equipped with
the metric $g\spa \rest_N$.
If $\lambda^{(N)}_M$ is the restriction of the distribution $\lambda_M \in \mD'(M\times M)$ to $C_0^\infty(N\times N)$, then 
 \beq
  WF(\lambda^{(N)}_M) = 
   \left\{ ((x,k_x),(y,-k_y)) \in T^*N\setminus {\bf 0} \times  T^*N\setminus {\bf 0} \:\:|\:\:
  (x,k_x) \sim (y,k_y)\:, k_x \vartriangleright 0\right\}\:. \label{QLHN}
  \eeq}
  
 \noindent{\em Proof}. The proof is given in the next section. $\Box$\\
  
\noindent We are now in place to take advantage of Radzikowski's results illustrated in Section \ref{Radres}. 
Since $\lambda^{(N)}_M$ determines a quasifree state for the Klein-Gordon field confined inside the globally hyperbolic
subspace $N$, the result established in (\ref{LHN}) entails that $\lambda^{(N)}_M$ is Hadamard on $N$ in view of Proposition \ref{GA}.
Therefore it verifies the local Hadamard condition (LH) in a open neighbourhood $\cG_p$ of every point $p\in N$.
Since the sets $N$ form a base of the topology of $M$ and the restriction of $\lambda^{(N)}_M$ to $\cG_p\times \cG_p$
is nothing but the restriction of  $\lambda_M$ to the same open neighbourhood, we conclude that $\lambda_M$ satisfies (LH)  in 
$\cG_p\times \cG_p$, for some open neighbourhood $\cG_p$ of  every point $p$ of $M$. By Proposition \ref{LA},
$\lambda_M$ is  Hadamard on $(M,g)$. $\Box$\\

\remark It is worth specifying that the  analysis of wavefront sets and their composition mentioned 
about the proof of Proposition \ref{propAA} leads to the inclusion $\subset$ only in (\ref{QLHN}). 
The other inclusion follows straightforwardly from H\"ormander's theorem about
propagation of singularities  \cite{Hor1} using  {\em only} the fact that $\lambda^{(N)}_M$ satisfies (Com) and (KG).
This sort of argument and result can be found in  the proof of Theorem 5.8 in \cite{SV} and in Proposition 6.1 in \cite{SVW}.\\

 \ssa{Proof of Proposition \ref{propAA}}
 Consider fixed Bondi frame $(u, \zeta,\bz)$ on  $\scri \equiv \bR\times  \bS^2$ and suppose that $\scri$
 is equipped with the measure $du \wedge \epsilon_{\bS^2}$,
 $\epsilon_{\bS^2}$  being  the standard volume form of the unit $2$-sphere referred to the coordinates 
 $(\theta,\varphi)$ with $\zeta = e^{i\varphi} \cot (\theta/2)$.
 In the following we view the measure $du\wedge \epsilon_{\bS^2}$ as that induced by the Riemannian metric
 given by $g_{\bS^2} \oplus g_\bR$, $g_{\bS^2}$ being the standard Riemannian metric 
 on the unit $2$-sphere  represented in coordinates $(\theta,\varphi)$ and $g_{\bR}$
 the usual Riemannian metric on $\bR$ referred to the coordinate $u$. In this way we can exploit 
 the definition of distribution on manifolds equipped with a nondegenerate  metric 
 as working on scalar fields. 
  One may fix a different nonsingular 
 smooth metric or define distributions as operating on scalar densities (see discussion on \cite{Hor})
 and it does not affect the wavefront sets:
 Different choices change distributions $u\in \mD'(M)$ by means of smooth {\em nonvanishing} factors $a$, however  
 from the definition of wavefront set, one has $\WF(au)\subset \WF(u) \subset \WF(au)$
 since both $a, 1/a \in C^\infty(M)$.  \\
 
\noindent {\bf General strategy}. Let us pass to the main statement of Proposition \ref{propAA}. 
Fix a Bondi frame $u,\z,\bz$, and the standard domain $N\subset M$.
For $f,g \in C_0^\infty(N)$, $\lambda_M^{(N)}(f,g)$  can be written via (\ref{formula}) as
 \beq
 \lambda^{(N)}_M(f,g) =  -\frac{1}{\pi}\int_{\bR^2\times \bS^2} \frac{\tilde{E}(\Omega^{-3}f)(u,\z,\bz)
\tilde{E}(\Omega^{-3}g)(u',\z,\bz)}{\omega_B(u,\z,\bz)\omega_B(u',\z,\bz)\left(u-u' - i0^+ \right)^2} \: du \wedge 
du' \wedge \epsilon_{\bS^2}(\z,\bz) \label{formulaN}\:.
 \eeq
  In the following we decompose the right-hand
side of (\ref{formulaN}) into four terms (see (\ref{formulaN'})) and we study the wavefront set of each term separately.
Each of those four distribution will be viewed as the composition of parts of the two distributions $\tilde{E} \in \mD'(\tM\times \tM)$  with one of the entries restricted to $\scri$, and the 
distribution $T$ 
\beq T = F \otimes D\:, \quad \mbox{with $\displaystyle F(u,u'):= \frac{1}{(u-u' - i0^+)^2}$ and 
$D(\omega,\omega'):=\delta(\omega, \omega')$.}\:,\label{T}\eeq
 As a result we shall achieve the inclusion $\subset$ in (\ref{QLHN}). The other inclusion
will be proved by a known propagation of singularity argument.\\

\noindent {\bf Preliminary constructions}. 
We need a preliminary construction in order to  extract the singular part of $\tilde{E}$.
The construction is based on the following lemma.\\

 \lemma\label{lemmacompact} {\em Assume that  the spacetime $(M,g)$ 
is an asymptotically flat vacuum spacetime with future time infinity $i^+$ (Definition \ref{dlast}) and that 
both $(M,g)$ and the unphysical spacetime 
 $(\tM,\tg)$ are globally hyperbolic. Consider a Bondi frame $u,\z,\bz$ on $\scri$.\\
 If $N \subset M$ is a standard domain, all the null geodesics (of $(\tM,\tg)$) joining points of $\overline{N}$ and points of $\scri$
 intersect $\scri$ in a set contained in the compact $[u_0, u_f]\times \bS^2$ for suitable numbers $u_0,u_f \in \bR$, where
 the former can be taken as the value $u_0$ determined in (b) of  Lemma \ref{lemmaLemme} for $K := \overline{N}$.}\\
 
 \noindent {\em Proof}. The proof is in the Appendix  \ref{AWF}. $\Box$\\

\noindent Consider two Cauchy surfaces: $S_1$ in the past of $\overline{N}$ such that, referring to  the compact set $S_1 \cap \scri$,
 $\max_{S_1 \cap \scri} u \leq u_0$, and $S_2$ in the future of $\overline{N}$ but in the past of $i^+$ and such that, considering
 the compact set $S_2 \cap \scri$, it holds $\min_{S_1 \cap \scri} u \geq u_1$. 
$u_0$ and $u_f$ are those individuated in  
Lemma \ref{lemmacompact} for the fixed standard domain $N$.
By construction no maximally extended future-directed null geodesics starting from 
$\overline{N}$ can meet $S_1$ and $S_2$ in $J^-(i^+;\tM)$
(concerning $S_1$ the proof is trivial, concerning $S_2$ we observe that if a future-directed maximally extended 
null geodesics starting from 
$\overline{N}$ intersect $S_2$ in $J^-(i^+;\tM)$ it must also meet $\scri$ in a forbidden point with $u>u_1$, since the geodesic
cannot remain confined in the compact $D^+(S_2; \tM) \cap J^-(i^+;\tM)$ as proved below (in the subsection entitled
{\em Analysis of the first term in the rhs of Eq. (\ref{formulaN'})}). Let $H$ be the compact subset 
of $J^-(i^+;\tM)$ bounded by $S_1$ in the past and by $S_2$
in the future and let
$\chi \in C^\infty_0(\tM)$ with $1 \geq \chi \geq 0$, and
$\chi= 1$ constantly in a neighbourhood of $H$ disjoint with $i^+$ and $\supp \chi \cap \scri \subset[U_0,U_f] \times \bS^2$. 
Define $\chi':= 1-\chi$.
If $\tilde{E}(x,y)$ is the Schwartz kernel of $\tilde{E}$, decompose $\tilde{E}$ as 
\beq
\tilde{E}(x,y) = \chi(x) \tE(x,y) + \chi'(x)\tE(x,y)\:.
\eeq 
$\null$ \\

\noindent{\bf Restriction of $\tilde{E}$ to $\scri$ and a bound for the $WF$ set of the restriction.}
As we said we aim to interpret the right-hand side of (\ref{formulaN}) in terms of a composition of distributions. To this end 
we notice that an entry of the two copies of $\tilde{E}$ appearing in (\ref{formulaN}) is constrained to stay on $\scri$.
Therefore we are first of all committed to focussing on the feasibility 
of such restrictions of distributions.\\
 By construction $\chi(x) \tE(x,y)$  has a nonempty singular support, 
 whereas $\chi'(x) \tE(x,y)$ is a smooth kernel when $y\in \overline{N}$ and $x \in J^+(\overline{N}; \tM)$. Therefore $\chi'\tE$
 can be restricted to $\scri \times N$ without problems and it determines a smooth function. Let us consider the same issue
 for $\chi\tE$. 
$\chi\tE$ can in fact be restricted to $\scri \times N$  producing distribution of $\mD'(\scri \times N)$. To show it
 define a local chart about $\scri$ given by coordinates $\Omega, u, \z,\bz$ \cite{Wald}. In these coordinates,
 exactly 
for $\Omega=0$, i.e. on $\scri$, the metric of $\tM$ reads
\beq -d\Omega \otimes du - du \otimes d\Omega + d\Sigma_{\bS^2}(\z,\bz)\:,\label{metricO}\eeq
 $d\Sigma_{\bS^2}(\z,\bz)$ being the standard metric on a
$2$-sphere.
Let $j : \scri \times N \to \tM \times \tM$ 
be  the immersion map of $\scri \times N$ in $\tM \times \tM$. With the used coordinate patch one has  $j : (\Omega, u, \z,\bz, y) 
\mapsto (0, u, \z,\bz, y)$
about $\scri$ and for $y\in N$. Hence the {\em set of normals of the map $j$} in the sense of
 Theorem 8.2.4 in \cite{Hor}) is (using notations of  Lemma \ref{propWFT})
$$N_j = \{ ((x, k_x),(y, k_y)) \in T^*\tM \times T^*\tM \:\:|\:\: (y, k_y) \in T^*N\:,\:\:   x \in \scri \:,\:\: k_x = 
(k_x)_\Omega
d\Omega  \:,\:\: (k_x)_\Omega \in \bR  \}\:.$$
On the other hand \cite{Rada}:
\beq WF(\tilde{E}) = \left\{ ((x,k_x), (y, -k_y)) \in T^*\tM \spa \setminus 0 \times T^*\tM \spa \setminus 0 \:\: | \: \:
(x,k_x) \sim (y, k_y) \right\} \label{WFE}\:.\eeq
If $N_j \cap  WF(\tilde{E}) = \emptyset$ then $\tilde{E}$ can be restricted to $\scri \times N$
as stated in Theorem 8.2.4 in \cite{Hor}. Let us prove that this is the case.
Suppose that $N_j \cap  WF(\tilde{E}) \ni ((x,k_x), (y, -k_y))$. In this case there would be 
a null geodesic $\gamma$ (with respect to $(\tM,\tg)$) intersecting both $y\in N$ and $x\in \scri$ with co-tangent vector $k_x$ at $x$ of the form $k_x = 
(k_x)_\Omega d\Omega \neq 0$. Using the expression of the metric on $\scri$ (\ref{metricO}) one sees that the {\em tangent}
vector of $\gamma$ at $x$ would be proportional to $\frac{\partial}{\partial u}$ and thus the remaining part of the geodesic after 
$x$ would coincide (up to re-parametrisation) with one of integral lines of $n$ emanating from $i^+$ and forming $\scri$, these are null
geodesics with respect to $(\tM,\tg)$ as said in section \ref{intro}. In other words $y \in M$ would be joined with $i^+$ by means of a null
geodesic of $(\tM,\tg)$. This is impossible as established in the proof of Lemma \ref{lemmacompact} (see the statement (A) therein).
We conclude that $N_j \cap  WF(\tilde{E}) = \emptyset$ and thus $\tilde{E}$ can safely be restricted to $\scri \times N$.
The presence of the smooth factor $\chi$ in front of $\tE$
does not affect the result by the very definition of wavefront set, so that $\chi \tE$ can be restricted to $\scri \times N$ in the same way. \\
 The above-mentioned Theorem 8.2.4 in \cite{Hor} implies also that (since $WF(\chi \tilde{E}\rest_{\scri \times N})
\subset WF(\tilde{E}\rest_{\scri \times N})$ which is known by (\ref{WFE})):
$$ WF(\chi \tilde{E}\rest_{\scri \times N})$$
\beq \subset \left\{\left.((x,k_x), (y, -k_y)) \in T^*\scri \spa \setminus 0 \times T^*N \spa \setminus 0 \quad \right| \quad (k_x,k_y) =\: 
{^t dj}(x,y) (h_x,h_y)\:, \: (x,h_x) \sim (y, h_y) \right\}\:.\label{AGGG} \eeq 
Let us give an explicit expression of that bound for $WF(\chi \tilde{E}|_{\scri \times N})$.
Using  the basis $d\Omega_x, du_x, d\z_x, d\bz_x$ of $T_x^*\tM$
and  $du_x, d\z_x, d\bz_x$ of $T^*_x\scri$, it holds:
\beq {^t dj}(x,y): ((h_x)_\Omega d\Omega+ (h_x)_u du +(h_x)_\z d\z + (h_x)_{\bz} d\bz ,h_y) \mapsto ((h_x)_u du + (h_x)_\z d\z + (h_x)_{\bz} d\bz ,h_y)\:.\label{jt}\eeq
The requirement (implicit in $(x,h_x) \sim (y, h_y)$ in (\ref{AGGG})) that $h_x$ is null, in view of the form  (\ref{metricO}) of the metric, means $$-2(h_x)^u (h_x)^\Omega + |h_x|^2_{\bS^2} =0\:, $$
 where the $|h_x|^2_{\bS^2}$ is the norm of $(h_x)^\z \frac{\partial}{\partial \z} + (h_x)^{\bz} \frac{\partial}{\partial \bz}$
 with respect to the {\em positive} metric on $\bS^2$. Notice that, if it were $(h_x)_u =0$
(i.e. $(h_x)^\Omega =0$), one would have  $|h_x|^2_\Sigma = 0$ so that also the components $(h_x)^\z,(h_x)^{\bz}$ would vanish 
and $h_x$ would reduce to $h_x = (h_x)_\Omega d\Omega$.
However this form for $h_x$ is not allowed when $h_x$ is, as in the considered case, the co-tangent vector of a null geodesic joining $x\in \scri$ and a point $y\in M$ as previously 
remarked. We conclude that $(h_x)_u \neq 0$ in (\ref{AGGG}). Summarising, taking (\ref{jt}) into account, we can say that:
\beq 
WF(\chi \tilde{E}|_{\scri \times N}) \subset \left\{ ((x,k_x), (y, -k_y)) \in T^*\scri \spa \setminus 0
 \times T^*N \spa \setminus 0 \:\: | \: \:
(x,\widehat{k_x}) \sim (y, k_y)\:,\: (k_x)_u \neq 0 \right\} \label{WFErest0}\:,\eeq
where, referring to the basis $d\Omega_x, du_x, d\z_x, d\bz_x$ of $T_x^*\tM$
and  $du_x, d\z_x, d\bz_x$ of $T^*_x\scri$,
the covector $\widehat{k_x} \in T^*_x\tM$ is  that uniquely determined  by $k_x \in T^*_x \scri \setminus \{0\}$
with $(k_x)_u \neq 0$
 imposing the 
condition $\tg(\widehat{k_x},\widehat{k_x}) =0$. $\widehat{k_x}$ is in fact the generic  co-tangent vector in $x$ of a null 
geodesic starting in $N \subset M$ and reaching  $\scri$ in $x$.\\

\remark The bound (\ref{WFErest0}) and the requirement $(k_x)_u \neq 0$ forbid the presence in
 $WF(\chi \tilde{E}|_{\scri \times N})$ of elements of the form $((x,0), (y,k_y))$, the geodesic joining $x\in \scri$ and $y \in N$ having $d\Omega_x$ as cotangent
vector at $x$.\\

\noindent {\bf Decomposition of $\lambda^{(N)}_M$}. Let us come back to (\ref{formulaN}) in order to study the wavefront set of the distribution in the right-hand side making use 
of the wavefront sets of the distributions  $E|_{\scri \times N}$ and $T$. 
It is convenient  to introduce the rearranged distributions $\bE \in \mD'(\scri\times N)$ 
and $\cE \in \mD'(\scri \times N) \cap C^\infty(\scri \times N)$
individuated via Schwartz kernel theorem by the operators $C_0^\infty(N) \to \mD'(\scri)$
\begin{align}
\bE(f)(u,\z,\bz) &:= \omega_B(u,\z,\bz)^{-1} \chi\tE\sp\rest_{\scri\times N} (\Omega^{-3} f)(u,\z,\bz) \:,
\quad \mbox{for $f\in C_0^\infty(N)$}\:,\\
\cE(f)(u,\z,\bz) &:= \omega_B(u,\z,\bz)^{-1} \chi'\tE\sp\rest_{\scri\times N} (\Omega^{-3} f)(u,\z,\bz)\:, 
\quad \mbox{for $f\in C_0^\infty(N)$.} \label{chi'E}
\end{align}
The wavefront set of $\cE$ is obviously empty, whereas as $\omega_B^{-1}$, $\Omega^{-3}$ being smooth,
 from (\ref{WFErest0}), we get again
\beq 
WF(\bE) \subset \left\{ ((x,k_x), (y, -k_y)) \in T^*\scri \spa \setminus 0
 \times T^*N \spa \setminus 0 \:\: | \: \:
(x,\widehat{k_x}) \sim (y, k_y)\:,\: (k_x)_u \neq 0 \right\} \label{WFErest}\:.\eeq
Indicating by $\omega$ the angular coordinates $\z,\bz$
on $\scri$, and with $du d\omega$ the measure on $\scri$, $du \wedge \epsilon_{\bS^2}(\z,\bz)$, by (\ref{formulaN})
 one has the decomposition (for $h\in C^\infty_0(N)$):
 \begin{align}
&\lambda^{(N)}_M(f,g) =  -\frac{1}{\pi}\int_{\bR^2\times \bS^2} 
 \frac{ \bE(f)(u,\omega)
\bE(g)(u',\omega)}{\left(u-u' - i0^+ \right)^2}\: du d\omega du' 
- \frac{1}{\pi}\int_{\bR^2\times \bS^2} 
 \frac{\cE(f)(u,\omega)
\bE(g)(u',\omega)}{\left(u-u' - i0^+ \right)^2}\:\: du d\omega du' \nonumber\\
 & -  \frac{1}{\pi}\int_{\bR^2\times \bS^2} 
 \frac{\bE(f)(u,\omega)
\cE(g)(u',\omega)}{\left(u-u' - i0^+ \right)^2}\:\: du d\omega du'-  \frac{1}{\pi}\int_{\bR^2\times \bS^2} 
 \frac{\cE(f)(u,\omega)
\cE(g)(u',\omega)}{\left(u-u' - i0^+ \right)^2}\:\: du d\omega du'   \label{formulaN'}\:.
 \end{align}
 In the following we compute the various contributions to $WF(\lambda_M^{(N)})$ due to each of the 
 terms in the right-hand side of (\ref{formulaN'}).\\

 \noindent{\bf The wavefront set of $T$.} Concerning $\WF(T)$ for the distribution $T$ in (\ref{T}), we have the following result.\\

 \lemma\label{propWFT}
 {\em Consider the distribution  $T\in  \mD'(\scri\times \scri)$ defined in (\ref{T}).
 defined as
\beq T = F \otimes D\:, \quad \mbox{with $\displaystyle F(u,u'):= \frac{1}{(u-u' - i0^+)^2}$ and 
$D(\omega,\omega'):=\delta(\omega, \omega')$.}\:,\nonumber \eeq
  where $u\in\bR$ with covectors $k \in T^*_u\bR$,
  $\omega\in \bS^2$ with covectors $\bk \in T^*_\omega\bS^2$ and
   similar notations are valid for primed variables.
 With those hypotheses it holds \beq \WF(T) = A \cup B \label{WFT}\eeq
 $$A:= \left\{ \spa\left.\left((u, \omega, k, \bk), (u',\omega', k', {\bf k}') \right) \in 
T^*\scri \spa \setminus 0 \times T^* \scri
 \spa \setminus 0
  \:  \right|\:
 u=u',  \omega =\omega', 0<k=-k', {\bf k} =-{\bf k}' \right\}$$
 $$B := \left\{\spa \left.\left((u, \omega, k, \bk), (u',\omega', k', {\bf k}')\right) \in T^*\scri \spa \setminus 0 \times T^* \scri
 \spa \setminus 0
  \: \: \right|\:\: 
 \omega =\omega', k = k' =0, {\bf k} = -{\bf k}'\right\}\:.$$}
 
 \noindent {\em Proof}. It is is a straightforward  consequence of the discussion after 
 Theorem 8.2.14 in \cite{Hor} and the known wavefront sets of the delta distribution and $1/(k\pm i0^+)$ (e.g. see \cite{RS}). $\Box$\\

\noindent{\bf Analysis of the first term in the rhs of Eq.} (\ref{formulaN'}). Let us focus on the first term in the right-hand side of (\ref{formulaN'}). First of all we notice that
it is possible to replace, {\em without affecting the final result}, the kernel $(u-u'-i\epsilon)^{-2}$ with 
the compactly supported kernel $\tchi (u,u')\:(u-u'-i\epsilon)^{-2}$
where $\tchi \in C^\infty_0(\bR^2)$ attains the value constant $1$ 
on  the compact $[U_0,U_f] \times [U_0,U_f]$ mentioned above defining $\chi$. 
The first term in the right-hand side of (\ref{formulaN'})  can be re-written as,
barring the factor
$-1/\pi$:
\beq \left\langle \tchi T  \:  ,\: \bE \otimes \bE  (f\otimes g) \right\rangle\label{lll}\eeq
where the tensor product of Schwartz kernels $\bE \otimes \bE$ is used as a map  
$C_0^\infty(N\times N)\to \mD'(\scri\times \scri)$
and $T\in \mD'(\scri \times \scri)$ has been introduced in Lemma \ref{propWFT}.
If $\{T_n\} \subset C_0^\infty (\scri\times \scri)$
is a sequence of functions which tends to $\tchi T$ in the topology of $\mD'(\scri \times \scri)$, we have trivially
\beq \left\langle \tchi T  \:  ,\: \bE \otimes \bE  (f\otimes g) \right\rangle
= \lim_{n\to +\infty }\left\langle  T_n  \:  ,\: \bE \otimes \bE  (f\otimes g) \right\rangle =
\lim_{n\to +\infty }\left\langle  ^t(\bE \otimes \bE)(T_n)  \:  ,\:  f\otimes g \right\rangle
\:,\label{ggg}\eeq
where $^t(\bE \otimes \bE): C_0^\infty(\scri \times \scri) \to \mD'(N\times N)$ indicates the adjoint of  $\bE\otimes \bE$:\\
$$\left(^t(\bE \otimes \bE)(T_n)\right)(x,x') := \int_{\bR \times \bS^2}\sp du\:du\: d\omega\: d\omega' T_n(u,\omega, u',\omega') \bE(u,\omega,x)\bE(u',\omega',x')\:.$$
Since $T_n \to \tchi T$,  one wonders if it is possible to re-write (\ref{ggg}) as:
\beq \left\langle \tchi T  \:  ,\: \bE \otimes \bE  (f\otimes g) \right\rangle = \left\langle  ^t(\bE \otimes \bE)(\tchi T)  \:  ,\:  
f\otimes g \right\rangle  \:, \label{LLL'}\eeq
where $ ^t(\bE \otimes \bE)(\tchi T)$ represents the action of  $^t(\bE \otimes \bE)$ on the {\em compact support distribution}
 $\tchi T$. 
By Theorem 8.2.13 in \cite{Hor} it is possible  and $ ^t(\bE \otimes \bE)(\tchi T)$ 
exists as an element of $\mD'(N\times N)$, 
provided that (a) $\tchi T$ has compact support -- and this is assured by the 
introduction of the function 
$\tchi$ which in turn may exist due to
Lemma \ref{lemmacompact} --  and (b):
\beq
WF'(^t(\bE \otimes \bE))_{\scri\times \scri} \bigcap WF(\tchi T) = \emptyset \label{condWWW}
\eeq
where, if $K \in \mD'(X\times Y)$, $WF'(K)_X:= \{(x,k_y)\:\: |\:\: ((x,-k_x), (y,0)) \in WF(K) \quad 
\mbox{for some $x\in X$}\}\:$, and $WF'(K)_Y$ is defined analogously.
To achieve  (\ref{LLL'}) from (\ref{ggg}),
the sequence of functions $T_n \to \tchi T$ has to tend to $\tchi T$ in the sense of H\"ormander pseudo topology in the domain
specified in Theorem 8.2.13 in \cite{Hor}. Existence of such a sequence is however guaranteed by  Theorem 8.2.3 in \cite{Hor}. 
(Notice that also  if the theorems above concern distributions defined on $\bR^n$, we can reduces to this case since $N$ is covered
by a single normal Riemannian coordinate patch, whereas $\scri$ is diffeomorphic to $\bR^3 \setminus \{0\}$.)
Let us prove that the condition (\ref{condWWW}) is fulfilled in our case.
By Theorem 8.2.9 in \cite{Hor}:
\beq WF(\bE \otimes \bE) \subset (WF(\bE) \otimes WF(\bE)) \bigcup \:((\supp \bE\times \{0\}) \times WF(\bE))
 \bigcup (WF(\bE) \times (\supp \bE\times \{0\}))\label{EE}\:.\eeq
Since there are no null geodesics with vanishing tangent vector in $y\in N$ joining $x\in \scri$, we have
$WF'(\bE \otimes \bE)_{\scri\times \scri} = \emptyset$ and so $WF'(^t(\bE \otimes \bE))_{\scri\times \scri} = \emptyset$, 
therefore (\ref{condWWW}) is fulfilled.  \\
We have obtained that the first term in the right-hand side of (\ref{formulaN'})
is nothing but the action of the distribution $^t(\bE \otimes \bE)(\tchi T) \in \mD'(N\times N)$ on $f\otimes g$. Now Theorem 8.2.13 in \cite{Hor} gives the inclusion:
\begin{align} &WF(^t(\bE \otimes \bE)(\tchi T)) \subset WF(^t(\bE \otimes \bE))_{N\times N}\nonumber\\ 
&\bigcup \left\{  ((x,k_x),(x', k_x')) \:\:|\:\: ((x,k_x), (x', k_x'), (u,\omega, -k_u, -\bk), 
  (u',\omega', -k_u', -\bk')) \in
WF(^t(\bE \otimes \bE))\:, \quad \right. \nonumber \\
&\left.\mbox{for some $(u,\omega, k_u, \bk), (u',\omega', k_u', \bk') \in WF( \tchi T)$}
\right\} \:. \label{lastlast} \end{align}
Similarly to $WF'(^t(\bE \otimes \bE))_{\scri\times \scri}$ one finds (with the same argument)
 $WF'(^t(\bE \otimes \bE))_{N\times N} = \emptyset$. Whereas the remaining part in the right hand side of 
 (\ref{lastlast}), taking into account the inclusions (\ref{EE}), the inclusion $WF(\tchi T) \subset WF(T)$
 and (\ref{WFErest}), exploiting (\ref{WFT}), 
 produces straightforwardly the  result:\\
\noindent {\em $WF(^t(\bE \otimes \bE)(\tchi T))$ is contained in the set $G$ 
of pairs $((x,k_x),(x',-k_x')) \in T^*N\setminus {\bf 0} \times  T^*N\setminus {\bf 0}$
such that:

 (a) $(x,k_x)$ and  $(x',k_x')$ are  points and associated cotangent vectors 
of the same maximal null geodesic $\gamma$ intersecting $\scri$ in some point $p$, and

 (b) $k_x$ is future directed.}\\
(Since the coordinate $u$ is future directed and (\ref{metricO}) holds,
 $k_x$ is future directed if and only if, considering the geodesic $\gamma$ with initial conditions $(x,k_x)\in T_x^*N$,
the opposite of the covector tangent to $\gamma$, in the point $p$ where $\gamma$ meets $\scri$, has component $(k_p)_u$ {\em positive}.
This agrees with the condition $k>0$  in the definition of the wavefront set of the distribution $T$ (\ref{WFT}) concerning the subset 
$A$, the subset $B$ gives no contribution to $G$.)\\
We can improve the obtained bound for $WF(^t(\bE \otimes \bE)(\tchi T))$ as follows.
Notice that, if $(x,k_x), (x,-k_y) \in G$, by construction it holds:
 $(x,k_x) \sim (x, k_y)$  with $k_x \vartriangleright 0$.
Conversely, consider a pair $(x,k_x) \sim (y,k_y)$ with $k_x \vartriangleright 0$. Let us prove that $((x,k_x), (y,-k_y))\in G$.
The maximal null geodesic  passing through $x$ and $y$ with respective cotangent vectors $k_x$ and $k_y$
must achieves $\scri$ in some point, since  it  cannot remain confined in the compact $D:= J^-(i^+;\tM) \cap D^+(S;\tM)$ 
where $S$ is a spacelike Cauchy surface of $\tM$ which
 intersects $x$ or $y$ and lies in the past of the other 
point. (Indeed, if the maximally extended geodesic $\gamma: (a,b) \to \tM$ were confined in $D$, there would be $c\in D$ with $
\gamma(t_k) \to c$ 
for some sequence of affine parameter points $(a,b) \ni t_k \to b$. 
In normal coordinates centred on $c$, the geodesics would assume standard form $t' \mapsto t' v^\mu$
for constants $v^\mu\in \bR$ and $t'\in (a',b') \ni 0$ being another parameter related to $t$ by means of a nonsingular affine transformation. 
This would imply, on a hand, that $b<+\infty$, and on the other hand that $\gamma$  admits an extension beyond $c$ and this 
is not possible by hypotheses.)
So $\gamma$ gets out intersecting $\partial (J^-(i^+;\tM) \cap D^+(S;\tM))$ in some point $p$. Since it cannot intersect twice $S$, the geodesic
has to meet $\partial J^-(i^+;\tM)$ somewhere. The point $i^+$ is forbidden as established in the proof of 
Lemma \ref{lemmacompact}. We conclude that the geodesics must intercept some point of $\scri$.
We have found that  if, for $x,y\in N$,  $(x,k_x) \sim (y,k_y)$ with $k_x \vartriangleright 0$, then
it also holds $(x,k_x), (y,-k_y) \in G$. We have finally obtained that {\em the contribution $ WF(^t(\bE \otimes \bE)(\tchi T))$ to the wavefront set of $\lambda_M^{(N)}$
due to first term in the rhs of (\ref{formulaN'}) fulfils the following bound}:
 \beq
  WF(^t(\bE \otimes \bE)(\tchi T)) \subset  \left\{ ((x,k_x),(y,-k_y)) \in T^*N\setminus {\bf 0} \times  T^*N\setminus {\bf 0} \:\:|\:\:
  (x,k_x) \sim (y,k_y)\:, k_x \vartriangleright 0\right\}\:. \label{QLHNfine}
  \eeq

\noindent{\bf Analysis of the last term in the rhs of Eq.} (\ref{formulaN'}).
To go on, we remind the reader that $\cE \in C^\infty(\scri \times N)$ by construction. 
Furthermore  $\cE(u,\omega,x) = 0$ smoothly for  $u< u_0$. 
Moreover by (\ref{chi'E})
recalling that $\chi'(x,y) \tE(x,y)$ has smooth kernel when $y\in N$ and $x \in J^+(\overline{N})$ so that it is smooth
for  $x$ varying in a neighbourhood of $i^+$ when $y\in N$,
we can control the behaviour as $u \to +\infty$ of $\partial^\alpha_x\cE(u,\omega,x)$ and $\partial^\alpha_x \partial_u 
 \cE(u,\omega,x)$
 by Lemma \ref{bastard} for every multi-index $\alpha$ and for any fixed $\beta \in [1,2)$: $\partial^\alpha_x\cE(u,\omega,x)$ 
 and $\partial^\alpha_x \partial_u 
 \cE(u,\omega,x)$  are bounded, respectively,
  by functions of the form $M_\alpha(x)/|u-b|$ and 
  $M_{\alpha\beta}(x)/|u-b|^\beta$.
The bounds $M_\alpha(x), M_{\alpha\beta}(x)$ can be made locally uniform in $x$ taking the $\sup$ in (\ref{Max0}) 
over $B\times B'$, $B'$ being a relatively compact neighbourhood 
 of every fixed point $x_0\in N$. \\
By integration by parts, the last term 
 in the right-hand side of (\ref{formulaN'})
 can be re-written (omitting a constant overall
 factor):
\beq\lim_{\epsilon \to 0^+} 
\int_{\bR^2 \times \bS^2\times N \times N} \sp\sp\sp\sp\sp\sp\sp\sp du\: du'\: d\omega\: d\mu_{\tg}(x) \:d\mu_{\tg}(x')\:\:\frac{\partial_u \cE(u,\omega,x)\cE(u',\omega,x')}{u-u'-i\epsilon}\: f(x)g(x')
\:.\label{last9}\eeq
The functional in (\ref{last9}) can be rearranged by using Fubini-Tonelli and Lebesgue's dominate convergence
and computing the limit under the symbol of $d\mu_{\tg}(x) \:d\mu_{\tg}(x')$ integration, obtaining
that the last term  in the right-hand side of (\ref{formulaN'}) is, in fact, up to an overall factor:
$$\int_{N\times N} K(x,x') f(x)g(x') \:d\mu_{\tg}(x)\:
 d\mu_{\tg}(x')$$
where the {\em smooth} kernel $K(x,x')$ reads:
$$i\pi \int_{\bR\times \bS^2} \sp\sp\sp  du\:d\omega\:\cE(u,\omega,x)\partial_u \cE(u,\omega,x') 
 - \int_{\bR\times \bR \times \bS^2} \sp\sp \sp\sp\sp du\: du'\:d\omega\:\rho(u-u')\:
\frac{\cE(u',\omega,x')- \cE(u,\omega,x')}{u-u'}\:\partial_u \cE(u,\omega,x) $$
$$ - \int_{\bR\times \bR \times \bS^2} \sp\sp \sp\sp\sp du\: du'\:d\omega\:
\rho'(u-u')\:\frac{\cE(u',\omega,x')}{u-u'}\:\partial_u \cE(u,\omega,x)\:,$$
$\rho':= 1-\rho$ and $\rho \in C_0^\infty(\bR)$ being any, arbitrarily fixed, function which attains the value 
$1$ constantly in a neighbourhood of $0$. 
Absolute convergence of the integrals and smoothness of $K(x,x')$ can be checked by direct inspection taking derivatives under the symbol of integration by standard 
theorems based on dominate convergence theorem together with the uniform bounds on the behaviour as $u\to +\infty$
mentioned above.\\
 {\em We conclude that the last term in right-hand side of (\ref{formulaN'}) gives no contribution to the wavefront set
of the two-point function of $\lambda^{(N)}_M$ since it is associate with a smooth kernel ($K(x,x')$)}.\\

\noindent{\bf Analysis of the second and third term in the rhs of Eq.} (\ref{formulaN'}).
Let us examine the third term in the right-hand side of (\ref{formulaN'}), the second  can be analysed with the 
same procedure obtaining the same result. As before this term can be re-arranged
and the limit can be explicitly computed obtaining that third term in the right-hand side of (\ref{formulaN'}) 
equals (up to a constant overall factor)
$$\lim_{\epsilon \to 0^+} \int_{\bR^2 \times \bS^2\times N } \sp\sp\sp\sp\sp\sp du\: du'\: d\omega \:d\mu_{\tg}(x')\:
\frac{\partial_{u} \bE(f)(u,\omega)\: \cE(u',\omega, x')}{u-u'-i\epsilon}\:  g(x') = \int_{N} \sp d\mu_{\tg}(x')\: 
 H(f,x')  g(x') \:,$$
with, for every fixed $f\in C_0^\infty(N)$,   the {\em smooth} function $H(f,x')$ given by:
$$i\pi \int_{\bR\times \bS^2} \sp\sp du\:d\omega \: \:\:\: 
\cE(u,\omega,x')\partial_u \bE(f)(u,\omega)
- \int_{\bR\times \bR \times \bS^2} \sp\sp \sp\sp\sp du\: du'\:d\omega\:\: \rho(u-u')
\frac{\cE(u',\omega,x')- \cE(u,\omega,x')}{u-u'}\partial_u \bE(f)(u,\omega)$$
$$- \int_{\bR\times \bR \times \bS^2} \sp\sp \sp\sp\sp du\: du'\:d\omega\:\: \rho'(u-u')
\frac{\cE(u',\omega,x')}{u-u'}\partial_u \bE(f)(u,\omega)\:.
$$ 
As before, the function $\rho \in C^\infty_0(\bR)$ is any function with $\rho=1$ in a neighbourhood of $0$ and $\rho' := 1-\rho$. 
Each of the three integrals in the expression of $H$ have form, with a corresponding ${\cal S} \in C^\infty(\scri \times N)$,
$$F(f)(x') := \int_{\bR\times \bS^2\times N} \sp\sp \sp \sp\sp\sp du\:d\omega \: \:\:\: 
{\cal S}(u,\omega, x')\partial_u \bE(f)(u,\omega)\:.$$
At least formally, one may think of $F : C^\infty_0(N) \to \mD'(N)$ as individuated by the Schwartz kernel $F(x,x')$ 
composition of Schwartz kernels:
\beq
F(x,x') = \int_{\bR\times \bS^2} \sp\sp\sp du\:d\omega \: \:\:\: {\cal S}(u,\omega, x') \partial_u \bE(u,\omega,x)\quad
\mbox{i.e.}\quad {^tF} = {^t{\cal S}} \circ \partial_u\bE\:.
\eeq
Similarly to what was concerned in the case of the first term in the right-hand side of (\ref{formulaN'}),
this interpretation makes rigorous sense in view of Theorem 8.2.14 of \cite{Hor} provided (a) the projection $\supp(\partial_u\bE) \ni (u,\omega, x) \mapsto x \in N$ is proper -- and this can be straightforwardly  verified true by  the properties of the support of $\bE$ --
and (b) $WF'(^t{\cal S})_{\scri} \cap WF (\partial_u\bE)_{\scri} =\emptyset$ -- and this is also true because
 $WF'(^t{\cal S})_{\scri}$ is empty since $^t{\cal S}$ is smooth,  whereas
  $WF (\partial_u\bE)_{\scri} \subset  WF (\bE)_{\scri}$ which is empty 
as can be found by direct inspection using (\ref{WFErest})
(there are no null geodesics from $N$ to $\scri$
 with zero tangent vector).
 The inclusion given in Theorem 8.2.14 in \cite{Hor}
states that $WF(^tF)$  is a subset of the union of the following sets: (1) $WF'(^t{\cal S}) \circ WF'(\partial_u\bE)$, which is empty because 
$WF'(^t{\cal S})$ is empty, (2) $WF(^t{\cal S})_N\times N \times \{0\}$, which is empty due to the same reason, and (3)
 $N \times \{0\} \times WF'(\partial_u\bE)_N$, which is empty because $WF'(\partial_u\bE)_N \subset WF'(\bE)_N $, 
 and referring to (\ref{WFErest}),
 there are no null geodesics from $N$ to $\scri$
 with zero tangent vector. Hence $WF(^tF)= \emptyset$ and thus $WF(F)= \emptyset$.\\
{\em We conclude that the second and the third term in right-hand side of (\ref{formulaN'}) give no contribution to the wavefront set
of the two-point function of $\lambda^{(N)}_M$.}\\

\noindent Collecting all the obtained results together, we realise that the only contribution to
$WF(\lambda^{(N)}_M)$ comes from the first term in right-hand side
of (\ref{formulaN'}) and thus,  due to (\ref{QLHNfine}):
 \beq
  WF(\lambda^{(N)}_M) \subset \left\{ ((x,k_x),(y,-k_y)) \in T^*N\setminus {\bf 0} \times  T^*N\setminus {\bf 0} \:\:|\:\:
  (x,k_x) \sim (y,k_y)\:, k_x \vartriangleright 0\right\}\:. \label{caso}
  \eeq
$\sim$ and $\vartriangleright$ in the right-hand side of (\ref{caso}) are refereed to the metric
 $\tg\spa \rest_N$. However  rescaling the by means of a
smooth  factor $\Omega^{-2}>0$ does not modify the right-hand side of (\ref{caso}).
One proves immediately it 
using the transformation rule of (co)tangent vectors of null geodesics
under local rescaling of the metric  (see the Appendix D in \cite{Wald}). Therefore (\ref{caso}) holds true also employing 
$\sim$ and $\vartriangleright$ associated with the metric $g\spa\rest_N$.\\

\noindent{\bf Propagation of singularity argument}. In view of (\ref{caso}), to conclude the proof of Proposition \ref{propAA} it is now sufficient to establish the other inclusion:
 \beq
  WF(\lambda^{(N)}_M) \supset \left\{ ((x,k_x),(y,-k_y)) \in T^*N\setminus {\bf 0} \times  T^*N\setminus {\bf 0} \:\:|\:\:
  (x,k_x) \sim (y,k_y)\:, k_x \vartriangleright 0\right\}\:. \label{LHN}
  \eeq
 Let us do it using an argument based on the theorem of singularity propagation.
 Let $E_N \in \mD'(N\times N)$ be the causal propagator associated with Klein-Gordon 
equation in the globally hyperbolic spacetime $(N, g\spa\rest_N)$ and, in the following we denote by $\mbox{sing supp}(S)$ the singular support of
a distribution $S$. {\em In this proof $p\sim q$ means that there is, in the considered spacetime, 
at least one null geodesic joining $p$ and $q$}.\\ 
We prove (\ref{LHN}) by means of {\em a reductio ad absurdum}.
Our {\em per absurdum} claim is that there are $p,q \in N$ with $p\sim q$ but 
$((p,k_p), (q, -k_q)) \not \in WF(\lambda^{(N)}_M)$, where $k_p$ and $k_q$ are the cotangent vectors to 
a  null geodesic joining $p$ and $q$ with $k_p\vartriangleright 0$. Actually that
geodesic is {\em uniquely determined}  -- for both $N$ and $M$ -- by $p$ and $q$, from the very definition of standard domain $N$.
(Notice also that the wavefront set is conic and thus the vectors $k_p,k_q$ are determined up to 
a common, strictly positive, factor completely irrelevant in our discussion.)
Since the singular support 
of a distribution of $\mD'(N\times N)$ is the projection on $N\times N$ the of the wavefront set of the distribution,
we must conclude that, in view of (\ref{caso}), $(p,q) \not \in \mbox{sing supp} (\lambda^{(N)}_M)$. 
However, as $p\sim q$, $(p,q)$ must belong to $\mbox{sing supp} (E_N)$ (this is because $E_N$ is the difference
of the advanced and the retarded fundamentals solutions whose known wavefronts and causal properties of supports 
\cite{Rada} entails that $\mbox{sing supp}(E_N)$ is made exactly by the pairs $(p,q)\in N\times N$ with $p\sim q$).
Since (Com) holds true, we conclude that $(q,p) \in \mbox{sing supp}(\lambda^{(N)}_M)$, and thus
 there are $k_p \in T^*_pN$ and $k_q\in T^*_qN$ such that
$((q,k_q), (p,-k_p)) \in WF(\lambda_M^{(N)})$, and so, via Proposition \ref{propAA}, $k_p$ and $k_q$ are vectors
cotangent to the null geodesic joining $p$ and $q$ 
(the same as before since it is unique) and, finally, 
$k_q \vartriangleright 0$.\\
The distribution $\lambda^{(N)}_M$ satisfies Klein-Gordon equation in both arguments
(in other words (KG) holds), therefore the {\em propagation of singularities theorem} \cite{Hor1} (see discussion after Theorem 4.6
in \cite{Rada}) implies that  the wavefront set  of $\lambda^{(N)}_M$ is the union of sets of the form
$B(x,k_x)\times B(y,k_y)$. Here $B(z,h_z)$ is the unique null maximal geodesic (viewed as a curve in $T^*N$) passing through
$z\in N$ with co-tangent vector $h_z\in T_z^*N$. As $((q,k_q), (p,-k_p)) \in WF(\lambda^{(N)}_M)$ 
and since $q$ and $p$ belong to the same null geodesic,
we are committed to conclude that $((p,k'_p), (q,-k'_q))  \in WF(\lambda^{(N)}_M)$
where the cotangent vector $k'_p$ is cotangent to the geodesic at $p$ and it has the same time orientation as $k_q$,
 so that $k'_p \vartriangleright 0$,
and the vector $k'_q$ is cotangent to the geodesic at $q$. In other words, changing the used names for
cotangent vectors:  $((p,k_p), (q,-k_q))  \in WF(\lambda^{(N)}_M)$ where $k_p\vartriangleright 0$. This is in contradiction
with our initial claim. $\Box$\\

\noindent We have finally established that, on $(N, g\spa\rest_N )$
 \beq
  WF(\lambda^{(N)}_M) = \left\{ ((x,k_x),(y,-k_y)) \in T^*N\setminus {\bf 0} \times  T^*N\setminus {\bf 0} \:\:|\:\:
  (x,k_x) \sim (y,k_y)\:, k_x \vartriangleright 0\right\}\:. \nonumber
  \eeq

$\Box$

 \section{Final Comments: summary and open issues.}
 Let us summarise the main results achieved in this work.
 We started from the unique, positive $BMS$-energy,  $BMS$-invariant, quasifree, pure state $\lambda$
 acting on a natural Weyl algebra defined on $\scri$.
That state is completely defined using the universal structure of the class of (vacuum) asymptotic flat 
spacetimes at null infinity, no reference to any particular spacetime is necessary. 
 In this sense $\lambda$ is {\em universal}. It is the vacuum state for a representation of BMS group 
 with vanishing BMS mass.
 Afterwards, we have seen that $\lambda$ induces
  in any fixed (globally hyperbolic) bulk spacetime $M$, a preferred
state  $\lambda_M$ for a conformally coupled massless real scalar field. This happens  if $M$ admits future time infinity $i^+$ (and the unphysical spacetime $\tM$
is globally hyperbolic as well). 
The induction of a state takes place by means of an injective isometric $*$ homomorphism $\imath : \cW(M) \to \cW(\scri)$
  which identifies Weyl observables of the field in the bulk with some Weyl observables 
 of the boundary $\scri$. 
 $$\lambda_M(a):= \lambda(\imath(a))\quad \mbox{for all $a\in \cW(M)$\:.}$$
Using a very inflated term, we may say that this is a {\em holographic correspondence}. \\
The picked out state $\lambda_M$ enjoys quite natural, as well as interesting, properties. These properties (barring the first one)
have been established in this paper:

(i) $\lambda_M$ coincides with Minkowski vacuum when $M$ is Minkowski spacetime, 

(ii) $\lambda_M$ is invariant under 
every isometry of $M$ (if any);

(iii) $\lambda_M$ fulfils the requirement of energy positivity with respect to every timelike Killing field in $M$
and, in the one-particle space, there are no zero modes for the self-adjoint generator of Killing-time displacements,

(iv)  $\lambda_M$ is Hadamard and therefore the state may be used as background 
for perturbative procedures (renormalisation in particular).

\noindent  
The statement (ii) holds as it stands replacing $\lambda_M$ with any other state $\lambda'_M$
uniquely defined by assuming that 
$\lambda'_M(a):= \lambda'(\imath(a))$ for all $a\in \cW(M)$
provided that $\lambda'$ be a BMS-invariant 
state (not necessarily quasifree or pure or satisfying some positivity-energy condition) defined on $\cW(\scri)$.\\
 The state $\lambda_M$ may have the natural interpretation of {\em outgoing scattering vacuum}, but also it provides 
 a natural and preferred notion of massless particle in the absence of Poincar\'e symmetry.
Indeed, all the construction works for massless conformally coupled scalar fields propagating in $M$.
Notice that the two notions of mass arising in our picture, that in the bulk based on properties of Klein-Gordon operator
(and on Wigner analysis if $M$ is Minkowski spacetime) and that referred to the extent on $\scri$
relying upon Mackey-McCarthy analysis of BMS group unitary representations, are in perfect agreement:
both vanish.
We do not see any obstruction to generalise all the results for other 
massless conformally invariant field equations.
However a natural question deserving future investigation is now: 
what about {\em massive} fields? How to connect, if possible,  massive particle defined in $M$ 
to fields on $\scri$ associated with known unitary BMS representations with positive BMS mass?
For an interesting attempt in that direction, specialized to Minkowski space, see recent Dappiaggi's paper \cite{DA07}.
Another technically interesting issue concerns the purity of the state $\lambda_M$: $\lambda$ is pure by definition,
but purity of $\lambda_M$ is not evident in the general case. Finally, it would be nice to describe interactions in the bulk, at least at perturbative level, by means of a theory
on $\scri$.\\
 
 \noindent {\bf Acknowledgements}. I would like to thank to R. M. Wald for
 technical suggestions (see footnote in the proof of Lemma \ref{lemmacompact}). I am grateful to A. Ashtekar for comments and
suggestions after the appearance of \cite{CMP5}. I am grateful to C. Dappiaggi, R. Brunetti and N. Pinamonti for useful comments. 
I would like to thank S. Hollands for having provided me with a copy of his unpublished Ph.D. thesis.
 
 \appendix

\section{Asymptotically flat spacetime with future time infinity}

 \definizione \label{dlast} {\em A time-oriented four-dimensional smooth spacetime $(M,g)$ 
is called {\bf asymptotically flat vacuum spacetime with future time infinity} $i^+$, if there is a smooth spacetime 
$(\tM,\tg)$ with a preferred point $i^+$, a diffeomorphism $\psi : M \to \psi(M) \subset \tM$  and a map $\Omega: \psi(M) \to [0,+\infty)$ so that
$\tg = \Omega^2 \psi^* g$ and the following facts hold. (We omit to write explicitly $\psi$ and $\psi^*$ in the following).\\
{\bf (1)} $J^-(i^+; \tilde{M})$ is closed and
$M = J^-(i^+)\setminus \partial J^-(i^+; \tilde{M})$.
(Thus $M= I^-(i^+; \tilde{M})$,  $i^+$ is in the future of
and time-like related with all the points of $M$ and $\scri \cap J^-(M; \tilde{M}) = \emptyset$.) 
Moreover $\partial M= \scri \cup \{i^+\}$ where
$\scri: = \partial J_-(i^+; \tilde{M}) \setminus \{i^+\}$ is the  {\bf future null infinity}. \\
{\bf (2)} $M$ is strongly causal and satisfies vacuum Einstein solutions in a neighbourhood of $\scri$ at least.\\
{\bf (3)} $\Omega$ can be extended to a smooth function on $\tM$.\\
{\bf (4)} $\Omega\spa \rest_{\partial J_-(i^+; \tilde{M})} =0$, but $d\Omega(x) \neq 0$ for $x\in \scri$, and 
  $d\Omega(i^+) = 0$, but $\tilde{\nabla}_\mu \tilde{\nabla}_\nu \Omega(i^+) = -2 \tg_{\mu\nu}(i^+)$.\\
{\bf (5)}  If $n^\mu:= \tg^{\mu\nu} \tilde{\nabla}_\nu \Omega$, for a strictly positive smooth function $\omega$, defined in a 
neighbourhood of $\scri$
 and satisfying $\tilde{\nabla}_\mu (\omega^4 n^\mu) =0$ on $\scri$,
the integral curves of $\omega^{-1}n$ are complete on $\scri$.}\\

\remark Notice that  $\omega$ in (5) can be fixed to be the factor $\omega_B$  mentioned in Section \ref{intro}.
The original definition due to Friedrich actually concerned the existence of the past time infinity $i^-$, our definition
is the trivial adaptation to the case of the existence of $i^+$.

  \section{Proofs  of some technical propositions.} \label{AWF}

\noindent {\bf Proof of (c) in Proposition \ref{prop2}}. Consider a one-parameter subgroup of $G_{BMS}$, $\{g_t\}_{t\in \bR} \subset \Sigma$. Suppose that
$\{g_t\}_{t\in \bR}$ arises from the integral curves of a complete smooth vector $\tilde{\xi}$ tangent to $\scri$.
In every Bondi frame $(u,\z,\bz)$ one finds:
$g_t : \bR \times \bS^2 \ni (u,\z,\bz) \mapsto \left(u+ f_t(\z,\bz), \z, \bz\right)\:,$
where, due to smoothness of $\tilde{\xi}$ and because of standard theorems of  ordinary differential equations, 
the function $(t, u,\z,\bz) \mapsto 
u+f_t(\z,\bz)$ is jointly smooth. In particular $f$ is jointly smooth and thus continuous in the parameter $t$, 
satisfies
 $f_t \in C^\infty(\bS^2)\equiv \Sigma$ and verifies
$f_{t}(\z,\bz) + f_{t'}(\z,\bz) =  f_{t+t'}(\z,\bz)$ for all $t,t'\in \bR$ and $(\z, \bz) \in \bS^2$.
The relation above entails
 $f_{\frac{p}{q}t}(\z,\bz)  = \frac{p}{q} f_{t}(\z,\bz)$ for all $t\in \bR$,
  $p,q\in \bZ$, $q\neq 0$ and $(\z,\bz) \in \bS^2$.
Using continuity in $t$ one finally gets:
$af_{t}(\z,\bz) + bf_{t'}(\z,\bz) =  f_{at+bt'}(\z,\bz)$ for all $t,t'\in \bR$,
 $a,b \in \bR$ and $(\z, \bz) \in \bS^2$.
 Therefore it holds:
 $f_{t}(\z,\bz) = tf_{1}(\z,\bz)\:.$
 We conclude that if the one-parameter sub-group 
 $\{g_t\}_{t\in \bR} \subset \Sigma \subset G_{BMS}$ arises from the complete integral curves of a smooth vector $\tilde{\xi}$ tangent 
 to $\scri$, in any fixed Bondy frame:
 $$g_t : \bR \times \bS^2 \ni (u,\z,\bz) \mapsto \left(u+ tf_1(\z,\bz), \z, \bz\right)\:,$$
where the function $f_1 \in C^\infty(\bS^2)\equiv \Sigma$ individuates completely the subgroup. $\Box$\\

   \noindent {\bf Proof of Proposition \ref{prop2bis}}. (a) is an immediate consequence of (a) in Proposition \ref{prop2} and the definition
  of asymptotic symmetry.  (b) Since the extension of $\xi$ to $\scri$, $\tilde{\xi}$,
  has to be tangent to $\scri$, referring to a fixed Bondi frame, it must hold
  $$\tilde{\xi} = \alpha \partial/\partial u + \beta \partial /\partial \z + \overline{\beta}  \partial /\partial \bz\:.$$
  Since the angular part of the degenerate metric on $\scri$ is positive, whereas that on the space spanned by
  $\partial/\partial u$ (which is orthogonal to the angular part) vanishes, one has
   $\tg(\tilde{\xi},\tilde{\xi}) \geq 0$ --
  with $\tg(\tilde{\xi},\tilde{\xi}) = 0$ if and only if $\beta=\overline{\beta}=0$. On the other hand we know that
  $g(\xi,\xi)\leq 0$ in $M$ by hypotheses and thus $\tg(\xi,\xi) \leq 0$ as well. 
Hence approaching $\scri$ it must be $\tg(\tilde{\xi},\tilde{\xi}) = 0$ by continuity. We have found that:
  $\tilde{\xi}(u,\z,\bz) = \alpha(u,\z,\bz) \partial/\partial u$. 
The (generally local) one-parameter group of transformations $g_t$ obtained by integration of $\tilde{\xi}$ acts only on the variable $u$:
$u \mapsto u_t$  and so it has to hold
\beq\frac{du_t(u,\z,\bz)}{dt}|_{t=0} = \alpha(u,\z,\bz)\:.\label{ora}\eeq
On the other hand this one-parameter group must coincide with a suitable one-parameter subgroup of BMS group
because $\tilde{\xi}$ is a one-parameter generator of such an action by (a) in Proposition \ref{prop2}.
By comparison 
  with the action (\ref{u})-(\ref{z}) of BMS group on coordinates $(u,\z,\bz)$,
  noticing that the subgroup leaves fixed the angular coordinates, the only possible action 
  is $u_t= u+ f(t,\z,\bz)$ for some smooth class of functions $\{f(t,\cdot,\cdot)\}_{t\in \bR}\subset C^\infty(\bS^2)$.
Therefore 
  $$\frac{du_t(u,\z,\bz)}{dt}|_{t=0} =\frac{\partial f(t,\z,\bz)}{\partial t}|_{t=0}\:.$$
  Comparing with (\ref{ora}) we conclude that $\alpha$ cannot depend on $u$.
   (b) in Proposition \ref{prop2} also entails that $\alpha$ cannot vanish identically
  on $\scri$. 
  In other words, $\tilde{\xi}$ is a generator of a nontrivial subgroup of $\Sigma$.
  Next, by (b) in Proposition \ref{prop2} we conclude that $\tilde{\xi}$ is a generator of a nontrivial subgroup of $T^4$.
  That is equivalent to say that $\alpha \in T^4\setminus\{0\}$.
  To conclude, as a consequence of by (c) of Proposition \ref{prop3},  it is sufficient to prove that $\alpha$ cannot assume both
   signs.
  Since $\xi$ is future directed with respect to $(M,g)$ and $(\tM,\tg)$,
   the limit values of $\xi$ toward $\scri$, $\alpha\partial/\partial u$ must either vanish or 
    be future directed. Since $\partial/\partial u$ is future directed 
  with respect to $(\tM,\tg)$ too, the factor given by the smooth function $\alpha$ cannot be negative anywhere. $\Box$\\

\noindent {\bf Proof of Proposition \ref{asympprop}}. In this proof $\Omega_B := \omega_B \Omega$. Under our hypotheses on $\xi$
and $\tilde{\xi}$, consider a smooth vector field 
$v$ defined on $\tM$
which reduces to $\xi$ in $M$ and reduces to $\tilde{\xi}$ on $\scri$. By construction the (jointly smooth in both arguments)
 one-parameter group of diffeomorphisms generated by $v$, $g^{(v)}$, reduces to those 
 generated by the relevant restrictions of $v$: $g^{(\xi)}$ and $g^{(\tilde{\xi})}$. The orbits of $v$ in $M \cup \scri$ are complete by construction.
 Indeed, if an orbit starts in $M$ it remains in $M$ and it is complete by hypotheses, 
if it starts on $\scri$ it must remain in $\scri$ and must be complete anyway,
since $\tilde{\xi}$ generates a (complete) one-parameter subgroup of $G_{BMS}$. This fact entails, in turn, that
the one-parameter group of diffeomorphisms generated by $v$ in $M \cup \scri$ is global and thus its pull-back
action on functions defined over $M \cup \scri$ is well defined.
 If $y\in \scri$ and $x\in M$ one has, by continuity of the flux of $v$: 
$$\lim_{x\to y} g^{(\xi)}_{\tau}(x) = \lim_{x\to y} g^{(v)}_{\tau}(x) = g^{(v)}_{\tau}(y) =g^{(\tilde{\xi})}_{\tau}(y)\:.$$
In the proof of Proposition 2.7 in \cite{DMP} (within a more generalised context) we have found that,
 referring to a Bondi-frame where $g^{(\tilde{\xi})}_{\tau} = \left(\Lambda^{(\tilde{\xi})}_{\tau},
f^{(\tilde{\xi})}_{\tau} \right)$ and $y \equiv (u,\z,\bz)$,
$$\lim_{x\to (u,\z,\bz)}\frac{\Omega_B\left(g^{(\xi)}_{-t}(x)\right)}{\Omega_B(x)} = K_{\Lambda^{(\tilde{\xi})}_{-t}}
\left(g^{(\tilde{\xi})}_{-t}(u,\z,\bz)\right)^{-1}\:.$$
Therefore one has trivially that $\Gamma(\phi \circ g^{(\xi)}_{-t})(y)$ coincides with
$$ \lim_{x\to y} \frac{\phi\left(g^{(\xi)}_{-t}(x)\right)}{\Omega_B(x)}
= \lim_{x\to y} \frac{\phi\left(g^{(\xi)}_{-t}(x)\right)}{\Omega_B\left(g^{(\xi)}_{-t}(x)\right)}
\lim_{x\to y}\frac{\Omega_B\left(g^{(\xi)}_{-t}(x)\right)}{\Omega_B(x)}=K_{\Lambda^{(\tilde{\xi})}_{-t}}
\spa \left(g^{(\tilde{\xi})}_{-t}(u,\z,\bz)\right)^{-1}\sp\sp \psi\left(g^{(\tilde{\xi})}_{-t}(y)\right)\:.$$
Comparing with (\ref{ABMS}), we finally find that:
$\Gamma(\phi \circ g^{(\xi)}_{-t}) = A_{g^{(\tilde{\xi})}_t}(\psi)$ and 
this concludes the proof. $\Box$\\

\noindent{\bf Proof of Lemma \ref{lemmaLemme}}. (a) Using the definition of $\Gamma_M$  (see Proposition \ref{holographicproposition}) and the fact that $E$ maps
 compactly-supported smooth functions to  smooth solutions of Klein-Gordon equation with compactly-supported Cauchy data, it arises:
 \beq {\psi_h}(u,\z,\bz) =  \omega_B(u,\z,\bz)^{-1} \left(\lim_{\to \scri}\Omega^{-1} E(h)\right)\rest_\scri(u,\z,\bz)\:.
  \label{psih}\eeq 
 On the other hand since also $(\tM,\tg)$ is globally hyperbolic,  the causal propagator $\tilde{E}$
 for the massless conformally coupled Klein-Gordon operator $\tilde{P}$ in $(\tM, \tg)$ is well defined. 
 Using the following facts: (1) that
  $E$ and $\tilde{E}$
 are the difference of the advanced and retarded fundamental solutions in the corresponding spaces $(M,g)$
 and $(\tM, \tg = \Omega^2 g)$, and (2) that the following identity holds
 $$\tilde{P} (\Omega^{-1}\phi) = \Omega^{-3} P \phi$$
 and (3) that the causality relations are preserved under (positive) rescaling of the metric, 
 one achieves the following identity valid on $M$
 $$\Omega^{-1} E(h) = \tilde{E}(\Omega^{-3}h) \:, \quad \mbox{if $h\in C_0^\infty(M)$\:.}$$
 The right-hand side is anyhow smoothly defined also in the larger manifold $\tM$ and on $\scri$ in particular.
Therefore, exploiting Eq. (\ref{psih}), the expression of ${\psi_h}(u,\z,\bz)$
found above can be re-written into a more suitable form given by (\ref{CENTRAL}).
The singularity of $\Omega^{-3}$ on $\scri$ is harmless because the support of $h$ does not intersect $\scri$
by construction and $\Omega >0$ in $M$. Notice that $\supp (\Omega^{-3} h) = \supp h$ if $h\in C_0^\infty(M)$. \\
(b) To prove the thesis take $h\in C_0^\infty(M)$ and a compact $K\subset M$ with $\supp h \subset K$. We use Definition \ref{dlast} from now on.
By definition  $\tilde{E}(\Omega^{-3}h)$ 
(equivalently $\tilde{E}(h)$ since $\supp (\Omega^{-3} h) = \supp h$)
is supported in $J^+(\supp h;\tM) \cup J^-(\supp h;\tM)$
and thus in $J^+(K;\tM) \cup J^-(K;\tM)$. However
$J^-(K;\tM)$ has no intersection with $\scri$ since $K \subset I^-(i^+;\tM)$ and $\scri \subset \partial J^-(i^+;\tM)$,
we conclude that the support of the solution $\tilde{E}(\Omega^{-3}h)$ intersects $\scri$ in a set completely included in 
$J^+(K; \tM)$ and thus in  $\scri \cap J^+(K; \tM)$. As a consequence 
\beq \mbox{the support of 
$\psi_h =  \omega_B(u,\z,\bz)^{-1}\tilde{E}(\Omega^{-3} h)\spa\rest_\scri$ is included in $\scri \cap J^+(K; \tM)$.} \label{included}\eeq 
Now consider a spacelike Cauchy surface $S$ of $(\tM,\tg)$ with $K$ completely contained in the chronological
future of $S$ (such a Cauchy surface does exists due to global hyperbolicity of $(\tM,\tg)$
and because  $K$ is compact, it is sufficient to use any Cauchy foliation of $\bR \times S \equiv \tM$
taking the value of the smooth global time function $t\in \bR$ far enough in the past). 
Notice that the set $C:= S \cap (\scri \cup \{i^+\}) =  S \cap \partial J^-(i^+;\tM)$
is compact because it is a closed subset of 
$J^-(i^+;\tM) \cap J^+(S;\tM)$  which is compact since $(\tM,\tg)$
is globally hyperbolic (e.g. see \cite{Wald}). $C$ cannot contain $i^+$ because $i^+ \in I^+(K; \tM)$, $K \subset I^+(S;\tM)$
and $S$ is achronal.
Let $u_0 = \min_{C} u$, which is finite because the coordinate $u: \scri \to \bR$ is smooth and $C\subset \scri$ is compact.
By construction it arises that $J^+(K; \tM) \subset I^+(S; \tM) \subset J^+(S;\tM)$ and
so $\scri \cap J^+(K; \tM) \subset  \scri \cap J^+(S;\tM) \subset J^+(C;\tM)$. Since $u$ increases toward the future, we have 
\beq u(\scri \cap J^+(K; \tM))  \subset [u_0,+\infty)\label{suppp}\:.\eeq
 Therefore, by (\ref{included}), we have that $\psi_h$ vanishes for  $u<u_0$ due to (\ref{suppp}). $\Box$\\

\noindent {\bf Proof of Lemma \ref{bastard}}. The proof is essentially  that given for Lemma 4.4 in  \cite{CMP5}. There the smooth 
function $\Psi \in C^\infty(\tM)$ was specialised to the case $\Psi = \Gamma_M(\phi)$ for some $\phi \in \cS(M)$, 
however such a restriction
can be removed without affecting the proof as it is evident from the proof of the cited  lemma.
The improvement concerning the exponent $\beta$ is obtained by noticing that in the last estimation before 
Eq. (44) in \cite{CMP5}, $e^{-\lambda(4+\epsilon)}$ can be replaced by the improved bound  $e^{-\beta\lambda(4+\epsilon)}$
for every $\beta \in [1,2)$ provided the free parameter $\epsilon>0$ fulfils $\epsilon < 4(2-\beta)/(\beta+4)$. $\Box$\\

\noindent {\bf Proof of Lemma \ref{lemmacompact}}.
We use here the geometric structure defined in Definition \ref{dlast}. 
  $K:= \overline{N} \subset M$ (the closure being referred to $M$) is compact. Using the same procedure as in the proof 
  of (b) in
 Lemma \ref{lemmaLemme}, we obtain that $J^+(K; \tM) \cap \scri$ is contained in a set  of the form $[u_0, +\infty) \times \bS^2$.
 Thus the null geodesics of $\tM$ joining $\overline{N}$ and $\scri$ must  intersect $\scri$ in a set contained in  
 $[u_0, +\infty)\times \bS^2$.
 Let us prove that $[u_0, +\infty)$ can actually  be restricted to a compact $[u_0, u_f]$.
First of all we notice that the following statement holds:\\
(A) {\em If $p\in M$, there is no null geodesic (with respect to $(\tM,\tg)$) joining $p$
and $i^+$.}\\
 Indeed, suppose that there is such a geodesic $\gamma$ for some $p\in M$. As is known from the general theory 
of causal sets in globally  hyperbolic spacetimes  and the structure of the boundary of $J^\pm(x)$ (e.g \cite{Wald}), after starting from $i^+$, $\gamma$ must belong to $\partial J^-(i^+;\tM)\setminus \{i^+\} = \scri$ till it encounters its cut locus $c\in \scri$ where $\partial J^-(i^+;\tM)$ terminates along the direction of $\gamma$. 
We  conclude, in particular, that $c$ is the end point on $\scri$ of one of the null geodesics
 forming $\partial J^-(i^+;\tM)$.
  After $c$, $\gamma$ leaves 
 $\partial J^-(i^+;\tM)$, enters $M$
and reaches $p$.  In the portion of its trip which lies on $\scri$, with a corresponding subset of the domain for
its affine parameter $t\in (0,b]$, one has $\Omega(\gamma(t)) =0$ for definition of $\scri$. Therefore $\dot{\gamma}^\mu(t) \nabla_\mu \Omega(\gamma(t)) = \dot{\gamma}^\mu(t)  n_\mu(\gamma(t))= 0$. Finally, since $\dot{\gamma}$ is null as $n$ (and both do not vanish anywhere), it has to be $\dot{\gamma}(t) = f(t) n(\gamma(t))$ for some non vanishing smooth function $f$. In other words, the portion of $\gamma$ contained in $\scri$ is, up to a re-parameterisation, an integral line of $n$. Therefore $c$ is the (past) end point on $\scri$ of one of the integral lines
of $n$ forming $\scri$. This is in contradiction with the requirement (5) in Definition \ref{dlast} which implies that
the integral lines of $n$ cannot have endpoints on $\scri$.\\
We pass to conclude the proof of existence of $u_f$.
Suppose {\em per absurdum} that,  for the  compact set $K := \overline{N} \subset M$, $u_f$ does not exist, so that 
the null geodesics starting from $K$
can intersect $\scri$ arbitrarily close to $i^+$. In this case we can consider a sequence $\{\gamma_n\}$ 
of null geodesics through $K$ which  intersect $\scri$ in the corresponding points
$\{p_n\}$ and $p_n \to i^+$ as $n\to +\infty$.
However the following statement
holds:\\
(B) {\em  If the mentioned sequence of geodesics $\{\gamma_n\}$ exists, there is a null geodesic $\gamma$ from $K \subset M$ to $i^+$.}\\
 Statement (B) is in contradiction with the statement (A), hence there is no sequence $\{\gamma_n\}$ with the claimed
 properties  and thus $u_f$ must exist. 
\\ To demonstrate the statement (B)\footnote{The kind of argument to prove the
 statement (B)  was suggested to the author by R. M. Wald.}
consider the sequence $\{\gamma_n\}$ where the geodesics are extended maximally after $i^+$ and before $K$. Choose a $(\tM,\tg)$ spacelike  Cauchy surface $C$ through $i^+$, and normalise the null-geodesic tangents so that they have unit inner product with the normal to $C$. Let $x_n$ denote the intersection point of the null geodesic with $C$ and and let $k_n$ denote the normalized tangent at $x_n$. Then $\{(x_n,k_n)\}$ is a sequence in a compact subset of the tangent bundle, so there is a subsequence that converges to a point $(x,k)$. Clearly $x=i^+$. Let $\gamma$ be the maximally extended null geodesic individuated by  $(p,k)$ and we assume that
all the used geodesics start from $C$ with the value of the affine parameter $s_0=0$. Moreover, since $\tM$ is globally hyperbolic,  rescaling the metric $\tg$
with a strictly positive smooth factor, we can make complete every null geodesic (Theorem 6.5 in \cite{BEE}),
 without affecting the causal structure of $\tM$.  In this way we ignore problems of domains of the parameters 
of the geodesics.
Let $C'$ be a second Cauchy surface in the past of $K$.
Since $\gamma$ is causal, one has  $\gamma(s_1) \in C'$ for some $s>0$.
Consider an auxiliary Riemannian smooth metric defined on $\tM$ and denote by $d$ the distance associated 
with that metric -- whose metric balls, as is known, form a base of the pre-existent topology of $\tM$ --.  
Using the jointly continuous dependence of maximal solutions of differential equations (in this case on $T\tM$)
from the parameter describing the curves  and the initial data, and exploiting the fact that continuous functions defined on a compact set 
are uniformly continuous, we get easily the following statement: For every $\epsilon >0$, there is 
a natural $N_\epsilon$ such that
$d(\gamma(s), \gamma_n(s)) < \epsilon$ for all $s\in [0,s_1]$ if $n>N_\epsilon$. It is clear that, in this way, if $\gamma$ does not
intersect $K$, one can fix $\epsilon$ in order that  no $\gamma_n$ meets $K$ if $n>N_\epsilon$.
 This is in contradiction with the hypotheses on the curves $\gamma_n$.   $\Box$\\

\section{Fourier-Plancherel transform on $\bR\times \bS^2$.} \label{APPfourier}
 Define $\mS(\scri)$ 
as the complex linear space of the smooth functions $\psi:\scri \to \bC$ such that,
in a fixed Bondi frame,
$\psi$ with all derivatives vanish as $|u|\to +\infty$, uniformly in $\z,\bz$,
faster than $|u|^{-k}$, $\forall k\in \bN$.  The space
$\mS(\scri)$ 
generalises straightforwardly Schwartz' function space on $\bR^n$, $\mS(\bR^n)$. 
$\mS(\scri)$ can be equipped with the Hausdorff topology induced from the countable class of seminorms --
{\em they depend on the Bondi frame but the topology does not} -- $p,q,m,n \in \bN$,
$$||\psi||_{p,q,m,n} := \sup_{(u,\z,\bz)\in \scri} \left||u|^p \partial^q_u\partial^m_{\z}\partial^n_{\bz}\psi(u,\z,\bz)\right|
\:.$$
$\mS(\scri)$ is dense in both $L^1(\bR\times \bS^2, du\wedge \epsilon_{\bS^2}(\z,\bz))$ and 
$L^2(\bR\times \bS^2, du\wedge \epsilon_{\bS^2}(\z,\bz))$ (with the topologies of these spaces
which are weaker than that of $\mS(\scri)$),
 because it includes the dense space 
$C_0^\infty(\bR\times \bS^2; \bC)$
of smooth compactly-supported complex-valued functions. We also define the space of {\em distributions}
$\mS'(\scri)$ containing all the linear functionals from $\bR\times \bS^2$ to $\bC$ 
which are weakly continuous with respect to the topology of  $\mS(\scri)$. Obviously
$\mS(\scri)\subset \mS'(\scri)$ and $L^p(\bR\times \bS^2, du\wedge \epsilon_{\bS^2}(\z,\bz))
\subset \mS'(\scri)$ for $p=1,2$.
We introduce the Fourier transforms $\mF_\pm[f]$ of $f\in \mS(\scri)$
$$\mF_\pm[f](k,\z,\bz) := \int_{\bR} \frac{e^{\pm i ku}}{\sqrt{2\pi}} f(u,\z,\bz) 
du \:, 
\quad (k,\z,\bz)\in \bR \times \bS^2\:.$$ 
$\mF_\pm$  enjoy the properties listed below which are straightforward extensions of the analogs
for standard Fourier transform in $\bR^n$. The proof of the following theorem is in the Appendix 
\ref{AWF}\footnote{The statement of (6) in Theorem C1 in
Appendix C of \cite{CMP5} is erroneous, but this fact affects by no means 
the results achieved in \cite{CMP5} since that statement did not enter the paper anywhere.}. \\

\teorema \label{fourier}{\em The maps $\mF_\pm$ satisfy the following properties.\\
{\bf (a)} for all $p,m,n \in \bN$ and every $\psi \in \mS(\scri)$ it holds 
$$\mF_\pm\left[\partial^p_u \partial^m_\z \partial^n_{\bz} \psi\right](k,\z,\bz) = (\pm i)^p k^p \partial^m_\z \partial^n_{\bz} 
\psi 
\mF_\pm[\psi](k,\z,\bz)\:.$$
{\bf (b)} $\mF_\pm$ are continuous bijections onto $\mS(\scri)$ and $\mF_-= (\mF_+)^{-1}$.\\
{\bf (c)} If $\psi,\phi \in \mS(\scri)$ one has
\begin{align} 
&\int_\bR \overline{\mF_\pm[\psi](k,\z,\bz)}\mF_\pm[\phi](k,\z,\bz) dk =
\int_\bR \overline{\psi(u,\z,\bz)}\phi(u,\z,\bz) du \:,\: \mbox{for all $(\z,\bz) \in \bS^2$}\:,\label{3a}\\
&\int_{\bR\times \bS^2} \overline{\mF_\pm[\psi](k,\z,\bz)}\mF_\pm[\phi](k,\z,\bz) dk\wedge \epsilon_{\bS^2}(\z,\bz) =
\int_{\bR\times \bS^2} \overline{\psi(u,\z,\bz)}\phi(u,\z,\bz) du\wedge \epsilon_{\bS^2}(\z,\bz)\:.\label{3b}
\end{align}
{\bf (d)} If $T\in \mS'(\scri)$ the definition
$\mF_\pm{T}[f]:= T\left(\mF_\pm[f]\right)\:
 \mbox{for all
 $f\in \mS(\scri)$}$,
 is well-posed, gives rise to  the unique weakly continuous linear extension of 
 $\mF_\pm$ to $\mS'(\scri)$ and one has, with the usual definition of derivative of a distribution,
 $$\mF_\pm \left[\partial^p_u \partial^m_\z \partial^n_{\bz} T\right] = 
 (\pm i)^p k^p \partial^m_\z \partial^n_{\bz}\mF_\pm[T]\:, \quad \mbox{for all $p,m,n \in \bN$}\:.$$
{\bf (e) Plancherel theorem}. 
  $\mF_\pm$ extend uniquely to unitary transformations 
   from 
  $L^2(\bR\times \bS^2, du\wedge \epsilon_{\bS^2}(\z,\bz))$ to $L^2(\bR\times \bS^2, du\wedge \epsilon_{\bS^2}(\z,\bz))$
 and the extension of $\mF_-$ is the inverse of that of $\mF_+$.
 These extensions coincide respectively with the restrictions
 to $L^2(\bR\times \bS^2, du\wedge \epsilon_{\bS^2}(\z,\bz))$ of the action of $\mF_\pm$ on distributions. \\
{\bf (f)} If $\tilde{\mF}_{\pm} : L^2(\bR,du) \to L^2(\bR,du)$ denotes the standard Fourier transform on the line,
 for every $\psi \in L^2(\bR\times \bS^2, du\wedge \epsilon_{\bS^2}(\z,\bz))$ it holds:
\beq \mF_\pm[\psi](k,\z,\bz) = \tilde{\mF}_\pm(\psi(\cdot,\z,\bz))(k) \:, \quad 
\mbox{almost everywhere on $\bR\times \bS^2$.} \label{ff}\eeq
As a consequence, if $\psi,\phi \in L^2(\bR\times \bS^2, du\wedge \epsilon_{\bS^2}(\z,\bz))$, one may say that
almost everywhere in $(\z,\bz) \in \bS^2$:
\begin{eqnarray} \int_\bR \overline{\mF_\pm[\psi](k,\z,\bz)}\mF_\pm[\phi](k,\z,\bz) dk &=& 
\int_\bR \overline{\psi(u,\z,\bz)}\phi(u,\z,\bz) du \:.\label{3a'}
\end{eqnarray}
{\bf (g)} If $m\in \bN$ and $T\in \mS'(\scri)$,  $\mF_+ [T]$ is a measurable function satisfying 
$$\int_{\bR\times \bS^2} (1+ |k|^2)^m |\mF_+[T]|^2 dk\wedge \epsilon_{\bS^2}(\z,\bz)<+\infty$$
if and only if the $u$-derivatives of $T$
 in the sense of distributions, are measurable functions with
$$\partial^n_u T \in L^2(\bR\times \bS^2, du\wedge \epsilon_{\bS^2})\:,\:\: \mbox{for $n=0,1,\ldots, m$}. $$} \\

\noindent{\em Proof.} 
(a) and (b) the statements can be proved with the same procedure used in $\bR^m$ in 
Theorem IX.1 in
\cite{RS} with trivial changes,
passing $\z,\bz$-derivatives under the relevant symbols of integration in $dk$ and $du$ since it is allowed
by compactness of $\bS^2$ and fast $\z,\bz$-uniform decaying for large $|u|$. (\ref{3a}) is a trivial consequence 
of the analogous statement in $\bR^1$ noticing that if $f \in \mS(\bR \times \bS^2)$
then, form fixed $\z,\bz$, the restriction  $u \mapsto f(u,\z,\bz)$ is a function of $\mS(\bR)$. Hence (\ref{3b})
follows from (\ref{3a}) via Fubini-Tonelli's theorem using the $\z,\bz$-uniform decaying for large $u$ of the integrands
in both sides of (\ref{3a})
and the fact that $\bS^2$ has finite measure.  (d) has the same proof as the analog in $\bR^n$ in  Theorem IX.2
\cite{RS}.
 (e) Has the same proof as in the $\bR^n$ case (Theorem IX.6 in \cite{RS}) noticing that
(\ref{3b}) holds true  and  that $\mS(\scri)$ is dense in 
the Hilbert space $L^2(\bR\times \bS^2, du\wedge \epsilon_{\bS^2}(\z,\bz))$. 
The identity (\ref{ff}) in (f) is trivially fulfilled 
for $\psi \in \mS(\bR\times \bS^2)$ by construction. Moreover, by Plancherel's theorem
on $\bR$, if  $\psi \in L^2(\bR\times \bS^2, du\wedge \epsilon_{\bS^2}(\z,\bz))$
(so that its restrictions at $\z,\bz$ fixed belongs to $L^2(\bR, du)$ by Fubini-Tonelli's theorem), one has
$$ \int_\bR |\tilde{\mF}_\pm[\psi(\cdot,\z,\bz)](k)|^2 dk = 
\int_\bR |\psi(u,\z,\bz)|^2 du$$
almost everywhere in $\z,\bz$. By Fubini-Tonelli's theorem the right-hand side, and thus also the left-hand side
is $\z,\bz$ integrable. By Fubini-Tonelli's theorem one finally has that the integrands
are $u,\z,\bz$ jointly integrable so that:
$$ \int_{\bR\times\bS^2} |\tilde{\mF}_\pm(\psi(\cdot,\z,\bz))(k)|^2 dk \wedge
 \epsilon_{\bS^2}(\z,\bz) = \int_{\bR\times\bS^2} |\psi(u,\z,\bz)|^2 du\wedge \epsilon_{\bS^2}(\z,\bz)\:.$$
We conclude that the map that associates  every $\psi \in L^2(\bR\times \bS^2, du\wedge \epsilon_{\bS^2}(\z,\bz))$
with the function (in the same space) $(k,\z,\bz) \mapsto \tilde{\mF}_\pm(\psi(\cdot,\z,\bz))(k)$ is continuous 
and isometric
and coincides with $\mF_{\pm}$ in the dense subspace $ \mS(\bR\times \bS^2)$, therefore it must coincide with $\mF_\pm$ extended to
$L^2(\bR\times \bS^2, du\wedge \epsilon_{\bS^2}(\z,\bz))$. In other words (\ref{ff})  holds true.
Now  (\ref{3a'}) can be re-written replacing $\mF_\pm$ by $\tilde{\mF}_\pm$ and in this form is nothing but 
Plancherel's theorem on the real line.
The proof of (g) is immediate from (d) and (e).
$\Box$\\


\begin{thebibliography}{999}


\bibitem[AD05]{AD} G.~Arcioni and C.~Dappiaggi,  
{\em ``Exploring the holographic principle in asymptotically flat space-times via the BMS group},
 Nucl.\ Phys.\ B {\bf 674},  553, (2003)


\bibitem[AH78]{AH} A.~Ashtekar and R.~O.~Hansen:
{\em ``A unified treatment of null and spatial infinity in general relativity. I. 
Universal structure, asymptotic symmetries and conserved quantities at spatial infinity''}
J. Math. Phys {\bf 19}, 1542, (1978),


\bibitem[AX78]{AX} A.~Ashtekar and B.~C.~Xanthopoulos,
{\em ``Isometries compatible with asymptotic flatness at null infinity: a complete description.''}
J. Math. Phys. {\bf 19}, 2216, (1978),


\bibitem[Ar99]{Araki} H.~Araki:
{\em ``Mathematical Theory of Quantum Fields''},
Oxford University Press,
Oxford (1999),

\bibitem[As80]{AO} A.~Ashtekar in
{\em ``General Relativity and Gravitation 2: One Hundred Years after the birth of Albert Einstein''} edited by
A. Held, (Plenum, New York, 1980), pages 37--70,


\bibitem[AS81]{AS} A.~Ashtekar, M.~Streubel:
{\it ``Symplectic geometry of radiative modes and conserved quantities at null infinity''},
Proc. R. Lond. A  {\bf 376} (1981) 585,

\bibitem[BGP96]{BEE} J. K. Beem, P.E. Eherlich and K.L. Easley:
{\em ``Global Lorentzian Geometry''}  Second Edition 
Marcel Dekker, Inc. , New York, (1996),



\bibitem[BGP96]{BGP} C. B\"ar, N. Ginoux, F. Pf\"affle:
{\em ``Wave equations on Lorentzian manifolds and quantization''} (2006)
in press in {\em ESI Lectures in
Mathematics and Physics} by the European Mathematical Society Publishing House.


\bibitem[BR021]{BR}
O.~Bratteli, D.~W.~Robinson:
{\em ``Operator Algebras And Quantum Statistical Mechanics. Vol. 1: C* And W* Algebras,
Symmetry Groups, Decomposition Of States''}, second edition, second printing,
Springer-verl. New York, Usa, (2002),


\bibitem[BR022]{BR2} O.~Bratteli, D.~W.~Robinson:
{\em ``Operator algebras and quantum statistical mechanics. Vol. 2: Equilibrium
states. Models in quantum statistical mechanics''}, second edition, second printing,
Springer Berlin, Germany  (2002),

\bibitem[BFK96]{BFK} R. Brunetti, K. Fredenhagen, M. Kohler 
{\em ``The Microlocal spectrum condition and Wick polynomials of free fields on curved space-times.''}
 Commun.\ Math.\ Phys. {\bf 180}, 633 , (1996). 

 \bibitem[BF00]{BF} R.Brunetti, K. Fredenhagen, 
{\em  ``Microlocal analysis and interacting quantum field theories: Renormalization on physical backgrounds.''}
 Commun.\ Math.\ Phys. {\bf 208}, 623, (2000).

\bibitem[BFV03]{BFV} R.Brunetti, K. Fredenhagen, R. Verch.
{\em `` The Generally covariant locality principle: A New paradigm for local quantum field theory.''}
\ Commun.\ Math.\ Phys. {\bf 237} 31 (2003). 
 
 
\bibitem[Da04]{Da04} C.~Dappiaggi, {\em ``BMS field theory and holography in asymptotically flat space-times.''},
 JHEP \ {\bf 0411} \ 011, (2004).

\bibitem[Da05]{Da05}  C.~Dappiaggi, {\em ``Elementary particles, holography and the BMS group''},
 Phys.\ Lett.\ B {\bf 615},  291, (2005).
 
 
 \bibitem[Da06]{Da06}  C.~Dappiaggi, {\em ``Free field theory at null infinity and white noise calculus: a BMS invariant dynamical system''},
 arXiv:math-ph/0607055,
 
 \bibitem[Da07]{DA07}  C.~Dappiaggi, 
{\em ``Projecting massive scalar fields to null infinity''},
 arXiv:0705.0284 [gr-qc],

\bibitem[DMP06]{DMP}
C.~Dappiaggi, V.~Moretti and N. Pinamonti:
\emph{``Rigorous Steps towards Holography in Asymptotically Flat Spacetimes''},
Rev.\ Math.\ Phy. 18,  349 (2006).\\
arXiv:gr-qc/0506069,


\bibitem[Di80]{Dimock}
J.~Dimock:
{\em ``Algebras of Local Observables on a Manifold''},
Commun.\ Math.\ Phys.\  {\bf 77}, 219 (1980).


\bibitem[Fr75]{friedlander}
F.~G.~Friedlander,
{\em ``The wave equation on a curved space-time''},
 Cambridge, UK: Univ. Pr. (1975).

 \bibitem[Fri86-88]{Friedrich} H.~Friedrich: {\em ``On Purely Radiative Space-Times''},
Commun.\ Math.\ Phys. {\bf 103}, 35 (1986);
 {\em ``On the Existence of $n$-Geodesically Complete or Future Complete Solutions of Einstein's Field Equations
with smooth Asymptotic Structure''}
Commun.\ Math.\ Phys. {\bf 107}, 585 (1986); {\em ``On Static and  Radiative Space-Times''},
Commun.\ Math.\ Phys. {\bf 119}, 51 (1988),

\bibitem[Ge77]{Geroch} R. Geroch, in: P. Esposito, L. Witten (Eds.) \emph{``Asymptotic Structure of Spacetime''}, 
Plenum, New York (1977),


\bibitem[HW01]{HW1} S. Hollands and R.M. Wald,
{\em ``Local Wick polynomials and time ordered products of quantum fields in curved space-time.''}
Commun.\ Math. \ Phys. {\bf 223}, 289 (2001). 


\bibitem[HW04]{HW2} S. Hollands and R.M. Wald,
{\em ``Conservation of the stress tensor in perturbative interacting quantum field theory in curved spacetimes.''}
\ Rev. \ Math. \ Phys. {\bf 17}, 227 (2005). 

  
  
  
\bibitem[Ho00]{Hollands} S. Hollands, {``Aspects of Quantum Field Theory in Curved Spacetime''},
 Ph.D.thesis (University of York, 2000), advisor B.S. Kay,  unpublished.


\bibitem[H\"o89]{Hor} L.~H\"ormander,
{``The Analysis of Linear Partial Differential Operators I''},
second edition,
Springer-Verlag, Berlin, Germany (1989).

\bibitem[H\"or71]{Hor1} L.~H\"ormander,
{\em ``Fourier integral operators. I''}
Acta \ Math. \ {\bf 127}, 79 (1971).


\bibitem[KW91]{KW} B.~S.~Kay, R.~M.~Wald,
{\em ``Theorems On The Uniqueness And Thermal Properties Of Stationary, Nonsingular,
Quasifree States On Space-Times With A Bifurcate Killing Horizon'',}
Phys.\ Rept.\  {\bf 207}, 49 (1991).

\bibitem[KND77]{KND} M. Ko, E.T. Newmann and K.T.Tod in: P. Esposito, L. Witten (Eds.) \emph{``Asymptotic Structure of Spacetime''}, 
Plenum, New York (1977).



\bibitem[Le53]{Leray} J.~Leray, {\em ``Hyperbolic Differential Equations'',}
Unpublished Lecture Notes, Princeton (1953),


\bibitem[MC72-75]{Mc} P.J. McCarthy: \emph{''Representations of the Bondi-Metzner-Sachs
group I''} Proc. R. Soc. London {\bf A330} (1972) 517; \emph{''Representations of the Bondi-Metzner-Sachs
group II''} Proc. R. Soc. London {\bf A333} (1973) 317; \emph{``The Bondi-Metzner-Sachs in the nuclear
topology''} Proc. R. Soc. London {\bf A343} (1975) 489,


\bibitem[Mo03]{stress} V.~Moretti:
{\em ``Comments on the stress-energy tensor operator in curved spacetime''}
Commun. \ Math.\ Phys. {\bf 232}, 189 (2003). 
arXiv: gr-qc/0109048


\bibitem[Mo06]{CMP5} V.~Moretti:
{\em ``Uniqueness theorems for BMS-invariant states of scalar QFT on the null boundary of asymptotically flat 
spacetimes and bulk-boundary observable algebra correspondence''},  
Commun.\ Math.\ Phys.  {\bf 268}, 727 (2006). arXiv:gr-qc/0512049,

\bibitem[O'N83]{O'Neill} B.~O'Neill,
 {``Semi-Riemannian Geometry with applications to Relativity''},
 New York, Academic Press, USA  (1983),


\bibitem[Pe63]{Penrose} R. Penrose: \emph{``Asymptotic Properties of Space and Time''} Phys. Rev. Lett. {\bf 10} (1963) 66,

\bibitem[Pe74]{Penrose2} R. Penrose, in: A.O. Barut (Ed.), \emph{``Group Theory in Non-Linear Problems''}, Reidel, 
Dordrecht (1974), p. 97 chapter 1,


\bibitem[Ra96a]{Rada}
M.~J.~Radzikowski:
{\em ``Micro-local approach to the Hadamard condition in quantum field theory on curved space-time''},
Commun.\ Math.\ Phys.\  {\bf 179}, 529 (1996),

\bibitem[Ra96b]{Radb}
M.~J.~Radzikowski:
{\em ``A Local to global singularity theorem for quantum field theory on curved space-time''},
Commun.\ Math.\ Phys.\  {\bf 180}, 1 (1996),

\bibitem[RS75]{RS} M.~Reed and B.~Simon: {\em ``Methods of Modern Mathematical Physics''} vol. II
Fourier Analysis, Self-Adjointness, Academic Press, New York (1975).

\bibitem[SV01]{SV} H.Sahlmann and R.Verch:  {\em ``Microlocal spectrum condition and Hadamard 
form for vector valued quantum fields in curved space-time''}.
 Rev. \ Math. \ Phys. {\bf 13} 1203, 2001. 

\bibitem[SVW02]{SVW}A.Strohmaier, R.Verch, M.Wollenberg: {\em `` Microlocal analysis of quantum fields on curved space-times: 
Analytic wavefront sets and Reeh-Schlieder theorems.''}.  J.\ Math.\ Phys. {\bf 43}, 5514, 2002. 

\bibitem[Wa84]{Wald} 
R.~M.~Wald: 
{\it ``General Relativity''},
Chicago University Press, Chicago (1984).


\bibitem[Wa94]{Wald2}
R.~M.~Wald:
{\it``Quantum field theory in curved space-time and black hole thermodynamics''},
 Chicago, USA: Univ. Pr. (1994).




\end{thebibliography}
\end{document}